\documentclass[article]{elsarticle}
\usepackage{amsmath}
\usepackage{amssymb}
\usepackage{graphicx}
\usepackage{subcaption}
\usepackage{booktabs}
\usepackage{multirow}
\usepackage[usenames, dvipsnames]{color}
\begin{document}
\journal{Frontiers in Physics}
\begin{frontmatter}
\title{Value of the axial-vector coupling strength in $\beta$ and $\beta\beta$ decays: A review}
\author[rvv]{Jouni Suhonen\corref{cor1}}
\ead{jouni.suhonen@phys.jyu.fi}
\cortext[cor1]{Corresponding author}
\address[rvv]{University of Jyvaskyla, Department of Physics, P.O. Box 35,
FI-40014, Jyvaskyla, Finland}

\date{}

\begin{abstract}
In this review the quenching of the weak axial-vector coupling strength, $g_{\rm A}$, is
discussed in nuclear $\beta$ and double-$\beta$ decays. On one hand, the 
nuclear-medium and nuclear many-body effects are separated,
and on the other hand the quenching is discussed from the points of view of different
many-body methods and different $\beta$-decay and double-$\beta$-decay processes. Both
the historical background and the present status are reviewed and contrasted against
each other. The theoretical considerations are tied to performed and planned measurements,
and possible new measurements are urged, whenever relevant and doable. Relation of
the quenching problem to the measurements of charge-exchange reactions and muon-capture
rates is pointed out.
\end{abstract}

%\pacs{21.10.Pc, 21.10.Re, 21.60.-n, 23.40.-s}
\end{frontmatter}
%\maketitle

%%%%%%%%%%%%%%%%%%%%%%%%%%%%%%%%%%%%%%%%%%%%%%%%%%%%

\section{Introduction \label{sec:intro}}

The neutrinoless double beta ($0\nu\beta\beta$) decays of atomic nuclei are
of great experimental and theoretical interest due to their implications of
physics beyond the standard model of electroweak interactions. Since
these processes occur in nuclei, nuclear-structure effects play an important
role and they may affect considerably the decay rates. The nuclear effects
are summarized as the nuclear matrix elements (NMEs) containing 
information about the initial and final states of the nucleus and the action
of the $0\nu\beta\beta$ transition operator on them. The NMEs, in turn, are computed
numerically using some nuclear-theory framework suitable for the nuclei under
consideration. The possible future detection of the $0\nu\beta\beta$ decay in 
the next generation of $\beta\beta$ experiments
constantly drives nuclear-structure calculations towards better 
performance. Accurate knowledge of the NMEs is required in order that the 
data will be optimally used to obtain information about the fundamental 
nature and mass of the neutrino 
\cite{REPORT,Vergados2012,Vergados2016,Engel2017,Deppisch:2012nb,Rodejohann:2011mu,Rodejohann2012}. 
In addition, the $0\nu\beta\beta$ decay relates also
to the breaking of lepton-number symmetry and the baryon asymmetry of the Universe 
\cite{Deppisch:2015yqa, Deppisch:2004kn}. A number of nuclear models, including
configuration-interaction based models like the 
interacting shell model (ISM), and various mean field models, have been adopted for the 
calculations. The resulting computed NMEs have been analyzed in the review articles 
\cite{Engel2017,Suhonen2012d,Vogel2012,Engel2015}. 
Most of the calculations have been done by the use  
of the proton-neutron quasiparticle random-phase approximation (pnQRPA) \cite{Suhonen2012c}.

The performed $0\nu\beta\beta$-decay calculations, as also those of the two-neutrino 
double beta ($2\nu\beta\beta$) decay, indicate that the following nuclear-structure
ingredients affect the values of NMEs: 
\begin{itemize}
\item{(a)} The chosen valence space of single-particle orbitals and their nucleon occupancies 
\cite{Suhonen2008b,Suhonen2010,Suhonen2011b}, 
\item{(b)} the effects stemming from the shell closures \cite{Suhonen2012d,Barea2009}. These
closures are formed by the bunching of single-particle orbitals in the nuclear
mean-field potential to form the so-called major shells that are separated by large 
energy gaps. The gaps occur at ``magic numbers'' of nucleons and have sometimes drastic
effects on nuclear properties.
\item{(c)} The nuclear deformation and seniority truncation. 
\cite{Sarriguren2001,Alvarez2004,Caurier2008b,Menendez2009a,Rodriguez2010}. In ground states
of even-even (even number of protons and neutrons) nuclei all nucleons are paired
to angular momentum zero and form a superfluid-like state with total angular momentum
zero. This is called seniority-zero state. If one pair is broken, extra angular momentum
is generated and this contributes to excited states of nuclei. These are called 
seniority-two states. Breaking more pairs generates higher-seniority states that can
mix with the lower-seniority states by the nuclear residual interaction. Cutting the
higher-seniority contributions, i.e., performing a seniority truncation, simplifies 
calculations considerably.
\item{(d)} Also, it has to be noted that the adopted closure approximation, i.e. omitting
the energy dependence of the involved energy denominator and
replacing the contributions coming from the intermediate virtual states by a unit
operator (for all other nuclear models, except for the quasiparticle random-phase 
approximation, QRPA), for the $0\nu\beta\beta$-decay calculations does not hold for
the calculations of the $2\nu\beta\beta$-decay rates \cite{REPORT,Tomoda1991,Faessler1998}.
\item{(e)} A further important aspect can be added to the list, namely the 
uncertain value of the weak axial-vector coupling strength 
$g_{\rm A}$, leading to an effective value of $g_{\rm A}$ in nuclear-model 
calculations. This deviation (usually quenching) from the free-nucleon value can arise from 
the \emph{nuclear medium effects} and
the \emph{nuclear many-body effects} described in more detail in the following
sections of this review.
\end{itemize}

At the nuclear level, $\beta$ decay can be considered as a mutual interaction of the 
hadronic and leptonic currents mediated by massive vector bosons $W^{\pm}$ 
\cite{Commins1983}. The leptonic and hadronic currents can be expressed as mixtures 
of vector and axial-vector contributions \cite{Feynmann1958a,Feynmann1958b,Theis1958}. 
The weak vector and axial-vector coupling strengths $g_{\rm V}$ and 
$g_{\rm A}$ enter the theory when the hadronic current is renormalized at the nucleon level 
\cite{Zuber2004}. The conserved vector-current hypothesis (CVC) \cite{Feynmann1958a}
and partially conserved axial-vector-current hypothesis (PCAC) 
\cite{Nambu1960,Gell-Mann1960} yield the free-nucleon values $g_{\rm V}=1.00$ and 
$g_{\rm A}=1.27$ \cite{Commins1983} but inside nuclear matter the value of  
$g_{\rm A}$ is affected by many-nucleon correlations and a \emph{quenched} or 
\emph{enhanced} value might be needed to reproduce experimental observations 
\cite{Blin-Stoyle1975,Khanna1978,Haxton1995,Suhonen2007}. Precise information 
on the effective value of $g_{\rm A}$ is crucial 
when predicting half-lives of neutrinoless double beta decays since the half-lives are 
proportional to the fourth power of $g_{\rm A}$ \cite{REPORT,Maalampi2013}. 

Since the vector bosons $W^{\pm}$ have large mass and thus propagate only a short distance, 
the hadronic current and the leptonic current can be considered to interact at a 
point-like weak-interaction vertex with an effective coupling strength $G_{\rm F}$, 
the Fermi constant. The parity non-conserving nature of the weak interaction forces 
the hadronic current to be written at the quark level 
(up quark $u$ and down quark $d$) as a mixture of vector and axial-vector parts: 
\begin{equation}
J_H^{\mu} = \bar{u}(x)\gamma^{\mu}(1-\gamma_5)d(x)\,,
\label{eq:h-current}
\end{equation} 
where $\gamma^{\mu}$ are the usual Dirac matrices and 
$\gamma_5=i\gamma^0\gamma^1\gamma^2\gamma^3$. 
Renormalization effects of strong interactions and energy scale of the processes
must be taken into account when moving from the quark level to the hadron level. 
Then the hadronic current between nucleons (neutron $n$ and proton $p$) takes the rather
complex form 
\begin{equation}
J_H^{\mu} = \bar{p}(x)\lbrack V^{\mu} - A^{\mu}\rbrack n(x)\,,
\label{eq:h-current-eff}
\end{equation} 
where the vector-current part can be written as
\begin{equation}
V^{\mu} = g_{\rm V}(q^2)\gamma^{\mu} + \textrm{i}g_{\rm M}(q^2)
\frac{\sigma^{\mu\nu}}{2m_{\rm N}}q_{\nu}
\label{eq:V-current}
\end{equation} 
and the axial-vector-current part as
\begin{equation}
A^{\mu} = g_{\rm A}(q^2)\gamma^{\mu}\gamma_5 + g_{\rm P}(q^2)q^{\mu}\gamma_5 \,.
\label{eq:A-current}
\end{equation} 
Here $q^{\mu}$ is the momentum transfer, $q^2$ its magnitude, $m_{\rm N}$ the
nucleon mass (roughly 1 GeV) and the weak couplings depend on the magnitude of the
exchanged momentum. For the vector and axial-vector couplings one usually adopts the
dipole approximation
\begin{align} 
\label{eq:dipole}
   g_{\rm V}(q^2) = \frac{g_{\rm V}}{\big(1+q^2/M_{\rm V}^2\big)^2} \ ;\ 
   g_{\rm A}(q^2) = \frac{g_{\rm A}}{\big(1+q^2/M_{\rm A}^2\big)^2} \,,
\end{align}
where $g_{\rm V}$ and $g_{\rm A}$ are the weak vector and axial-vector 
coupling strengths at zero momentum transfer ($q^2=0$), respectively. For the vector and 
axial masses one usually takes $M_{\rm V}=84\,\textrm{MeV}$ \cite{Bodek2008}
and $M_{\rm A}\sim 1\,\textrm{GeV}$ \cite{Bodek2008,Bhattacharya2011,Amaro2015} coming
from the accelerator-neutrino phenomenology.
For the weak magnetism term one can take $g_{\rm M}(q^2)=(\mu_p-\mu_n)g_{\rm V}(q^2)$ and for
the induced pseudoscalar term it is customary to adopt the Goldberger-Treiman relation
\cite{Goldberger1958a}
$g_{\rm P}(q^2)=2m_{\rm N}g_{\rm A}(q^2)/(q^2+m_{\pi}^2)$, where $m_{\pi}$ is the pion mass and
$\mu_p-\mu_n=3.70$ is the anomalous magnetic moment of the nucleon.
It should be noted that the $\beta$ decays and $2\nu\beta\beta$ decays are low-energy
processes (few MeV) involving only the vector [first term in Eq.~(\ref{eq:V-current})] and 
axial-vector [first term in Eq.~(\ref{eq:A-current})] parts at the limit $q^2 = 0$
so that the $q$ dependence of Eq.~(\ref{eq:dipole}) does not play any role in the treatment
of these processes in this review. Contrary to this, the $0\nu\beta\beta$ decays and
nuclear muon-capture transitions involve momentum transfers of the order of
$100\,\textrm{MeV}$ and the full expression (\ref{eq:h-current-eff}) is active with slow
decreasing trend of the coupling strengths according to Eq.~(\ref{eq:dipole}).

\section{Effective values of $g_{\rm A}$: Preamble \label{sec:preamble}}

The effective value of $g_{\rm A}$ can simply be characterized by a 
\emph{renormalization factor} $q$ (in case of quenching of the value of $g_{\rm A}$ it is
customarily called \emph{quenching factor}):
\begin{align} 
\label{eq:q}
	q = \frac{g_{\rm A}}{g^{\rm free}_{\rm A}} \,,
\end{align}
where 
\begin{align} 
\label{eq:g-free}
	g^{\rm free}_{\rm A} = 1.2723(23)
\end{align}
is the free-nucleon value of the axial-vector coupling measured in 
neutron beta decay \cite{Patrignani2016} and $g_{\rm A}$ is the value of the axial-vector
coupling derived from a given theoretical or experimental analysis. This derived $g_{\rm A}$
can be called the \emph{effective} $g_{\rm A}$ so that from (\ref{eq:q}) one obtains for
its value
\begin{align} 
\label{eq:g-eff}
	g^{\rm eff}_{\rm A} = qg^{\rm free}_{\rm A} \,.
\end{align}
Eqs.~(\ref{eq:q})$-$(\ref{eq:g-eff}) constitute the basic definitions used in this review.

The effective value of $g_{\rm A}$ can be derived from several different experimental and
theoretical analyses. In these analyses it is mostly impossible to separate the different
sources of renormalization affecting the value of $g_{\rm A}$: (i) the meson-exchange currents
(many-body currents) that are beyond the one-nucleon impulse approximation (only one 
nucleon experiences the weak decay without interference from the surrounding nuclear medium), 
usually assumed in the theoretical calculations, (ii) other nuclear medium effects like
interference from non-nucleonic degrees of freedom, e.g. the $\Delta$ isobars
and (iii) the deficiencies in the nuclear many-body approach
that deteriorate the quality of the wave functions involved in the decay processes.

The effects (i) and (ii) can be studied by performing calculations using meson-exchange models
and allowing non-nucleonic degrees of freedom in the calculations. These calculations
that go beyond the nucleonic impulse approximation are described in Sec.~\ref{sec:medium-eff}
in the context of Gamow-Teller $\beta$ decays for which the related effects are measurable. 
The calculations yield a fundamental quenching factor 
$q_{\rm F}$ and the related fundamentally renormalized effective $g_{\rm A}$ for the space 
components ($\mu =1,2,3$) of the axial current (\ref{eq:A-current}) via the effects of the
virtual pion cloud around a nucleon. The time component of $\mu=0$, the axial charge $\rho_5$, 
is, however, fundamentally enhanced by, e.g., heavy meson exchange and the corresponding
effective coupling $g_{\rm A}^{\rm eff}(\gamma_5)$ is discussed in Sec.~\ref{subsec:FF-nu},
in the context of first-forbidden $0^+\leftrightarrow 0^-$ transitions for which the 
effect is measurable.

The interacting shell model has the longest history behind it in studies of the axial quenching
in Gamow-Teller $\beta$ decays. The reason for this is the success of the ISM to describe
nuclear spectroscopy of light nuclei and the rather large amount of data on these type
of allowed $\beta$ decays. The results of these studies are presented in 
Sec.~\ref{sec:GT}. In the same section the ISM results are compared with those obtained
by the use of the pnQRPA. In 
Sec.~\ref{subsec:FF-u} the effective value of $g_{\rm A}$ is analyzed for the first-forbidden
unique $\beta$ decays for which there are some experimental data available. In 
Sec.~\ref{sec:high-forb-u} this study is extended to higher-forbidden unique $\beta$ decays
where no experimental data are available and one has to resort to mere theoretical
speculations. In Sec.~\ref{sec:F-nu} the forbidden non-unique $\beta$ decays are discussed.
Experimentally, there are available data for the above-mentioned first-forbidden non-unique
$0^+\leftrightarrow 0^-$ and other $\beta$ transitions. For the higher-forbidden non-unique
transitions, discussed in Sec.~\ref{subsec:high-forb-nu},
there are scattered half-life and $\beta$-spectrum data but more measurements
are urgently needed, in particular for the shapes of the $\beta$ spectra. Unfortunately,
in all these studies it is not possible to completely disentangle the nuclear-medium effects (i) 
and (ii) from the nuclear-model effects (iii). 

In the last two sections, 
Sec.~\ref{sec:spin-multipole} and Sec.~\ref{sec:mucapture}, more exotic methods to
extract the in-medium value of $g_{\rm A}$ are presented: The spin-multipole strength
functions and nuclear muon capture. Measurements of the spin-multipole strength
functions, in particular the location of the corresponding giant resonances, help 
theoretical calculations fine-tune the parameters of the model Hamiltonians such that
the low-lying strength of, say $2^-$ states, is closer to reality. Hence, more such 
measurements are called for. The nuclear muon capture probes the axial current (\ref{eq:A-current})
at 100 MeV of momentum transfer and thus suits perfectly for studies of the renormalization
of the NMEs related to $0\nu\beta\beta$ decays. This means that muon-capture experiments 
for medium-heavy nuclei are urgently needed.

The renormalization of $g_{\rm A}$ which stems from the nuclear-model effects (iii) depends on
the nuclear-theory framework chosen to describe the nuclear many-body wave functions
involved in the weak processes, like $\beta$ and $\beta\beta$ decays. This is why the
effective values of $g_{\rm A}$ can vary from one nuclear model to the other.
On the other hand, the different model frameworks can give surprisingly similar results
as witnessed in Sec.~\ref{subsec:high-forb-nu} in the context of the comparison
of the measured $\beta$ spectra with the computed ones. The renormalization of $g_{\rm A}$
can also depend on the process in question. For the zero-momentum-exchange ($q^2 = 0$) 
processes, like $\beta$ and $2\nu\beta\beta$ decays, the renormalization can be different
from the high-momentum-exchange ($q^2\sim 100$ MeV) processes, like $0\nu\beta\beta$
decays (in Sec.~\ref{subsec:high-forb-nu} the related $g_{\rm A}$ is denoted as
$g^{\rm eff}_{{\rm A},0\nu}$) or nuclear muon captures.

This introduction to the many-faceted renormalization of the axial-vector coupling is
supposed to enable a ``soft landing'' into the review that follows. As can be
noticed, the renormalization issue is far from being solved and lacks a unified picture
thus far. There is not yet a coherent effort to solve the issue, but rather some
sporadic attempts here and there. The most critical issue may be the nuclear many-body
deficiencies (iii) that hinder a quantitative assessment of the nuclear-medium effects (i)
and (ii) in light, medium-heavy and heavy nuclei. Only gradually this state of affairs will 
improve with the progress in the ab-initio nuclear methods extendable to nuclei beyond
the very lightest ones. Hence, the lack of perfect nuclear many-body theory is reflected
in this review as a wide collection of different effective $g_{\rm A}$ variants, 
different for different theory frameworks and processes and not necessarily connected to 
each other (yet). The
hope is that in the future the different studies would point to one common low-energy
renormalization of $g_{\rm A}$ for the $\beta$ and two-neutrino $\beta\beta$ decays and that we
would have some idea about the renormalization mechanisms at work in the case of 
the neutrinoless $\beta\beta$ decays.

On the other hand, there are some attempts to disentangle the nuclear medium effects from
the nuclear many-body effects. Examples are the fundamental quenching elaborated in
Sec.~\ref{sec:medium-eff} and the nuclear-medium-independent quenching factor $k$
introduced in Sec.~\ref{subsec:QRPA} for the Gamow-Teller $\beta$ decays, and in 
Sec.~\ref{subsec:FF-u} and Sec.~\ref{sec:high-forb-u} for the unique-forbidden
$\beta$ transitions. This factor is designed to give hints about the impact of
the changes in the complexity of the nuclear model on the value of the effective axial 
coupling. Also the previously mentioned method based on the examination of $\beta$ spectra in 
Sec.~\ref{subsec:high-forb-nu} is largely nuclear-model independent and seems to be
a reasonable measure of the nuclear-medium effects (i) and (ii). More measurements of
the $\beta$ spectra are thus urgently called for.

\section{Nuclear-medium effects \label{sec:medium-eff}}

Based on the early shell-model studies of Gamow-Teller $\beta$ decays, effects of
$\Delta$ resonances and meson-exchange currents on the weak axial-vector coupling strength 
of the space part, $\mathbf{A}$, of the axial current $A^{\mu}$ (\ref{eq:A-current}) 
is expected to be \emph{quenched} in nuclear medium and finite nuclei. 
Contrary to this, the coupling strength of the 
time part, $A^0$, of (\ref{eq:A-current}) is expected to be \emph{enhanced} 
by, e.g., the contributions coming from exchanges of heavy mesons. Many of these 
modifications in the strengths of the axial couplings stem from processes beyond the 
\emph{impulse} approximation where only one nucleon at a time is experiencing a weak
process, e.g. $\beta$ decay, without interference from the surrounding nuclear medium.
In fact, based on general arguments concerning soft-pion amplitudes \cite{Kubodera1978}
the space part of $A^{\mu}$ is quenched and the time part of $A^{\mu}$ is enhanced 
relative to the single-particle processes of the impulse approximation.

The origin of the quenching of the space part of $A^{\mu}$ is not completely known
and various mechanisms have been proposed for its origin: studied have been the
$\Delta$-isobar admixture in the nuclear wave function \cite{Suhonen1991},
shifting of Gamow-Teller strength to the $\Delta$-resonance region, and renormalization
effects of meson-exchange currents. The $\beta^-$ and $\beta^+$ 
Gamow-Teller strengths were related to the $\Delta$-isobar region e.g. in 
\cite{Delorme1982} and sizable $\Delta$-resonance effects on $\beta$ decays of
low-lying nuclear states by tensor forces were reported in \cite{Oset1979,Bohr1981}.
In \cite{Towner1979,Towner1983} very simple nuclear systems were used to
study the tensor force and related effects in order to minimize the impact of nuclear
many-body complexities. Studied were the tensor effects and their interference with the
$\Delta$-isobar current and meson-exchange currents in building up corrections to
the Gamow-Teller matrix elements. Also relativistic corrections to the Gamow-Teller
operator were included. Large cancellations among the various contributions were 
recorded and corrections below some 20\% were obtained for the light (simple) nuclei.
However, recent experimental studies of (p,n) and (n,p) reactions 
\cite{Ichimura2006} report that the 
$\Delta$-nucleon-hole admixtures into low-lying nuclear states play only a minor role
in the quenching of $g_{\rm A}$, in line with the results of \cite{Suhonen1991}.
Also extended sum rules have been derived for relating $g_{\rm A}$ to pion-proton total
cross sections \cite{Adler1965,Weisberger1966,Kim1966}, or the method
of QCD sum rules has been utilized \cite{Henley1992}. 

In \cite{Wilkinson1974} the renormalization of the $\beta$-decay operator 
by the two- or many-nucleon correlations, in terms of inter-nucleonic and intra-nucleonic
mesonic currents, leads to the notion effective ``\emph{fundamentally}'' renormalized axial 
coupling $g_{\rm AeF}$. The quenching of $g_{\rm A}$ is then described by the fundamental
quenching factor $q_{\rm F}$ such that $q_{\rm F}>q$ since $q$ contains, in addition, the 
quenching stemming from the inadequate treatment of the nuclear many-body problem. From
here on the above notation is adopted for the renormalization of $g_{\rm A}$ stemming from the 
(fundamental) mesonic-current effects. 

In the early study of M. Ericson \cite{Ericson1971} of the sum rule for Gamow-Teller 
matrix elements a (fundamental) quenching of roughly
\begin{align} 
\label{eq:q-Ericson}
	q_{\rm F} = 0.9 
\end{align}
was obtained for very light nuclei ($A\le 17$) by the examination of the effects
of meson-exchange currents on the pion-nucleon interaction vertex and extending the
result to a sum rule for Gamow-Teller matrix elements. This (practically) 
model-independent study produces the following (fundamentally) renormalized value 
of the axial coupling strength
\begin{align} 
\label{eq:g-eff-Ericson}
	g_{\rm AeF} = 0.9\times 1.27 = 1.1 \,.
\end{align}
The above result does not necessarily apply to individual Gamow-Teller transitions
between low-energy nuclear states. 

The work of \cite{Ericson1971} was followed by the works \cite{Ericson1973,Rho1974} where
it was found that the renormalization should be universal for all transitions, in particular 
applicable to the mentioned Gamow-Teller transitions at low nuclear excitations. The
procedure bases on the fact that the partially conserved axial current (PCAC) hypothesis
\cite{Nambu1960,Gell-Mann1960} enables one to calculate the full axial-current 
matrix element in terms of a pion-nucleus vertex \cite{Blin-Stoyle1967}. At the 
low-momentum-exchange limit, relevant for the nuclear $\beta$ decays, the PCAC leads 
to the Goldberger-Treiman relation \cite{Goldberger1958a,Goldberger1958b} 
which relates the effective value of $g_{\rm A}$ to the effective
value of the pionic coupling constant $g_{\pi}$ by \cite{Ericson1971}
\begin{align} 
\label{eq:gee-pi}
	\frac{g^{\rm eff}_{\rm A}}{g^{\rm eff}_{\pi}} = \frac{g^{\rm free}_{\rm A}}{g^{\rm free}_{\pi}}
= \frac{f_{\pi}}{\sqrt{2}m_{\rm N}} \,,
\end{align}
where $m_{\rm N}$ is the nucleon mass and $f_{\pi}=0.932 m_{\pi}$ is the pion decay constant,
$m_{\pi}$ being the pion mass. The pionic coupling constant is, in turn, renormalized
by the effects on the virtual pion field by the presence of other nucleons. For large
nuclei (surface effects can be omitted) the renormalization arises from nucleonic 
short-range correlations leading to voids between nucleons and the renormalization 
can be understood via an electromagnetic analog: an
electric dipole in a correlated dielectric medium is renormalized in a similar way as the
pionic coupling constant. There is also a connection to the low-energy scattering of
pions on nuclei: the short-range correlations quench the p-wave pion-nucleon
amplitude by the same amount as the dielectric effect. For finite nuclei a model-dependent
surface factor has to be taken into account \cite{Ericson1971}. The size renormalization 
emerges from the nuclear surface layer of a thickness of the order of the pion Compton 
wavelength and thus the quenching of $g_{\rm A}$ increases with increasing 
nuclear radius and, as a consequence, with increasing nuclear mass.

In \cite{Rho1974} the pion-nucleus
vertex was calculated and the related quenched $g_{\rm A}$ agreed with the one of 
\cite{Ericson1971} to leading order. In \emph{infinite nuclear matter} This
quenching turns out to be \cite{Rho1974}
\begin{align} 
\label{eq:q-Rho}
	q^{\infty}_{\rm F} = 0.76 \quad (\textrm{infinite nuclear matter})
\end{align}
leading to the quenched effective axial coupling strength
\begin{align} 
\label{eq:g-eff-Rho}
	g^{\infty}_{\rm AeF} = 0.76\times 1.27 = 0.96 \,. \quad (\textrm{infinite nuclear matter})
\end{align}
in infinite nuclear matter.

The works of Ericson \cite{Ericson1971,Ericson1973} and Rho \cite{Rho1974} were used by 
Wilkinson \cite{Wilkinson1974} to bridge the gap between the infinite nuclear matter and
finite nuclei. In \cite{Wilkinson1974} it was argued that the fundamental quenching can
be described by the formula
\begin{align} 
\label{eq:g-finite-A}
	q_{\rm F} = \sqrt{(q^{\infty}_{\rm F})^2 + \left[ 1-(q^{\infty}_{\rm F})^2\right]/A^{0.17}}
\end{align}
for finite nuclei of mass number $A$. This formula includes the short-range correlation
effect and the finite-size factor \cite{Ericson1973,Rho1974} and
gives for the fundamental quenching,
using (\ref{eq:q-Rho}), between $A=50-150$ the value $q_{\rm F}=0.88$. This
means that the fundamental quenching is practically constant over the range of
nuclei of interest to the double beta decay. The corresponding fundamentally quenched 
value of the axial-vector strength is plotted in Fig.~\ref{fig:gA-ranges1}, and its
value is practically 1.1 through the whole range of interest.

In \cite{Siiskonen2001} the renormalization of the axial current (and vector and
induced pseudoscalar terms of the nucleonic current) was studied for
several nuclear systems as a function of transition energy by including
effective transition operators up to second order in perturbation theory. Thus
the renormalization of $g_{\rm A}$ contains both the fundamental and nuclear many-body
aspects. It was found that the renormalization was practically constant up to 60 MeV 
in transition energy, in agreement with the $q$ dependence of $g_{\rm A}$ in relation
(\ref{eq:dipole}). The obtained quenchings are as follows
\begin{align} 
\label{eq:g-eff-Siis}
	g^{\rm eff}_{\rm A} = 1.0\ (1s0d \textrm{ shell})\ ;\ 0.98\ (1p0f \textrm{ shell})\ ;
  \ 0.71\ (^{56}\textrm{Ni})\ ;\ 0.52\ (^{100}\textrm{Sn}) \,.
\end{align}
The results (\ref{eq:g-eff-Siis}), obtained by using the nuclear-medium-corrected 
transition operators have been repeated in Table~\ref{tab:ISM} of
Sec.~\ref{subsec:ISM} and Fig.~\ref{fig:gA-ranges2} of Sec.~\ref{subsec:QRPA}
in order to compare them with the more phenomenological shell-model results.
Effective operators have also been used in the connection with the calculations
for the double beta decays in a solvable model \cite{Engel2004} and for the
nucleus $^{92}$Mo \cite{Suhonen1997b} and the nuclei
$^{76}$Ge and $^{82}$Se \cite{Engel2009,Holt2013} in the framework of the interacting shell
model.

As speculated in \cite{Wilkinson1974}, the mesonic effects (meson-exchange currents)
show up as effective two-body contributions to the $\beta$-decay operators. These 
two-body currents quench $g_{\rm A}$ and this quenching was first estimated in
\cite{Menendez2011}, in the framework of the chiral effective field theory (cEFT) where
both the weak currents and nuclear forces can be described on the same footing
and to a given order of approximation (leading order, next-to-leading order, etc.) In
\cite{Menendez2011} the two-body currents were replaced by an effective one-body current 
derived from the cEFT, leading to a momentum-dependent effective coupling 
$g^{\rm eff}_{\rm A}(q^2)$, renormalized with respect to the bare axial coupling of
(\ref{eq:dipole}). It turned out that the additional quenching is caused by 
the short-range nucleon-nucleon coupling present in the original two-body current. 
The additional quenching decreases with increasing $q$, being
the strongest at the zero-momentum-transfer limit, affecting mostly the nuclear
$\beta$ and $2\nu\beta\beta$ decays. In fact, the strength of the short-range nucleon-nucleon
coupling in the two-body current can be adjusted such as to reproduce the empirical quenching 
of the Gamow-Teller $\beta$ decays discussed in Sec.~\ref{sec:GT}. As the $0\nu\beta\beta$ 
decay is a high-momentum-transfer process ($q\sim 100\,\mathrm{MeV}$) it is expected that
the two-body currents have not such a drastic effect on the one-body current 
(\ref{eq:A-current}) for the $0\nu\beta\beta$ decay. Here it should be noted that
the one-body current (\ref{eq:h-current-eff}) has been fully 
taken into account in \emph{all} $0\nu\beta\beta$-decay calculations and the two-body 
currents introduce a renormalization, $g^{\rm eff}_{\rm A}(q^2)$, that deviates from the
one-body dipole $g_{\rm A}(q^2)$ of (\ref{eq:dipole}) the less the higher the momentum
exchange $q$ is. The quenching caused by the two-body currents could probably be measured
by using charge-exchange reactions \cite{Ichimura2006} 
in advanced nuclear-physics infrastructures.

In \cite{Menendez2011} it was estimated, by using the interacting shell model 
(ISM) many-body framework in the mass range $A=48-136$, that the effect of the two-body 
currents on the value of the $0\nu\beta\beta$ NME is between -35\% and 10\% 
depending on the (uncertain) values of the cEFT parameters, the smallest corrections
occurring for $A=48$. In \cite{Engel2014} the effect of the two-body currents was 
studied in the framework of the proton-neutron quasiparticle random-phase
approximation (pnQRPA) in the mass range $A=48-136$, and a quenching effect of 
$10-22$\% was obtained for the $0\nu\beta\beta$ NMEs, the 10\% effect pertaining to
the case of $^{48}$Ca. A more complete calculation, including three-nucleon forces and 
consistent treatment of the two-body currents and the nuclear Hamiltonian, was performed
in \cite{Ekstrom2014}. Application to the Gamow-Teller $\beta$ decays in $^{14}$C and
$^{22,24}$O nuclei yielded the quenching $q=0.92-0.96$ by comparison of the computed
strengths to that of the Ikeda $3(N-Z)$ sum rule \cite{Suhonen2007,Ikeda1963}. 
This less than 10\% quenching is in
line with the trend observed in the studies \cite{Menendez2011,Engel2014} where the quenching
approaced the 10\% limit for light nuclei. It should be noted that the
two-body meson-exchange currents appear also in 
neutrino-nucleus scattering \cite{Katori2016} but at energies where two nucleons
are ejected as a result of the scattering (the so-called two-particle-two-hole
exchange currents). The higher energy evokes considerable difficulties in handling
the two-body meson-echange currents, as demonstrated in \cite{Simo2016}.

The meson-exchange currents can cause also enhancement phenomena, like in the case of the
renormalization of the one-body weak axial charge density $\rho_5$ 
[time part of $A^{\mu}$ in (\ref{eq:A-current})] in the case of the 
$0^-\leftrightarrow 0^+$ nuclear $\beta$ transitions \cite{Kubodera1978,Kirchbach1988}. 
In this case the $\gamma^5$ operator mediates the first-forbidden non-unique $\beta$ 
transition and the corresponding axial-vector coupling strength is enhanced quite
strongly. In the work of \cite{Kirchbach1988} the effects of a pionic two-body part of 
$\rho_5$ was studied for 4 nuclear masses and the corresponding leading single-particle 
transitions. This work was extended by Kirchbach \textit{et al.} \cite{Kirchbach1992} and
Towner \cite{Towner1992} by taking into account also the heavy-meson exchanges. 
In \cite{Towner1992} 6 nuclear masses and a number of single-particle transitions 
were computed by using nuclear wave functions from the ISM.
An interesting investigation of the role of the two-particle-two-hole excitations in
the $A=16$ nuclei was performed in \cite{Towner1981}: The renormalization of
the weak axial charge by the meson-exchange currents had to be taken into account in
order to explain the measured rates of both the $0^-\to 0^+$ $\beta$ decay and the 
$0^+\to 0^-$ muon capture. The axial-charge enhancement is elaborated further,
quantitatively, in Sec.~\ref{subsec:FF-nu}.

Very recently break-through results in the calculations of the axial charge and
axial-vector form factors have been achieved in the lattice QCD (quantum chromodynamics) 
calculations \cite{Bhattacharya2016,Berkowitz2017,Gupta2017}. In the work 
\cite{Berkowitz2017} the result
\begin{align} 
\label{eq:g-lattice}
	g^{\rm free}_{\rm A} = 1.278(21)(26) \quad\quad  (\textrm{lattice calculation})
\end{align}
was obtained, where the first uncertainty is statistical and the second comes from the
extrapolation systematics. This computed value is quite compatible with the measured free
value of $g_{\rm A}$ in (\ref{eq:g-free}). Also the lattice QCD calculations of the
double beta decay are advancing in the two-nucleon (toy) systems, see \cite{Tiburzi2017}.

\section{Nuclear-model effects \label{sec:model-eff}}

The studies on the effective value of the axial-vector coupling
strength, $g_{\rm A}$, have mainly been performed for $\beta$ decays in established 
nuclear many-body frameworks. Also the magnetic moments of nuclei have been studied
\cite{Shimizu1974,Hyuga1980} for simple one-particle and one-hole nuclei in order
to pin down the effects of the \emph{tensor force} in shifting low-energy strength
of Gamow-Teller type to higher energies, and thus effectively quenching the 
spin-isospin operator for Gamow-Teller decays. 
The used many-body frameworks encompass the interacting
shell model (ISM) \cite{Caurier2005} and the proton-neutron quasiparticle 
random-phase approximation (pnQRPA) \cite{Suhonen2012c,Ring1980}. Also the frameworks
of the microscopic interacting boson model (IBM-2) 
\cite{Iachello1987} and the interacting boson-fermion-fermion model, 
IBFFM-2 \cite{Iachello1991}, have been used.
Let us discuss next the various many-body aspects of these models that may affect 
the (apparent) renormalization of the magnitude of $g_{\rm A}$. It is appropriate to note 
here that in all these studies the nuclear many-body framework can be considered
more or less deficient and thus the many-body effects cannot be disentangled
from the nuclear-medium effects, discussed in Sec.~\ref{sec:medium-eff}.

\subsection{Many-body aspects of the ISM \label{subsec:model-ISM}}

The ISM is a many-body framework that uses a limited set of single-particle states,
typically one harmonic-oscillator major shell or one nuclear major shell, to describe nuclear
wave functions involved in various nuclear processes. The point of the ISM is to
form all the possible many-nucleon configurations in the given single-particle space, 
each configuration described by one Slater determinant, and diagonalize the nuclear 
(residual) Hamiltonian in the basis formed by these Slater determinants. In this way the
many-body features are taken into account exactly but only in a limited set of
single-particle states. The problem is to extend the single-particle space beyond
the one-shell description due to the factorially increasing size of the sparse Hamiltonian
matrix to be diagonalized. In this way only the low-energy features of a nucleus can be
described, leaving typically the giant-resonance region out of reach. The other problem
with the ISM is to find a suitable (renormalized) nucleon-nucleon interaction to match
the limited single-particle space. Since this space is small, the renormalization
effects of the two-body interaction become substantial. Typically, mostly in the early 
works, all the matrix elements of the two-body interaction were fitted such that
the computed observables, energies, electromagnetic decays, etc., are as close
as possible to the corresponding measured ones (see Sec.~\ref{subsec:ISM}). In some
works also perturbative approaches through particle-hole excitations from the
valence to the excluded space have been considered 
(see, e.g., \cite{Kuo1990,Hjorth-Jensen1995,Engeland2014} and the references therein).

From early on there have been difficulties for the ISM to reproduce the measured
$\beta$-decay rates \cite{Towner1997}. This has lead to a host of
investigations of the effective (quenched) value of $g_{\rm A}$ in the ISM framework
(see Sec.~\ref{subsec:ISM} below). The main limitation of the ISM is its confinement
to small single-particle spaces, typically comprising one oscillator major shell
or a magic shell, leaving one or two spin-orbit partners out of the model space.
From, e.g., pnQRPA calculations \cite{Suhonen2010,Suhonen2011b} and 
perturbative ISM calculations \cite{Holt2013,Kwiatkowski2014}
one knows that inclusion of all spin-orbit partners in the single-particle model 
space is quite essential. This has been noticed also in the extended ISM
calculations where the missing spin-orbit partners have been included at least
in an effective way \cite{Caurier2008,Horoi2013}. Even extension of the ISM to include
two harmonic-oscillator shells ($1s0d$ and $1p0f$ shells) has been done
for the calculation of the $0\nu\beta\beta$ decay of $^{48}$Ca \cite{Iwata2016}.

Several advanced shell-model methods have been devised in order to include larger
single-particle spaces into the calculations. One can try to find clever ways to select
the most important configurations affecting the observables one is interested
in. Such an established algorithm is the Monte Carlo shell model (MCSM) where
statistical sampling of the Slater determinants is used \cite{Abe2012,Togashi2016}.
One can also use importance-truncation schemes \cite{Stumpf2016} or very advanced
\textit{ab initio} methods, like the coupled-cluster theory, where the two- and
three-body interactions can be derived from the chiral effective field theory (cEFT)
\cite{Jansen2014}. One can also use the in-medium similarity renormalization group 
(IM-SRG) method, like in \cite{Bogner2014}, where an \textit{ab initio} construction of a
nonperturbative $1s0d$-shell Hamiltonian, based on cEFT two- and three-body forces, has
been done.
Another new method is the density matrix renormalization group (DMRG) algorithm
\cite{Legeza2015}, which exploits optimal ordering of the proton and neutron 
single-particle orbitals and concepts of quantum-information theory. 

All the new methods extend the traditionally used ISM model spaces and the 
future $\beta$-decay calculations using these methods will either confirm or reduce 
the amount of quenching of $g_{\rm A}$ observed in the older ISM calculations, described 
in Sec.~\ref{subsec:ISM} below. The \textit{ab initio} methods are already available for 
the light nuclei, occupying the $0p$ and $1s0d$ shells, and later for the medium-heavy
and heavy nuclei dwelling in the higher oscillator shells. The quenching problem
can only be solved by using many-body methods with error estimates, including a 
systematic way to improve their accuracy. At the same time the two- and three-body
forces used in the calculations should be produced on the same footing as the many-body
framework itself, preferably from \textit{ab initio} principles. One should not forget
that also the operators used in the computations should be made effective operators
that match the adopted single-particle valence spaces. Using these prescriptions one
can eliminate the deficiencies of the nuclear many-body framework and obtain
information about the quenching of $g_{\rm A}$ in the nuclear medium
(see Sec.~\ref{sec:medium-eff}), beyond the effects caused by the deficiencies 
of a nuclear model.

\subsection{Many-body aspects of the pnQRPA \label{subsec:model-pnQRPA}}

The random-phase approximation (RPA) is an extension of the Tamm-Dancoff model (TDM) in 
the description of magic nuclei (at closed major shells) by particle-hole excitations 
across the magic gaps between closed nuclear major shells
\cite{Suhonen2007,Rowe1970}. In the RPA the simple particle-hole vacuum, with the
single-particle orbitals fully occupied up to the Fermi surface at the magic gap, is
replace by the correlated vacuum, containing two-particle--two-hole, 
four-particle--four-hole, etc. excitations across the magic gap. The use of the
correlated vacuum in the RPA enhances the strength of collective transitions 
\cite{Suhonen2007,Rowe1970}. Its quasiparticle version, quasiparticle RPA (QRPA)
describes open-shell nuclei, outside the closures of magic shells, by
replacing the particle-hole excitations by two-quasiparticle excitations. Usually
these quasiparticles are generated by the use of the Bardeen-Cooper-Schrieffer (BCS)
theory \cite{Lane1964} from the short-range interaction part of the nuclear Hamiltonian
in an even-even \emph{reference nucleus}.
The quasiparticles can be viewed as partly particles and partly holes, inducing
fractional occupancies of the nuclear single-particle orbitals and leading to a
smeared Fermi surface for protons and/or neutrons for open-shell nuclei. The
proton-neutron version of the QRPA (pnQRPA) uses two-quasiparticle excitations that
are built from a proton and a neutron quasiparticle. This enables description of
odd-odd nuclei starting from the even-even BCS reference nucleus.

The strong point of the pnQRPA theory is that it can include large single-particle
valence spaces in the calculations. There are no problems associated with
leaving spin-orbit-partner orbitals out of the computations. On the other hand,
the pnQRPA has a limited configuration space, essentially including two-quasiparticle
excitations on top of a correlated ground state \cite{Suhonen2007}. Deficiencies
of the pnQRPA formalism have been analyzed against the ISM formalism, e.g., in
\cite{Menendez2009a} by using a seniority-based scheme (seniority was defined 
earlier, at point (c) in Sec. \ref{sec:intro}). In that work the pnQRPA
was considered to be a low-seniority approximation of the ISM. But one the other
hand, the ground-state correlations of the pnQRPA introduce higher-seniority
components to the pnQRPA wave functions and the deficiencies stemming from the
incomplete seniority content of the pnQRPA should not be so bad \cite{Escuderos2010}.
Also the renormalization problems of the two-body interaction are not so severe as
in the ISM due to the possibility to use large single-particle model spaces. On the
other hand, it is harder to find a perturbative scheme for the effective Hamiltonian
due to the incompleteness of the available many-body configuration space. Due to
this, schematic or G-matrix-based boson-exchange Hamiltonians have widely been used
(see Sec.~\ref{subsec:QRPA}).

In any case, the configuration content of the pnQRPA is limited and extensions
and improvements of the theory framework are wanted in order to see how the quenching 
problem of $g_{\rm A}$ evolves with these extensions and improvements. Such
extensions have been devised, including, e.g., the renormalized QRPA (RQRPA)
\cite{Toivanen1995,Toivanen1997} and similar ``fully'' renormalized schemes
\cite{Raduta1998,Raduta2010,Raduta2011}. Another possible improvement of the pnQRPA is
the relativistic quasiparticle time-blocking approximation (RQTBA), in particular its
proton-neutron version, the pn-RQTBA, advocated in \cite{Robin2016}. It
shows good promise for improvements over the $\beta$-decay calculations of the 
ordinary pnQRPA the use of which clearly points out to need for a quenched value
of $g_{\rm A}$ in $\beta$-decay calculations, as discussed in Sec.~\ref{subsec:QRPA}.

The (charge-conserving) QRPA framework, with linear combinations of proton-proton 
and neutron-neutron quasiparticle pairs, \emph{phonons} \cite{Suhonen2007}, can 
be used to describe (collective) excitations of even-even nuclei (collectivity is where
the name phonon stems from). These, in turn, 
can be used as \emph{reference} nuclei in building the excitations of the 
neighboring odd-mass (odd-proton or odd-neutron) nuclei by coupling the
QRPA phonons with proton or neutron quasiparticles. This phonon-quasiparticle
coupling can be carried out in a microscopic way, based on a realistic effective
residual Hamiltonian. This has been achieved, e.g. in the microscopic 
quasiparticle-phonon model (MQPM) \cite{Toivanen1995b,Toivanen1998} where a microscopic
effective Hamiltonian based on the Bonn G matrix has been used to produce the
one- and three-quasiparticle states in odd-mass nuclei. This extension of the QRPA has
been used to describe $\beta$ decays, and in particular in connection with the 
renormalization problem of $g_{\rm A}$, as discussed in Sec.~\ref{subsec:high-forb-nu}. 

It should be noted that odd-mass nuclei can also be described by starting from an
odd-odd reference nucleus, described by the pnQRPA phonons \cite{Suhonen2007}. 
By coupling either proton or neutron quasiparticles with pnQRPA phonons one can, again,
create the states of either a neutron-odd or a proton-odd nucleus. This approach
was coined the proton-neutron MQPM (pnMQPM) and was used to describe forbidden
beta decays in \cite{Mustonen2007}. Although the pnQRPA-based phonons better take into
account the Ikeda sum rule \cite{Suhonen2007,Ikeda1963} and the Gamow-Teller giant-resonance
region of the $\beta^-$-type strength function, the pnMQPM lacks the important 
three-proton-quasiparticle and three-neutron-quasiparticle contributions, essential
for good reproduction of the low-energy spectra of odd-mass nuclei. This is why
its use in $\beta$-decay calculations has been very limited.

\subsection{Many-body aspects of the IBM \label{subsec:model-IBA}}

In its simplest version, the interacting boson model (IBM), the theory framework
consists of $s$ and $d$ bosons which have as their microscopic paradigms
the $0^+$ and $2^+$ coupled collective Fermion pairs present in nuclei. Even a mapping
of the collective Fermion pairs to these bosons can be devised \cite{Iachello1987}.
An extension of the IBM is the microscopic IBM (IBM-2) where the proton and neutron
degrees of freedom are explicitly separated. The IBM and IBM-2 are sort of
phenomenological versions of the ISM, containing the seniority aspect and the 
restriction to one magic shell in terms of the single-particle valence space. The
Hamiltonian and the transition operators are constructed from the $s$ and $d$ bosons 
as lowest-order boson expansions with coupling coefficients to be determined by
fits to experimental data or by relating them to the underlying fermion valence space
through a mapping procedure \cite{Otsuka1978,Otsuka1996}. Thus the IBM and its 
extensions use more or less phenomenological operators mimicking the renormalized 
operators used in the ISM (see Sec.~\ref{subsec:model-ISM}).

The two versions of the IBM can be extended to include higher-multipole bosons, like
$g$ bosons, as well. Further extension concerns the description of odd-mass nuclei
by the use of the interacting boson-fermion model (IBFM) and its extension, the
microscopic IBFM (IBFM-2) \cite{Iachello1991}. The IBM concept can also be used to
describe odd-odd nuclei by using the interacting boson-fermion-fermion model (IBFFM)
and its proton-neutron variant, the proton-neutron IBFFM (IBFFM-2) \cite{Brant1988}.
Here the problems arise from the interactions between the bosons and the one or
two extra fermions in the Hamiltonian, and from the transition operators containing 
a host of phenomenological parameters to be determined in some way.
The IBM-2 and the IBFFM-2 have been used to access the renormalization of $g_{\rm A}$, as
described in Sec.~\ref{subsec:IBM-bb}.

\section{Effective value of $g_{\rm A}$ in allowed Gamow-Teller $\beta$ decays \label{sec:GT}}

\begin{figure}[htbp!]
%\begin{center}
\begin{subfigure}{.33\textwidth}
\centering
\includegraphics[width=1.0\linewidth]{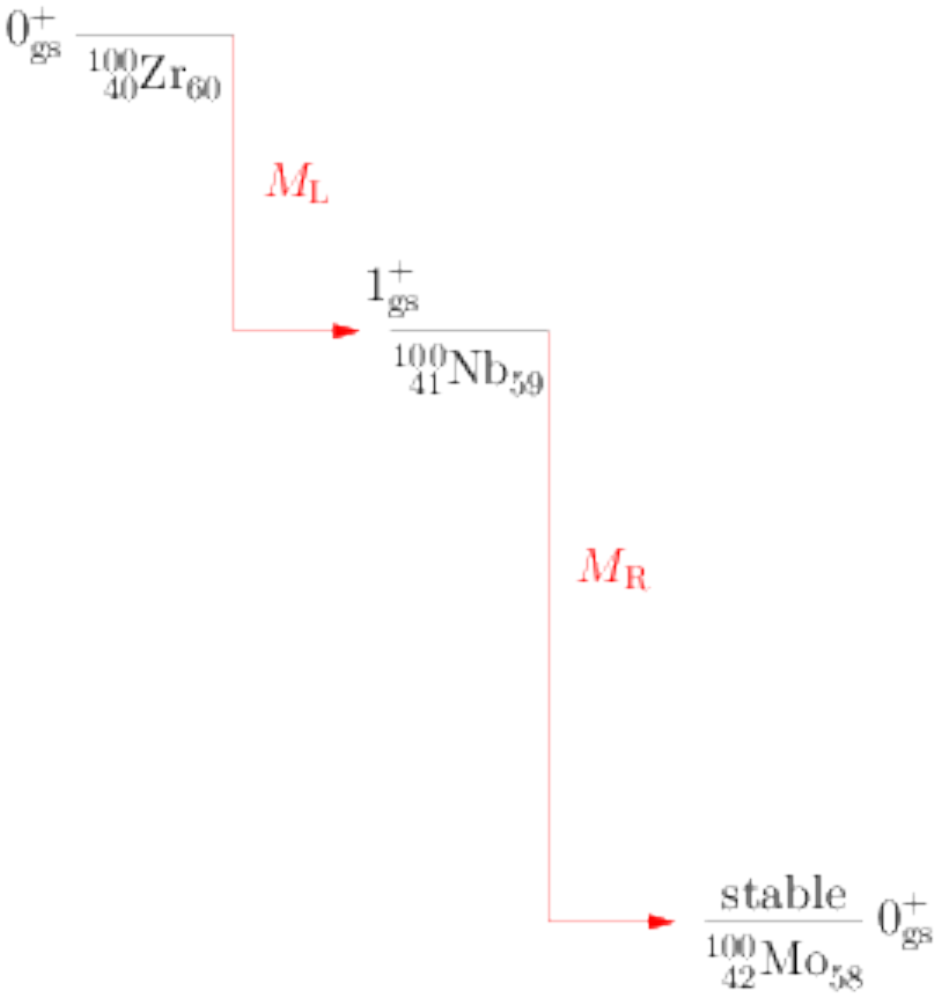}
\end{subfigure}
\begin{subfigure}{.33\textwidth}
\centering
\includegraphics[width=1.0\linewidth]{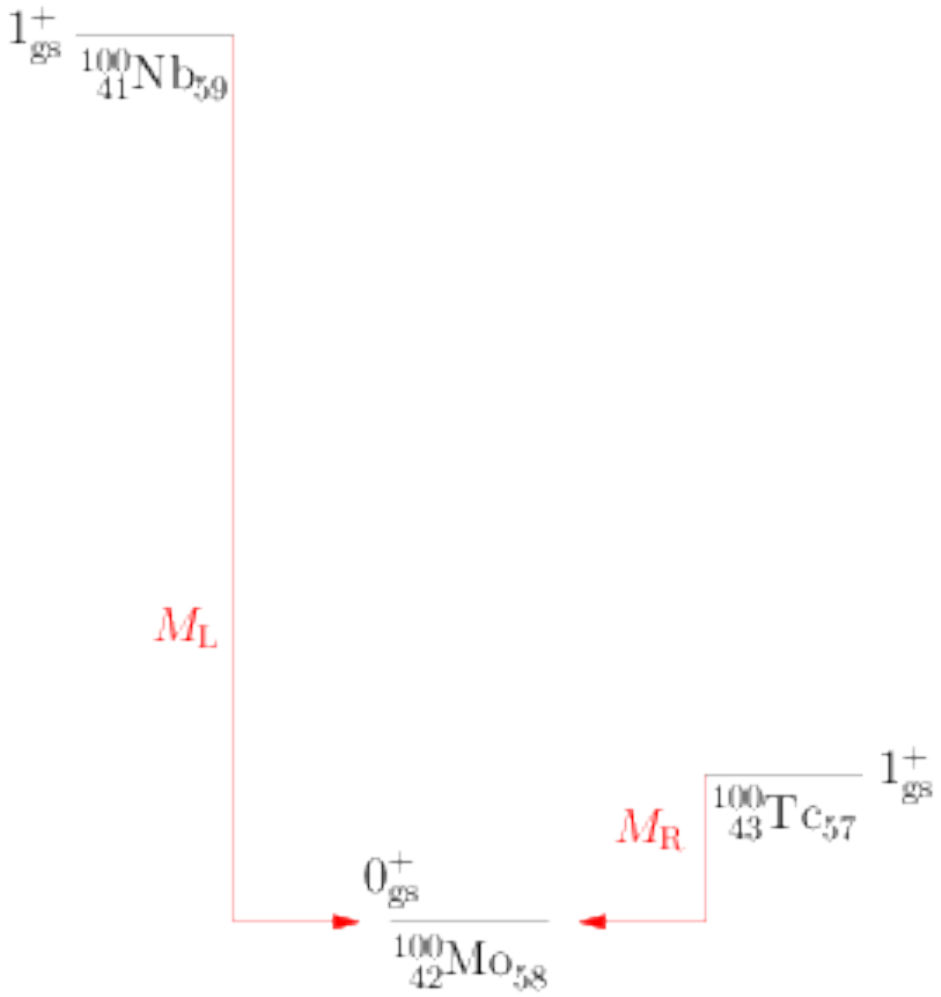}
\end{subfigure}
\begin{subfigure}{.33\textwidth}
\centering
\includegraphics[width=1.0\linewidth]{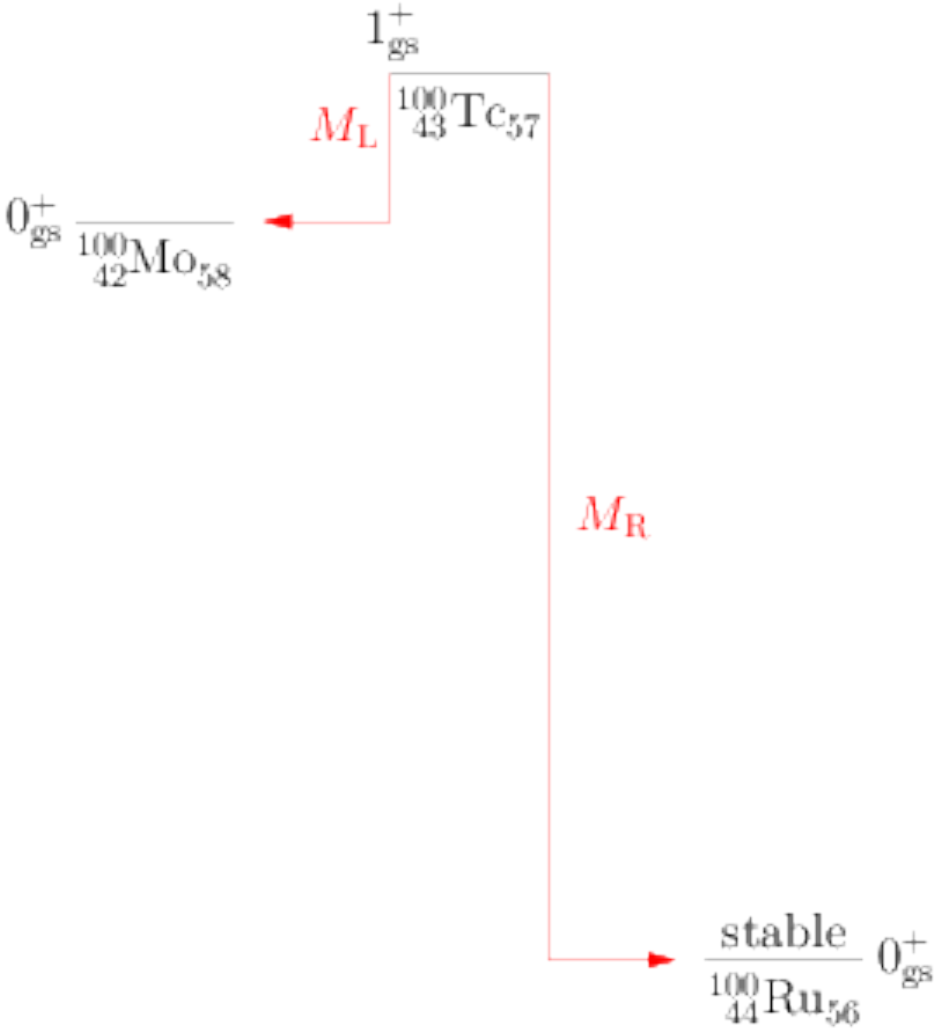}
\end{subfigure}
\caption{Gamow-Teller beta decays in the $A=100$ Zr-Nb-Mo-Tc-Ru region.}
\label{fig:GT-transitions}
%\end{center}
\end{figure}

Gamow-Teller decays are mediated by the Pauli spin operator 
$\mbox{\boldmath{$\sigma$}}$ and they are thus able to change the initial nuclear
spin $J_i$ by one unit. In the renormalization studies the simplest Gamow-Teller
transitions are selected, namely the ground-state-to-ground-state ones. In 
Fig.~\ref{fig:GT-transitions} are depicted Gamow-Teller 
ground-state-to-ground-state $\beta^-$ and $\beta^+$/EC 
transitions between even-even $0^+$ and odd-odd $1^+$
ground states in the $A=100$ Zr-Nb-Mo-Tc-Ru region. Shown are three different
situations with a \emph{cascade} pattern (left panel), lateral feeding \emph{to} 
a middle nucleus (middle panel), and lateral feeding \emph{from} a middle nucleus
(right panel). All these transitions are mediated by a Gamow-Teller NME, 
$M_{\rm GT}$, of the Pauli spin operator, defined, e.g. in \cite{Suhonen2007}.
The corresponding $\beta$-decay data can be obtained from \cite{ENSDF}.
In the figure this NME is denoted by $M_{\rm L}$ ($M_{\rm R}$) in the case it is to the
left (right) of the central nucleus.
The corresponding reduced transition probability $B_{\rm GT}$ can be written as
\begin{align} 
\label{eq:nmeGT}
	B_{\rm GT} = 
	\frac{g_{\rm A}^2}{2J_i+1}\left\vert M_{\rm GT}\right\vert^2 \,,
\end{align}
where $J_i$ is the spin of the ground state of the initial nucleus, $g_{\rm A}$ is 
the weak axial-vector coupling strength, substituted by the effective coupling 
strength $g^{\rm eff}_{\rm A}$ of Eq.~(\ref{eq:g-eff}) in practical calculations 
of the $\beta$-decay rates involving nuclear levels of low excitation energy [Hence,
the coupling strength $g_{\rm A}$ is probed at the $q^2\to 0$ limit in (\ref{eq:dipole})]. 
It is worth noting that the Gamow-Teller decays probe only $g_{\rm A}$, not $g_{\rm V}$
which is carried by the vector part (Fermi spin-zero operator) of the $\beta$ 
transitions, not active for the here discussed $1^+\leftrightarrow 0^+$ transitions
due to the conservation of angular momentum.

The comparative half-lives ($\log ft$ values) of the $1^+\leftrightarrow 0^+$ 
Gamow--Teller transitions are given in terms of the reduced transition probabilities
as \cite{Suhonen2007}
\begin{align} 
\label{eq:logft}
	\log ft = \log_{10} (f_0 t_{1/2}[s]) = 
\log_{10} \left(\frac{6147}{B_{\rm GT}}\right)
\end{align}
for the $\beta^+$/EC or $\beta^-$ type of transitions. The 
half-life of the initial nucleus, $t_{1/2}$, has been given in seconds.  

Next we inspect the evolution of the quenching
concept, based on (\ref{eq:nmeGT}) and (\ref{eq:logft}), in nuclear-structure 
calculations performed during the last four decades.

\subsection{Interacting shell model \label{subsec:ISM}}

Traditionally the renormalization of the axial-vector coupling strength
has been addressed in the context of the interacting shell model (ISM) in a wealth of
calculations pertaining to Gamow-Teller $\beta$ decays of very light ($p$-shell),
light ($sd$-shell), and medium-heavy ($pf$-shell and $sdg$-shell) nuclei. In these
calculations it appears that the value of $g_{\rm A}$ is quenched. As
indicated by the ISM results below, the quenching factor (\ref{eq:q}) is roughly a 
decreasing function of the nuclear mass number $A$, implying stronger quenching with
increasing nuclear mass. The studies can be grouped
according to the mass regions as follows.

\subsubsection{Results for the $0p$-shell nuclei \label{subsubsec:p-shell}}

A thorough study of the Gamow-Teller $\beta$ decays of the $0p$-shell nuclei was 
performed in \cite{Chou1993}. A $0p-1s0d$ cross-shell Hamiltonian derived by 
Warburton and Brown \cite{Warburton1992} was used in the calculations. 
The thus derived phenomenological (the fundamental, Sec.~\ref{sec:medium-eff}, and 
nuclear-model induced renormalization
cannot be disentangled) quenching factor [see Eq.~(\ref{eq:q})] 
(from a least-squares fit, with one standard deviation error) assumed the value
\begin{align} 
\label{eq:q-Chou}
	q = 0.82\pm 0.02 \,,
\end{align}
when using the then adopted value $g^{\rm free}_{\rm A} = 1.26$ in contrast to the
presently adopted value of Eq.~(\ref{eq:g-free}). Since the presently adopted 
free value of $g_{\rm A}$ is a bit larger, the quenching increases slightly and for
the effective value (\ref{eq:g-eff}) of $g_{\rm A}$ we have to use 
\begin{align} 
\label{eq:g-eff-Chou}
	g^{\rm eff}_{\rm A} = (0.82\pm 0.02)\times\frac{1.26}{1.27}\times 1.27 =
                           1.03^{+0.03}_{-0.02} \,,
\end{align}
leading to an effective quenched value of $g_{\rm A}$ close to unity.

\subsubsection{Results for the $1s0d$-shell nuclei \label{subsubsec:sd-shell}}

A pioneering early work of Wilkinson \cite{Wilkinson1973} investigated Gamow-Teller $\beta$
decays in the $0p$ shell and lower $1s0d$ shell for the quenching of $g_{\rm A}$. In this
work Wilkinson obtained a quenching factor which was slightly
corrected in \cite{Wilkinson1974} based on new experimental data. The corrected
value reads (from a least-squares fit, with one standard deviation error)
\begin{align} 
\label{eq:q-Wilk}
	q = 0.899\pm 0.035 \,,
\end{align}
when using the then adopted value $g^{\rm free}_{\rm A} = 1.25$. Using again the
correction for $g^{\rm free}_{\rm A}$ we have  
\begin{align} 
\label{eq:g-eff-Wilk}
	g^{\rm eff}_{\rm A} = (0.899\pm 0.035)\times 1.25 =
                           1.12^{+0.05}_{-0.04} \,.
\end{align}
The same quenching was obtained in \cite{Brown1978} by using a different ISM
effective Hamiltonian indicating that the quenching is not very sensitive to
the detailed aspects of the shell-model analysis. In \cite{Wilkinson1974} the
empirical result (\ref{eq:q-Wilk}) was combined with relativistic corrections to
yield
\begin{align} 
\label{eq:q-Wilk2}
	q = 0.927\pm 0.038 \,. \quad (\textrm{with relativistic corrections})
\end{align}
This yields
\begin{align} 
\label{eq:g-eff-Wilk2}
	g^{\rm eff}_{\rm A} = (0.927\pm 0.038)\times 1.25 =
                           1.18 \pm 0.05 \,.
\end{align}
when including the relativistic corrections.

In \cite{Wilkinson1973} and
\cite{Wilkinson1974} it was speculated that the renormalization effects of the Gamow-Teller
transitions at low nuclear excitation are of the order expected from
fundamental mesonic effects \cite{Ericson1971,Ericson1973,Rho1974} (nuclear medium effect,
see Sec.~\ref{sec:medium-eff}) or from the lifting of Gamow-Teller strength to higher 
energies by the nuclear tensor force \cite{Shimizu1974,Hyuga1980}
(nuclear model effect, see Sec.~\ref{sec:model-eff}). Indeed, by using sum-rule
arguments of \cite{Ericson1971} the expected quenching by the meson-exchange effects
would be around $q=0.93$ for nuclei in the vicinity of $A=16$. This is in very good
agreement with the relativistically corrected empirical result (\ref{eq:q-Wilk2}).

A full $sd$-shell analysis of the quenching was performed in \cite{Wildenthal1983} with
a new set of wave functions derived from a Hamiltonian reproducing the global
spectroscopic features of the $1s0d$-shell nuclei. The least-squares study 
(with one standar deviation error) yielded the (empirical) quenching factor 
$q=0.77\pm 0.02$ and thus leads to the global 
$g^{\rm free}_{\rm A}$-corrected $1s0d$-shell effective axial-vector coupling of 
\begin{align} 
\label{eq:g-eff-Wild}
	g^{\rm eff}_{\rm A} = (0.77\pm 0.02)\times 1.25 =
                           0.96^{+0.03}_{-0.02} \,,
\end{align}
which is notably smaller than (\ref{eq:g-eff-Wilk}) obtained for the lower $1s0d$ shell.
In the least-squares-fit studues, like this and the one of \cite{Chou1993} 
(see Sec.~\ref{subsubsec:p-shell}), the separation of the fundamental quenching 
(see Sec.~\ref{sec:medium-eff}) from the total quenching is impossible.

\subsubsection{Results for the $1p0f(0g_{9/2})$-shell nuclei \label{subsubsec:pf-shell}}

In the work \cite{Martinez-Pinedo1996} 64 Gamow-Teller $\beta$ decays for the
nuclear mass range $A=41-50$ were studied. This mass range covers the lower part
of the $1p0f$ shell. The shell-model work was based on \cite{Caurier1994} and
KB3 two-body interaction was adopted. In \cite{Martinez-Pinedo1996} the experimental
values of Gamow-Teller matrix elements (extracted by using the free value of $g_{\rm A}$)
were compared with their computed values by plotting them against each other in an $xy$
plane. The plot was well described by a line with the slope giving a phenomenological 
quenching factor. From the slope and its error the quenching factor
\begin{align} 
\label{eq:q-MaPi}
	q = 0.744\pm 0.015 
\end{align}
was derived, when using the their adopted value $g^{\rm free}_{\rm A} = 1.26$. Then the
$g^{\rm free}_{\rm A}$-corrected lower $pf$-shell quenching amounts to 
\begin{align} 
\label{eq:g-eff-MaPi}
	g^{\rm eff}_{\rm A} = (0.744\pm 0.015)\times 1.26 =
                           0.937^{+0.019}_{-0.018} \,.
\end{align}
It is interesting to note that with this value of $g^{\rm eff}_{\rm A}$ the half-life
of the $2\nu\beta\beta$ decay of $^{48}$Ca could be predicted \cite{Caurier1990}
in perfect agreement with the later measured value \cite{Balysh1996}. In the
work \cite{Horoi2007} it was confirmed that the value $q=0.77$ reasonably describes
the quenching in the $A=48$ region. The quenching in the $1s0d$ and $1p0f$ shells
was also studied in \cite{Auerbach1993} for the nucleus $^{26}$Mg ($1s0d$ model
space) and for the nuclei $^{54}$Fe and $^{56}$Ni ($1p0f$ model space) by using
both the random-phase approximation and the ISM. The computed $\beta^+$ Gamow-Teller
strengths were compared with those derived from the (n,p) charge-exchange reactions.
This comparison implied a phenomenological quenched value of $g^{\rm eff}_{\rm A}\sim 0.98$, 
not far from the value (\ref{eq:g-eff-Wild}), extracted in the $1s0d$ shell by 
\cite{Wildenthal1983} and the value (\ref{eq:g-eff-MaPi}), extracted in the 
$1p0f$ shell.

The upper $1p0f(0g_{9/2})$-shell Gamow-Teller transitions were analyzed in 
\cite{Honma2006} in the $0f_{5/2}1p0g_{9/2}$ valence space using a renormalized 
G-matrix-based two-body interaction, fitted in the mass region $A=63-96$. 
A rough phenomenological quenching factor
\begin{align} 
\label{eq:q-Honma}
	q = 0.6
\end{align}
was adopted in the subsequent calculations of the $2\nu\beta\beta$-decay rates 
of $^{76}$Ge and $^{82}$Se. This, in turn, leads to an upper $1p0f(0g_{9/2})$-shell 
effective coupling strength of 
\begin{align} 
\label{eq:g-eff-Honma}
	g^{\rm eff}_{\rm A} = 0.6\times 1.26 = 0.8 \,,
\end{align}
which is considerably smaller than (\ref{eq:g-eff-MaPi}) obtained for the lower 
$1p0f$ shell.

\subsubsection{Results for the $0g_{7/2}1d2s0h_{11/2}$-shell nuclei 
\label{subsubsec:sdg-shell}}

In \cite{Caurier2012} an analysis of the Gamow-Teller $\beta$ decays in
the (incomplete) $sdg$ shell (for $A=128-130$) was performed using
the $0g_{7/2}1d2s0h_{11/2}$ single-particle space. A model Hamiltonian
based on a renormalized Bonn-C G-matrix with a subsequent fitting of about 300
energy levels of some 90 nuclei in the $0g_{7/2}1d2s0h_{11/2}$ shell was used in
the calculations. The resulting phenomenological quenching factor was
\begin{align} 
\label{eq:q-Caurier}
	q = 0.57 \,,
\end{align}
implying a $0g_{7/2}1d2s0h_{11/2}$-shell effective coupling strength of 
\begin{align} 
\label{eq:g-eff-Caurier}
	g^{\rm eff}_{\rm A} = 0.57\times 1.26 = 0.72 \,,
\end{align}
which is a bit smaller than those obtained in the $1p0f(0g_{9/2})$ shell.

In \cite{Caurier2012} also the case of $A=136$ was discussed for the $2\nu\beta\beta$
decay of $^{136}$Xe using the above-mentioned single-particle space. Comparing 
the experimentally available \cite{Puppe2011} 
(p,n) type of strength function on $^{136}$Xe (up to excitation energies of 
3.5 MeV in $^{136}$Cs) with the computed one, the authors concluded a phenomenological
quenching factor
\begin{align} 
\label{eq:q-Caurier2}
	q = 0.45 
\end{align}
for $A=136$. This leads to a heavily quenched effective axial-vector coupling
strength of
\begin{align} 
\label{eq:g-eff-Caurier2}
	g^{\rm eff}_{\rm A}(A=136) = 0.45\times 1.26 = 0.57 \,,
\end{align}
for the $A=136$ region of the $0g_{7/2}1d2s0h_{11/2}$ shell. On the other hand, more
recent calculations by M. Horoi \textit{et al.} \cite{Neacsu2015,Horoi2016} for
the $2\nu\beta\beta$ NMEs of $^{130}$Te and $^{136}$Xe
suggest a milder quenching and a larger value 
$g^{\rm eff}_{\rm A}(A=130-136)=0.94$ \cite{Horoi2016} for the effective 
coupling strength. This is in a rather sharp tension with the results 
(\ref{eq:g-eff-Caurier}) and (\ref{eq:g-eff-Caurier2}) of \cite{Caurier2012}.

In \cite{Juodagalvis2005} a cross-shell study for the mass region $A=90-97$ was
performed in the single-particle space $1p_{1/2}0g_{9/2}$ for protons and
$0g_{7/2}1d0s0h_{11/2}$ for neutrons by using a Bonn-CD-based potential with 
perturbative renormalization. Again, lack of the full space of spin-orbit partners 
lead to a strong phenomenological Gamow-Teller quenching
\begin{align} 
\label{eq:q-Juoda}
	q = 0.48 \,,
\end{align}
leading to a cross $pf-sdg$-shell effective coupling strength of 
\begin{align} 
\label{eq:g-eff-Juoda}
	g^{\rm eff}_{\rm A} = 0.48\times 1.26 = 0.60 \,.
\end{align}
The above-derived quenching is not far from the quenching $q=0.5$ derived in 
\cite{Brown1994} for nuclei in the $^{100}$Sn region using a $0f_{5/2}1p0g_{9/2}$ 
proton-hole space and $0g_{7/2}1d0s0h_{11/2}$ neutron-particle space.

A quite recent ISM analysis of the nuclei within the mass range $52\le A\le 80$ was
performed \cite{Kumar2016}. There the $1p0f$-shell nuclei, $52\le A\le 67$,
were treated by using the KB3G interaction, and the comparison with the experimental
$\beta^-$-decay half-lives produced a phenomenological quenching factor leading to
the effective coupling strength
\begin{align} 
\label{eq:g-eff-Kumar1}
	g^{\rm eff}_{\rm A} =  0.838^{+0.021}_{-0.020} \quad (52\le A\le 67) \,.
\end{align}
The $0f_{5/2}1pg_{9/2}$-shell nuclei, $67\le A\le 80$, were computed by using the 
JUN45 interaction, producing the effective coupling strength
\begin{align} 
\label{eq:g-eff-Kumar2}
	g^{\rm eff}_{\rm A} =  0.869\pm 0.019 \quad (67\le A\le 80) \,.
\end{align}
In this work the error estimation is given by the slopes-of-the-lines method 
\cite{Martinez-Pinedo1996}, discussed in the context of Eq.~(\ref{eq:q-MaPi}) above.

\begin{table}[t!]
\centering
\begin{tabular}{lcl}
\toprule
Mass range & $g^{\rm eff}_{\rm A}$ & Reference \\
\midrule
Full $0p$ shell & $1.03^{+0.03}_{-0.02}$ & W. T. Chou \textit{et al.} 1993 \cite{Chou1993} \\
$0p-\mathrm{low\ }1s0d$ shell & $1.12^{+0.05}_{-0.04}$ & D. H. Wilkinson 1974 
\cite{Wilkinson1974} (no RC) \\
 & $1.18\pm 0.05$ & D. H. Wilkinson 1974 \cite{Wilkinson1974} (with RC) \\
Full $1s0d$ shell & $0.96^{+0.03}_{-0.02}$ & B. H. Wildenthal \textit{et al.} 
1983 \cite{Wildenthal1983} \\
 & $1.0$ & T. Siiskonen \textit{et al.} 2001 \cite{Siiskonen2001} \\
$A=41-50$ ($1p0f$ shell) & $0.937^{+0.019}_{-0.018}$  
& G. Mart{\' \i}nez-Pinedo \textit{et al.} 1996 \cite{Martinez-Pinedo1996} \\
$1p0f$ shell & $0.98$ & T. Siiskonen \textit{et al.} 2001 \cite{Siiskonen2001} \\
$^{56}$Ni & $0.71$ & T. Siiskonen \textit{et al.} 2001 \cite{Siiskonen2001} \\
$A=52-67$ ($1p0f$ shell) & $0.838^{+0.021}_{-0.020}$ 
& V. Kumar \textit{et al.} 2016 \cite{Kumar2016} \\ 
$A=67-80$ ($0f_{5/2}1p0g_{9/2}$ shell) & $0.869\pm 0.019$ 
& V. Kumar \textit{et al.} 2016 \cite{Kumar2016} \\ 
$A=63-96$ ($1p0f0g1d2s$ shell) & $0.8$ & M. Honma \textit{et al.} 2006 \cite{Honma2006} \\ 
$A=76-82$ ($1p0f0g_{9/2}$ shell) & $0.76$ & E. Caurier \textit{et al.} 
2012 \cite{Caurier2012} \\
$A=90-97$ ($1p0f0g1d2s$ shell) & $0.60$ & A. Juodagalvis \textit{et al.} 
2005 \cite{Juodagalvis2005} \\
$^{100}$Sn & $0.52$ & T. Siiskonen \textit{et al.} 2001 \cite{Siiskonen2001} \\
$A=128-130$ ($0g_{7/2}1d2s0h_{11/2}$ shell) & $0.72$  & E. Caurier \textit{et al.} 
2012 \cite{Caurier2012} \\
$A=130-136$ ($0g_{7/2}1d2s0h_{11/2}$ shell) & $0.94$  & M. Horoi \textit{et al.} 
2016 \cite{Horoi2016} \\
$A=136$ ($0g_{7/2}1d2s0h_{11/2}$ shell) & $0.57$  & E. Caurier \textit{et al.} 
2012 \cite{Caurier2012} \\
\bottomrule
\end{tabular}
\caption{Mass ranges and effective values of $g_{\rm A}$ extracted from the
works of the last column. RC in lines 2 and 3 denotes relativistic corrections.}
\label{tab:ISM} 
\end{table}

All the results of the ISM analyses have been collected in Table~\ref{tab:ISM}. There
the mass range (magic shell), value of $g^{\rm eff}_{\rm A}$, and the author information
are given. Also the results of \cite{Siiskonen2001}, from
Sec.~\ref{sec:medium-eff}, obtained by the use of effective operators in the 
nuclear medium, have been given for comparison.
In addition, the ISM results (adding the $^{100}$Sn results of \cite{Siiskonen2001}) 
for masses $60\le A\le 136$ have been visualized in Fig.~\ref{fig:gA-ranges2} of 
Sec.~\ref{subsec:QRPA}. In the figure the results 
of \cite{Honma2006}, \cite{Caurier2012}, \cite{Horoi2016}, \cite{Juodagalvis2005}, 
\cite{Kumar2016}, and \cite{Siiskonen2001} (see the
discussions above) have been plotted against the background (the hatched 
region of Fig.~\ref{fig:gA-ranges2}) of the results of the pnQRPA analyses 
performed in Sec.~\ref{subsec:QRPA}. Looking at the figure makes it obvious that
the ISM results of the aforementioned references are commensurate with the
results of the (global) analyses of Gamow-Teller transitions performed in the 
framework of the proton-neutron quasiparticle random-phase approximation (pnQRPA).

Finally, it is of interest to point out to the recent work \cite{Konieczka2016} where
no-core-configuration-interaction formalism, rooted in multireference density
functional theory, was used to compute the Gamow-Teller NMEs
for $T=1/2$ mirror nuclei (pairs of nuclei where either a neutron or a proton is added to an
even-even $N=Z$ core nucleus) in the $1s0d$ and $1p0f$ shells. The computations were
performed in a basis of 10 or 12 spherical harmonic-oscillator shells by using 
two different Skyrme forces. The computed quenching factors coincide surprisingly closely
with those of the ISM quoted in (\ref{eq:g-eff-Wild}) [Ref.~\cite{Wildenthal1983}] for 
the $1s0d$ shell and in (\ref{eq:q-MaPi}) [Ref.~\cite{Martinez-Pinedo1996}] for the 
$1p0f$ shell, despite the big differences in the two nuclear models. This would point to
the possibility that the quenching in the $1s0d$ and $1p0f$ shells is not so much
related to the deficiencies of the nuclear models but rather to omission of effects
coming from the nuclear medium, like from the two-body currents and other mesonic effects
discussed in Sec.~\ref{sec:medium-eff}.

\subsection{Quasiparticle random-phase approximation \label{subsec:QRPA}}

Only recently the important aspect of the effective value of $g_{\rm A}$ has been 
addressed within the framework of the proton-neutron quasiparticle random-phase
approximation (pnQRPA). The situation with pnQRPA is more involved than in
the case of the ISM since the adopted schematic or realistic interactions are
usually renormalized separately in the particle-hole ($g_{\rm ph}$ parameter)
and particle-particle ($g_{\rm pp}$ parameter) 
\cite{Vogel1986,Civitarese1987,Suhonen1988a,Suhonen1988b} channels. Typically the
particle-hole parameter, $g_{\rm ph}$, is fitted to reproduce the centroid of
the Gamow-Teller giant resonance (GTGR) obtained from the semi-empirical
formula \cite{Suhonen1988a,Suhonen1988b}
\begin{equation}
\label{eq:GTGR}
\Delta E_{\rm GT} = E(1^+_{\rm GTGR}) - E(0^+_{\rm gs}) = 
\Big[ 1.444\left(Z+\frac{1}{2}\right)
A^{-1/3} - 30.0\big( N-Z-2\big)A^{-1} + 5.57\Big]\,\textrm{MeV} .
\end{equation}
The above formula indicates that the difference $\Delta E_{\rm GT}$ between the GTGR and 
the ground state of the  neighboring even-even reference nucleus depends on the 
proton and neutron numbers $(Z,N)$ of the reference nucleus, as well as on its 
mass number. For the particle-particle parameter, $g_{\rm pp}$, there is no unique way 
to fix its value, as criticized in \cite{Suhonen2005}. Furthermore, the exact value of
$g_{\rm pp}$ depends on the size of the active single-particle model space.
In this review several ways how this can be done are discussed. As a result of the
$g_{\rm pp}$ problems and problems with systematic renormalization of the two-body
interactions, the fundamental quenching (see Sec.~\ref{sec:medium-eff}) cannot be
disentangled from the nuclear-model effects, discussed in Sec.~\ref{sec:model-eff}.

The first pnQRPA attempts were inspired by a simultaneous
description of $\beta$ and $2\nu\beta\beta$ decays, as elaborated more in
Sec.~\ref{sec:2n-beta-beta}. In \cite{Delion2014} 9 isobaric systems, with 
$A=70,78,100,104,106,110,116,128,130$, of the 
type displayed in the right panel of Fig.~\ref{fig:GT-transitions} were analyzed
by using a spherical pnQRPA with schematic particle-hole and particle-particle
forces. The pnQRPA calculations were performed in the even-even reference nuclei.
For each GTGR-fixed $g_{\rm ph}$ the value of $g_{\rm pp}$ was varied in order
to reproduce the experimentally known ratio $M_{\rm R}/M_{\rm L}$ which is independent
of the value of $g_{\rm A}$. The value of $g_{\rm A}$ was then determined by requiring
$M_{\rm R}({\rm th})/M_{\rm R}({\rm exp})=1$. This produced the mean value
\begin{align} 
\label{eq:g-eff-Delion}
	g^{\rm eff}_{\rm A} = 0.27 
\end{align}
and the approximate mass dependence $g_{\rm pp}\approx 0.5/\sqrt{A}$. By using this
dependence of $g_{\rm pp}$ and the above value (\ref{eq:g-eff-Delion}) for $g_{\rm A}$
the experimental $\beta^+$/EC and $\beta^-$ NMEs of 218 Gamow-Teller transitions
were quite well reproduced in \cite{Delion2014}. The quite low value obtained for
$g^{\rm eff}_{\rm A}$ implies that a larger quenching is required than in the ISM due
to the simple schematic form of the adopted Hamiltonian in the pnQRPA calculations.
in other words, the quenching coming from the many-body effects is stronger for
the pnQRPA calculation than for the ISM calculation which is more realistic in
terms of two-body interactions and configuration space. In this analysis the effects
coming from the nuclear medium (Sec.~\ref{sec:medium-eff}) cannot be disentangled from
the many-body effects, unfortunately.

In \cite{Pirinen2015} an analysis of 26 $\beta^-$ and 22 $\beta^+$/EC
Gamow-Teller transitions of the type depicted
in Fig.~\ref{fig:GT-transitions} in the  mass range $A=100-136$ was performed. In
this study the geometric mean
\begin{align} 
\label{eq:G-mean}
	\bar{M}_{\rm GT} = \sqrt{M_{\rm L}M_{\rm R}}
\end{align}
of the extracted experimental NMEs was compared with that computed by the use of
the pnQRPA with realistic effective forces based on the $g_{\rm ph}$- and 
$g_{\rm pp}$-renormalized Bonn-A G matrix. The use of the geometric mean of the left and
right NMEs stabilizes the values of the mean NMEs and smoother trends can be obtained.
This is based on the fact that the NME for the $\beta^-$ branch is a decreasing
function of $g_{\rm pp}$ and the NME for the $\beta^+$/EC branch is an increasing 
function of $g_{\rm pp}$. Thus, the product of the NMEs of these branches remains 
essentially constant over a wide range of $g_{\rm pp}$ values (see the figures in 
\cite{Ejiri2015}).

\begin{figure}[htbp!]
\centering
\includegraphics[width=0.6\columnwidth]{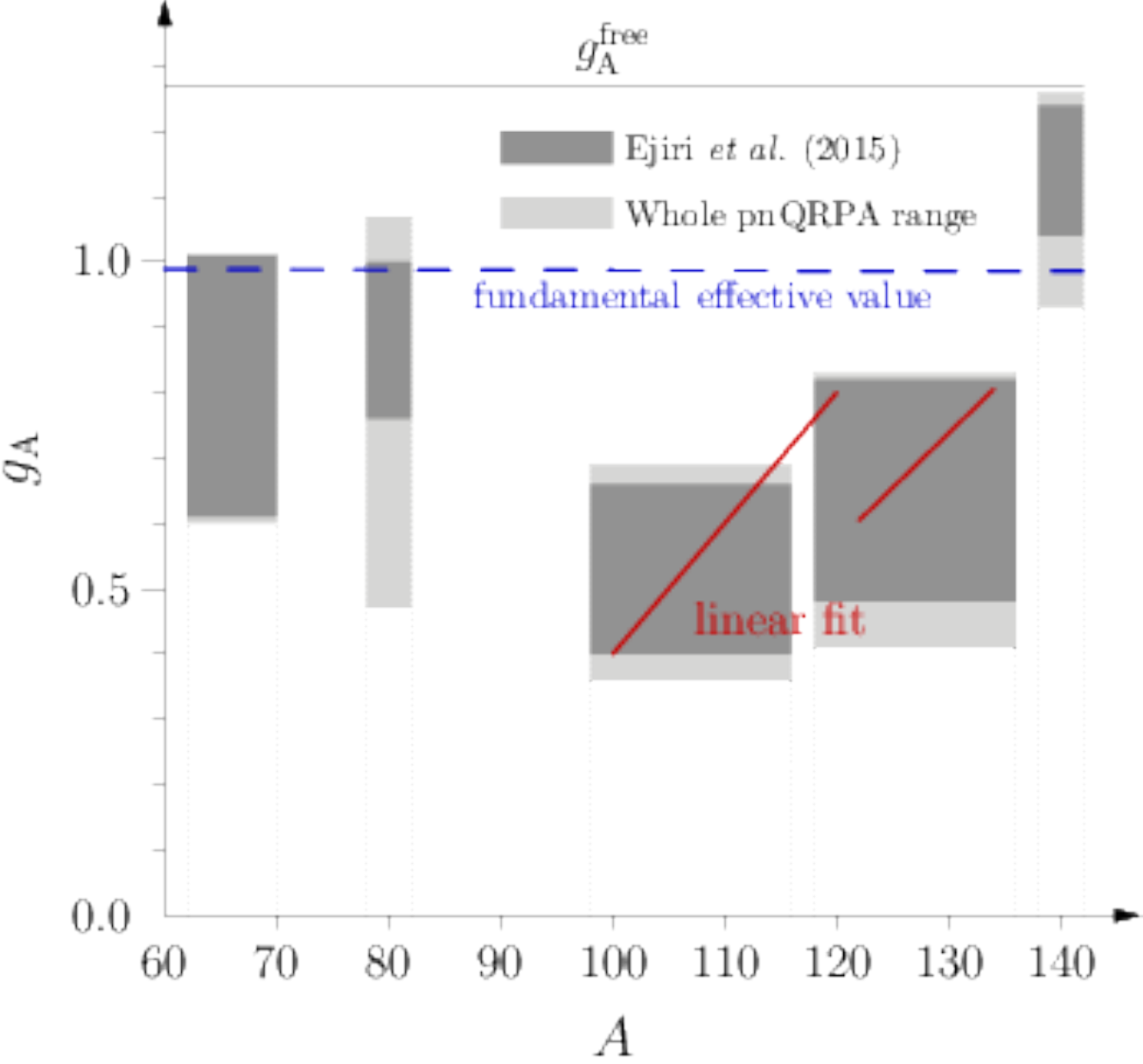}
\caption{Averaged effective values of $g_{\rm A}$ [$g^{\rm eff}_{\rm A}$ of 
(\ref{eq:k-bar-g-eff})] in the 5 different mass ranges, plotted from
the numbers of Table~\ref{tab:g-bar-eff}. The legends inside the figure correspond
to the following references: Ejiri \textit{et al.} (2015) \cite{Ejiri2015} 
(dark-hatched boxes);
Whole pnQRPA range: combined results of \cite{Pirinen2015},  
\cite{Deppisch2016}, and \cite{Ejiri2015} (light-hatched boxes) illustrate
the total range of $g^{\rm eff}_{\rm A}$ for each mass region.
Also the linear fit (\ref{eq:g-eff-lin}) and the ``fundamentally'' quenched $g_{\rm A}$,
Eq.~(\ref{eq:g-finite-A}), are plotted for comparison.}
\label{fig:gA-ranges1}
\end{figure}

Like in \cite{Delion2014}, the pnQRPA calculations of \cite{Pirinen2015} were 
performed in the even-even reference nuclei. The value of $g_{\rm ph}$ was fixed by the
phenomenological centroid (\ref{eq:GTGR}) of the GTGR separately for each nucleus.
In the calculations it turned out that the value $g_{\rm pp}=0.7$ represents
a reasonable global value for the particle-particle interaction strength in
the model spaces used in the calculations: at least one oscillator major shell above
and below those oscillator shells where the proton and neutron Fermi surfaces lie. 
Furthermore, an average piece-wise linear behavior 
\begin{align} 
\label{eq:g-eff-lin}
	g^{\rm eff}_{\rm A} = \left\{\begin{array}{ll} 
        0.02A-1.6 & \mathrm{for\ } A\le 120 \\
        \frac{1}{60}A-\frac{43}{30} & \mathrm{for\ } A\ge 122 
        \end{array} \right.
\end{align}
of $g_{\rm A}$ was found in the calculations. These derived values of $g^{\rm eff}_{\rm A}$,
plotted in Fig.~\ref{fig:gA-ranges1},
were used, in turn, to describe the Gamow-Teller and $2\nu\beta\beta$ decay rates to 
the ground state and lowest excited states in the even-even reference nucleus 
in the $A=100-136$ mass region. These results were compared with those
obtained by the use of the average value
\begin{align} 
\label{eq:g-eff-ave-Pirinen}
	g^{\rm eff}_{\rm A}(\mathrm{ave}) = 0.6 
\end{align}
for $g^{\rm eff}_{\rm A}$. The average value reproduced surprisingly well the 
experimentally known $2\nu\beta\beta$ half-lives in this mass region.

The work of \cite{Pirinen2015} was extended in \cite{Deppisch2016} to a wider
range of nuclei ($A=62-142$) and to a more refined statistical analysis of the
results. The same renormalized Bonn-A G matrix as in \cite{Pirinen2015} was adopted
for the pnQRPA calculations, along with the scaling with the $g_{\rm ph}$ 
and $g_{\rm pp}$ parameters. A Markov
chain Monte Carlo statistical analysis of 80 Gamow-Teller transitions in 47 
isobaric decay triplets of the kind depicted in Fig.~\ref{fig:GT-transitions} was
performed. The analysis was also extended to 28 longer isobaric chains and the results
were compared with those obtained for the isobaric triplets. Also the measured
half-lives of $2\nu\beta\beta$ decays occurring in the isobaric chains were analyzed.
A roughly linearly increasing trend of $g^{\rm eff}_{\rm A}$ as a function of the 
mass number $A$ could be extracted from the analysis of the isobaric triplets for
$A\ge 100$, in accordance with the result of \cite{Pirinen2015}. Similar features
were seen also in the fits to longer multiplets. For the range $100\le A\le 136$ the
average (\ref{eq:g-eff-ave-Pirinen}) was roughly obtained in both analyses.

In contrast to \cite{Pirinen2015} also the value of $g_{\rm pp}$ was kept as a free
parameter, the same for the left and right NMEs of transitions in the triplets like in
Fig.~\ref{fig:GT-transitions}, and different for each even-even reference nucleus in
the longer chains. Both types of analysis yield a rough average of $g_{\rm pp}\approx 0.7$
for the particle-particle strength parameter in the mass range $100\le A\le 136$
(see the last column of Table~\ref{tab:g-bar-eff}), in accordance with the value used in
the analysis of \cite{Pirinen2015}. At this point it should be noted that
the adopted single-particle model spaces used in the calculations correspond to those
of \cite{Pirinen2015} for $100\le A\le 136$: at least one oscillator major shell above
and below those oscillator shells where the proton and neutron Fermi surfaces lie. 

A slightly different analysis of the Gamow-Teller transitions in the mass range
$62\le A\le 142$ was carried out in \cite{Ejiri2015}. This is the same mass
range as analyzed in \cite{Deppisch2016}. Again the $g_{\rm ph}$- and 
$g_{\rm pp}$-renormalized Bonn-A G matrix was used in a pnQRPA framework, and 
the geometric mean (\ref{eq:G-mean}) was used in the analysis to smooth the 
systematics. The mass range was divided in 5 sub-ranges according to the leading
proton-neutron ($pn$) configuration influencing the Gamow-Teller decay rate. The reduction
of the NME in the chain $\bar{M}_{\rm qp}\to\bar{M}_{\rm pnQRPA}\to\bar{M}_{\rm exp}$ 
was followed, where $\bar{M}_{\rm qp}$ is the mean two-quasiparticle NME (\ref{eq:G-mean})
for the leading $pn$ configuration, $\bar{M}_{\rm pnQRPA}$ is the pnQRPA-computed mean NME, 
and $\bar{M}_{\rm exp}$ is the mean experimental NME, extracted from 
the experimental decay-half-life data by using $g^{\rm free}_{\rm A} = 1.27$. The ratio
\begin{align} 
\label{eq:k-ratio-GT}
k = \frac{\bar{M}_{\rm pnQRPA}}{\bar{M}_{\rm qp}}
\end{align}
is a measure of the quenching of the NME when going from a rudimentary many-body approach
towards a more sophisticated one. This ratio is independent of the nuclear-matter effects
and is usually nuclear-mass dependent (the results of the analysis \cite{Ejiri2015} are
quoted in the second column of Table~\ref{tab:k_k} in Sec.~\ref{sec:high-forb-u}).
The quenching of $g_{\rm A}$ by the nuclear-medium and many-body (inseparable!) effects 
was incorporated in the ratio
\begin{align} 
\label{eq:k-bar}
	k_{\rm NM} = \langle \bar{M}_{\rm exp}/\bar{M}_{\rm pnQRPA} \rangle \,,
\end{align}
representing an average of the ratio $\bar{M}_{\rm exp}/\bar{M}_{\rm pnQRPA}$ 
of the experimental NME and the pnQRPA-computed NME over each sub-range of masses. 
The resulting effective $g_{\rm A}$ can be extracted from $k_{\rm NM}$ by using 
the simple relation
\begin{align} 
\label{eq:k-bar-g-eff}
	g^{\rm eff}_{\rm A} = g^{\rm free}_{\rm A}k_{\rm NM} = 1.27k_{\rm NM} \,.
\end{align}
The resulting values of $g^{\rm eff}_{\rm A}$, along with the mass ranges and
leading $pn$ configurations are listed in Table~\ref{tab:g-bar-eff}. The pnQRPA
results were obtained by fitting the $g_{\rm ph}$ parameter to the phenomenological
centroid (\ref{eq:GTGR}) of the GTGR separately for each nucleus, and by adopting
$g_{\rm pp}=0.67$, in line with the analyses of \cite{Pirinen2015,Deppisch2016}. Again,
the adopted single-particle model spaces correspond to those
of \cite{Deppisch2016}: at least one oscillator major shell above
and below those oscillator shells containing the proton and neutron Fermi surfaces.

\begin{table}[htb]
\centering
\begin{tabular}{ccccccc}
\toprule
$A$ & $pn$ configuration & \multicolumn{4}{c}{$g^{\rm eff}_{\rm A}$} & 
$g_{\rm pp}$ \cite{Deppisch2016}  \\ 
\cmidrule(lr{0.75em}){3-6}
  &  & \cite{Ejiri2015} & \cite{Deppisch2016} (tripl.) & \cite{Deppisch2016} (mult.) &
  \cite{Pirinen2015} & \\
\midrule
$62-70$ & $1p_{3/2}-1p_{1/2}$ & $0.81\pm 0.20$ & $0.80\pm 0.20$ & $0.84\pm 0.15$ 
& - & $0.71\pm 0.34$ \\
$78-82$ & $0g_{9/2}-0g_{9/2}$ & $0.88\pm 0.12$ & $0.77\pm 0.30$ & ($0.87\pm 0.74$) 
& - & $0.53\pm 0.33$ \\
$98-116$ & $0g_{9/2}-0g_{7/2}$ & $0.53\pm 0.13$ & $0.54\pm 0.15$ & $0.53\pm 0.14$ 
& $0.52\pm 0.16$ & $0.74\pm 0.17$ \\
$118-136$ & $1d_{5/2}-1d_{5/2}$ & $0.65\pm 0.17$ & $0.65\pm 0.16$ & $0.59\pm 0.18$ 
& $0.67\pm 0.16$ & $0.56\pm 0.24$ \\
$98\le A\le 136$ & & $0.72\pm 0.16$ & $0.69\pm 0.12$ & $0.71\pm 0.17$ 
& $0.60\pm 0.11$ & $0.63\pm 0.11$ \\
$138-142$ & $1d_{5/2}-1d_{3/2}$ & $1.14\pm 0.10$ & $1.13\pm 0.13$ & $1.07\pm 0.14$ 
& - & $0.59\pm 0.11$ \\
\bottomrule
\end{tabular}
\caption{Mass ranges, the corresponding leading $pn$ configurations and average
effective values (\ref{eq:k-bar-g-eff}) of $g_{\rm A}$ extracted from three 
different works. The numbers of column 4 (5) are obtained from fits to isobaric 
triplets (multiplets). The last column shows the averaged values of $g_{\rm pp}$ 
deduced from \cite{Deppisch2016}. In all studies the same single-particle model spaces
have been used (see the text). The second last line shows the averages in the 
mass interval $98\le A\le 136$. In all analyses the arithmetic mean with a standard
deviation from it has been given.}
\label{tab:g-bar-eff} 
\end{table}

In Table~\ref{tab:g-bar-eff} also the averaged results of \cite{Deppisch2016} and
\cite{Pirinen2015} are shown for comparison. For \cite{Deppisch2016} are shown the
results of both the isobaric triplet (tripl.) and multiplet (mult.) fits, as also
the averaged $g_{\rm pp}$ values, extracted from the analysis of
the triplet fits of \cite{Deppisch2016}. In all the analyses the same single-particle 
model spaces were used: at least one oscillator major shell above and below those 
oscillator shells containing the proton and neutron Fermi surfaces. The triplet
and multiplet fits of \cite{Deppisch2016} are quite consistent, excluding the multiplet
fit of mass range $78-82$ (the result in parenthesis) which has two fitted multiplets,
the other rendering an ambiguous result. The results of \cite{Deppisch2016} are
very close to those of \cite{Pirinen2015} and \cite{Ejiri2015}. 
Most of the (quite small) differences between the various
calculations stem from the different ways of treating the value of the
particle-particle strength $g_{\rm pp}$, which for the studies of \cite{Ejiri2015}
and \cite{Pirinen2015} was kept constant ($g_{\rm pp}=0.67$ and $g_{\rm pp}=0.7$, 
respectively) but was allowed to vary in the work of \cite{Deppisch2016} (see
the last column of Table~\ref{tab:g-bar-eff}).

The numbers of Table~\ref{tab:g-bar-eff} have been visualized in
Fig.~\ref{fig:gA-ranges1}. Also the linear fit (\ref{eq:g-eff-lin}) and the 
``fundamentally'' quenched $g_{\rm A}$, Eq.~(\ref{eq:g-finite-A}), are plotted 
for comparison. The plot reveals quite a simple structure of the 
ranges of $g^{\rm eff}_{\rm A}$ within different mass regions. The numbers
of Ejiri \textit{et al.} \cite{Ejiri2015} are given as dark-hatched regions 
while the light-hatched regions contain the results of Ejiri \textit{et al.} 
plus the results of Pirinen \textit{et al.} \cite{Pirinen2015} and Deppisch
\textit{et al.} \cite{Deppisch2016}. A general decreasing trend of the ranges
of $g^{\rm eff}_{\rm A}$ (the hatched boxes) can be seen, except for the 
heaviest masses $A\ge 138$. It is noteworthy that there is a small shift in
the values of $g^{\rm eff}_{\rm A}$ at $A=120$ indicated by all the pnQRPA analyses
(both light and hatched boxes). Also the linear fit (\ref{eq:g-eff-lin}) indicates 
a discontinuity close to this mass number. The most probable cause for this
displacement is the change in the nuclear wave functions from
the $0g$-orbital dominated to the $1d$-orbital dominated proton-neutron configuration,
as seen in Table~\ref{tab:g-bar-eff}. A similar, even more drastic, displacement is
seen between $A=70-78$ where the dominating proton-neutron configuration of the
nuclear wave functions shifts from the $1p$ orbitals to the $0g$ orbitals.

The obtained pnQRPA ranges can be compared with results obtained
by performing combined $g^{\rm eff}_{\rm A}$ analyses of $\beta$ and $2\nu\beta\beta$ 
decay rates in the pnQRPA and other models: The
light-hatched regions of Fig.~\ref{fig:gA-ranges1} have been plotted in
Fig.~\ref{fig:gA-plus-curves} for comparison with the results of 
Sec.~\ref{sec:2n-beta-beta}. The result of the linear fit (\ref{eq:g-eff-lin}) is
not included in that plot since the hatched regions are a better way to
describe the (large) spread of the pnQRPA results for different masses $A$. This large
spread is not perceivable in the linear fit.

\begin{figure}[htbp!]
\centering
\includegraphics[width=0.6\columnwidth]{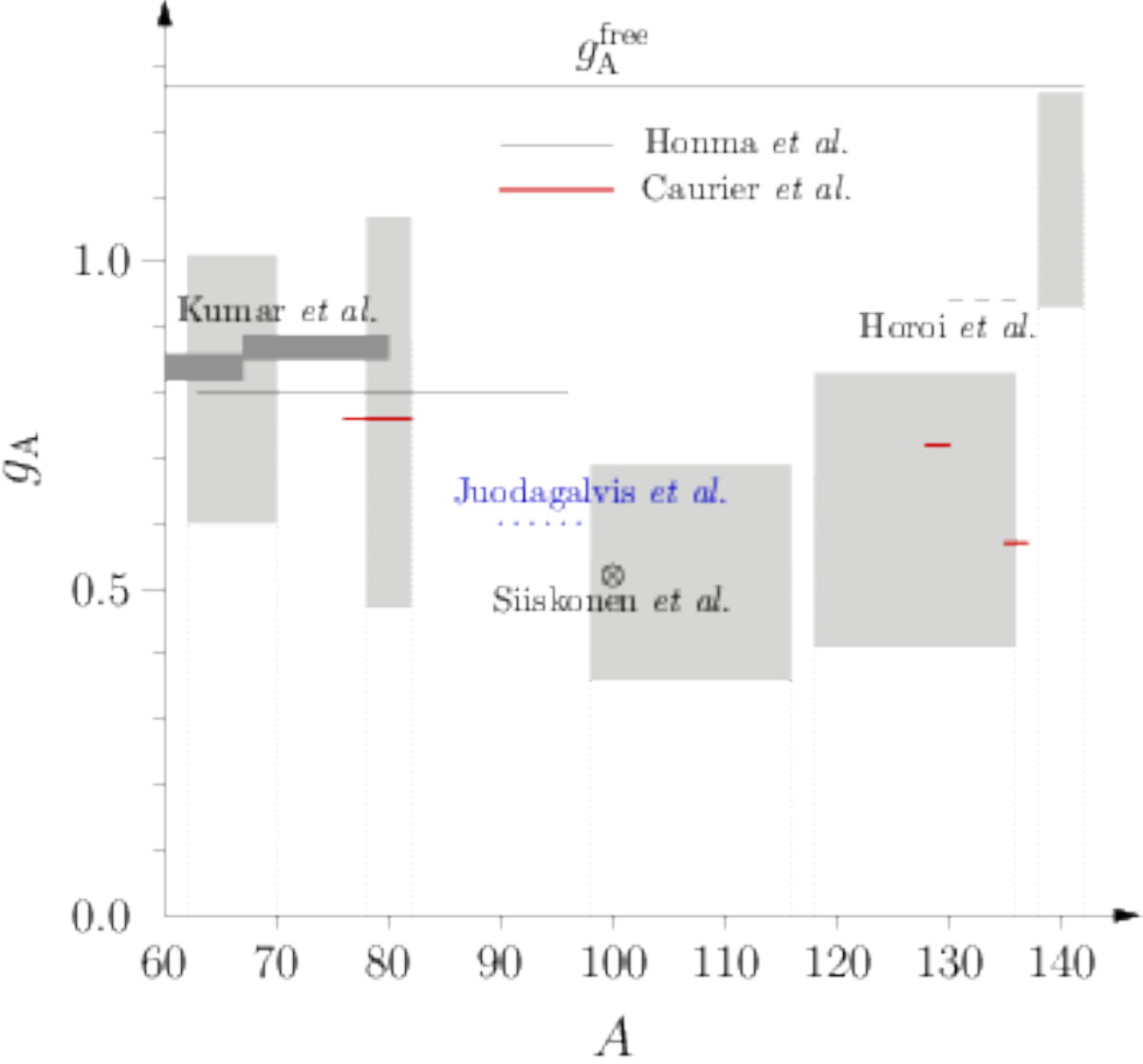}
\caption{Whole ranges of averaged effective values of $g_{\rm A}$ from Fig.~\ref{fig:gA-ranges1}
(light-hatched regions) plotted against the ISM results of Sec.~\ref{subsec:ISM}. 
The ISM results come from \cite{Honma2006} (Honma \textit{et al.}),
\cite{Caurier2012} (Caurier \textit{et al.}), 
\cite{Juodagalvis2005} (Juodagalvis \textit{et al.}), 
\cite{Kumar2016} (Kumar \textit{et al.}, dark-hatched regions), and
\cite{Siiskonen2001} (Siiskonen \textit{et al.}).}
\label{fig:gA-ranges2}
\end{figure}

In Fig.~\ref{fig:gA-ranges2} the light-hatched regions of Fig.~\ref{fig:gA-ranges1} 
(combined results of the pnQRPA analyses) have 
been plotted as a background against the results of the ISM (interacting shell model)
of Sec.~\ref{subsec:ISM}. As can be seen in the figure, the ISM results and the
pnQRPA results are in excellent agreement with each other. This is a non-trivial
result considering the quite different premises of these two different
calculation frameworks. For the masses $A\ge 138$ there is no comparison
between the two approaches since mid-shell heavy nuclei, with increasing deformation, 
are hard to access by the ISM due to an overwhelming computational burden. 

\section{Quenching of $g_{\rm A}$ in forbidden unique $\beta$ decays \label{sec:F-u}}

The forbidden unique $\beta$ transitions are the simplest ones that mediate $\beta$
decays between nuclear states of large angular-momentum difference $\Delta J$. In
particular, if one of the states is a $0^+$ state, then for
a $K^{th}$ forbidden ($K=1,2,3,\ldots$) unique beta decay 
the angular momentum of the other involved state is $J = K+1$. At the same time the 
parity changes in the odd-forbidden and remains the same in the even-forbidden 
decays \cite{Suhonen2007}. The change in angular momentum and parity for different 
degrees of forbiddenness is presented in Table~\ref{tab:forbiddenness}, and they obey
the simple rule 
\begin{align}
\label{eq:fu-rule}
(-1)^{\Delta J}\Delta\pi = -1 \,. \quad \textrm{(Forbidden unique decays)}
\end{align}
Here it is interesting to note that also the Gamow-Teller decays obey the rule
(\ref{eq:fu-rule}) if one of the involved nuclear states has the multipolarity $0^+$.

\begin{table}[t!]
\centering
\begin{tabular}{lrrrrrrr}
\toprule
   K & 1 & 2 & 3 & 4 & 5 & 6 & 7\\
  \midrule
$\Delta J$ & 2 & 3 & 4 & 5 & 6 & 7 & 8 \\
$\Delta\pi = \pi_i\pi_f$ & -1 & +1 & -1 & +1 & -1 & +1 & -1 \\
\bottomrule
\end{tabular}
  \caption{Change in angular momentum and parity in $K^{th}$ forbidden 
unique $\beta$ decays with a $0^+$ state as an initial or final nuclear state.} 
\label{tab:forbiddenness}
\end{table}

\subsection{Theoretical considerations \label{subsec:F-u-theory}}

The theoretical half-lives $t_{1/2}$ of $K^{th}$ forbidden unique $\beta$ decays 
can be expressed in terms of reduced transition probabilities $B_{K\mathrm{u}}$ 
and phase-space factors $f_{K\mathrm{u}}$. The $B_{K\mathrm{u}}$ is given by the NME, 
which in turn is given by the single-particle NMEs and one-body transition 
densities. Then (for further details see \cite{Suhonen2007})
\begin{equation}
t_{1/2} = \frac{\kappa}{f_{K\mathrm{u}}B_{K\mathrm{u}}}\ ; \quad
B_{K\mathrm{u}}=\frac{g_\mathrm{A}^2}{2J_i+1}|M_{K\mathrm{u}}|^2 \,,
\label{eq:half-life}
\end{equation}
where $J_i$ is the angular momentum of the mother nucleus and $\kappa$ is a 
constant with value \cite{Hardy1990}
\begin{equation}
\kappa = \frac{2\pi^3\hbar^7\mathrm{ln \  2}}{m_e^5c^4(G_{\rm F}
\cos \theta_{\rm C})^2}= 6147 \ \mathrm{s} \,,
\label{eq:kappa}
\end{equation}
with $G_{\rm F}$ being the Fermi constant and $\theta_{\rm C}$ being the Cabibbo angle.
The phase-space factor $f_{K\mathrm{u}}$ for the $K^{th}$ forbidden unique 
$\beta^{\pm}$ decay can be written as
\begin{equation}
f_{K\mathrm{u}} =\left( \frac{3}{4} \right)^K \frac{(2K)!!}{(2K+1)!!} 
\int_1^{w_0} C_{K\mathrm{u}}(w_e)p_ew_e(w_0-w_e)^2F_0(Z_f,w_e) \mathrm{d} w_e \,,
\label{eq:phase-space}
\end{equation}
where $C_{K\mathrm{u}}$ is the shape function for $K^{th}$ forbidden unique $\beta$ 
decays which can be written as (see, e.g., \cite{Suhonen2007,Gove1971})
\begin{equation}
C_{K\mathrm{u}}(w_e) = \sum_{k_e+k_{\nu}=K+2}\frac{\lambda_{k_e}p_e^{2(k_e-1)}(w_0-w_e)^{2(k_{\nu}-1)}}
{(2k_e-1)!(2k_{\nu}-1)!} \,,
\label{eq:shape-Ku}
\end{equation}
where the indices $k_e$ and $k_{\nu}$ (both $k = 1,2,3... $) come from the 
partial-wave expansion of the electron ($e$) and neutrino ($\nu$) wave functions. 
Here $w_e$ is the total energy of the emitted electron/positron, $p_e$ is the 
electron/positron momentum, $Z_f$ is the charge number of the daughter nucleus and 
$F_0(Z_f,w_e)$ is the Fermi function taking into account the coulombic 
attraction/repulsion of the electron/positron and the daughter nucleus 
\footnote{For positron emission the change $Z_f\to -Z_f$ has to be performed in
$F_0(Z_f,w_e)$ and $F_{k_e-1}(Z_f,w_{e})$, Eq.~(\ref{eq:lambda}) below.}. The factor
$\lambda_{k_e}$ contains the generalized Fermi function $F_{k_e-1}$ \cite{Behrens1982}
as the ratio
\begin{equation}
\lambda_{k_e} = \frac{F_{k_e-1}(Z_f,w_{e})}{F_{0}(Z_f,w_{e})}\,.
\label{eq:lambda}
\end{equation}
The integration is performed over the total (by electron rest-mass) scaled energy of the 
emitted electron/positron, $w_0$ being the endpoint energy, corresponding to the maximum 
electron/positron energy in a given transition.

The NME in (\ref{eq:half-life}) can be expressed as
\begin{equation}
M_{K\mathrm{u}} =
\sum_{ab} M^{(K\mathrm{u})}(ab)(\psi_f|| [c_a^{\dagger}\tilde{c}_{b}]_{K+1} ||\psi_i) \,,
\label{eq:NME-Ku}
\end{equation}
where the factors $M^{(K\mathrm{u})}(ab)$ are the single-particle matrix elements and the 
quantities $(\psi_f|| [c_a^{\dagger}\tilde{c}_{b}]_{K+1} ||\psi_i)$ are the one-body 
transition densities with $\psi_i$ being the initial-state wave function and $\psi_f$ 
the final-state wave function. The operator $c_a^{\dagger}$ is a creation operator for a 
nucleon in the orbital $a$ and the operator $\tilde{c}_{a}$ is the corresponding annihilation
operator. The single-particle matrix elements are given (in the Biedenharn-Rose phase 
convention) by
\begin{equation}
M_{K\mathrm{u}}(ab) = \sqrt{4\pi}\left(a||r^K\lbrack Y_K\mbox{\boldmath{$\sigma$}}
\rbrack_{K+1} i^K||b\right) \,,
\label{eq:spNME-Ku}
\end{equation}
where $Y_K$ is a spherical harmonic of rank $K$, $r$ the radial coordinate, and
$a$ and $b$ stand for the single-particle orbital quantum numbers. The NME
(\ref{eq:spNME-Ku}) is given explicitly in \cite{Suhonen2007}.

\subsection{First-forbidden unique $\beta$ decays \label{subsec:FF-u}}

The first-forbidden unique $\beta$ transitions are mediated by a 
rank-2 (i.e. having angular-momentum content 2) parity-changing spherical 
tensor operator [a special case of the operator (\ref{eq:spNME-Ku})], 
schematically written as $\mathcal{O}(2^-)$.
For these decays it is customary to modify the
general structure of Eqs.~(\ref{eq:half-life})--(\ref{eq:phase-space}) by replacing
the phase-space factor $f_{K=1,\mathrm{u}}$ of (\ref{eq:phase-space})
by a 12 times larger phase-space factor $f_{1\mathrm{u}}$, i.e.
\begin{equation}
f_{1\mathrm{u}} = 12f_{K=1,\mathrm{u}} \,,
\label{eq:f1-phspace}
\end{equation}
yielding a factor $\log 12 = 1.079$ times larger comparative half-lives 
(\ref{eq:logft}) than in the standard definition (\ref{eq:half-life}).

\begin{figure}[htbp!]
\centering
\includegraphics[width=0.6\columnwidth]{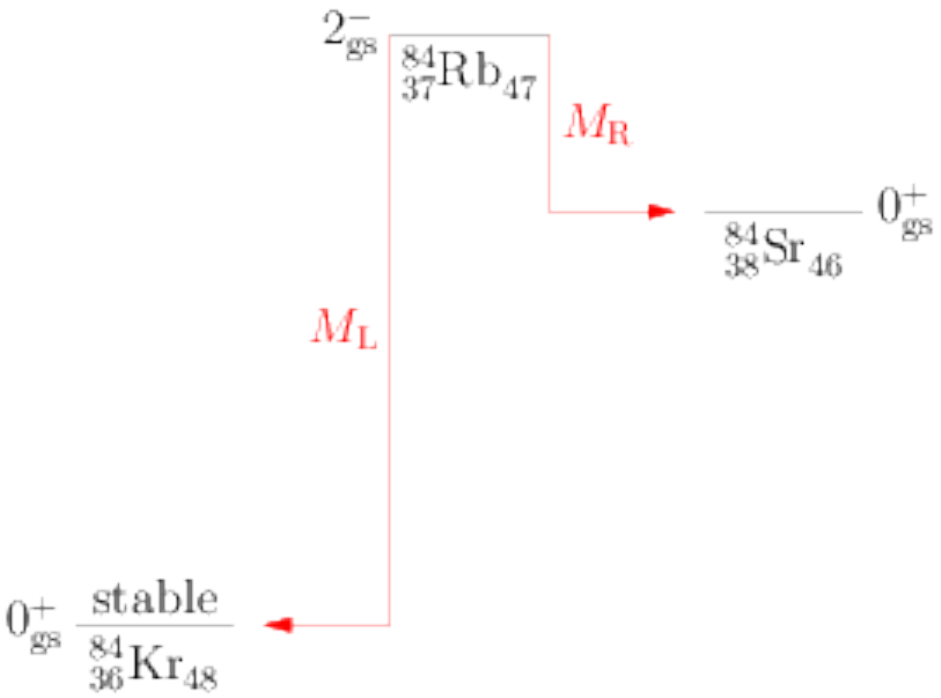}
\caption{First-forbidden unique beta decays in the $A=84$ Kr-Rb-Sr isobaric chain.}
\label{fig:FF-transitions}
\end{figure}

In the quenching studies it is advantageous to use the simplest first-forbidden
transitions, namely the ground-state-to-ground-state ones. In 
Fig.~\ref{fig:FF-transitions} are depicted the first-forbidden unique 
ground-state-to-ground-state $\beta^-$ and $\beta^+$/EC 
transitions between even-even $0^+$ and odd-odd $2^-$
ground states in the $A=84$ Kr-Rb-Sr isobaric chain. Shown is the 
lateral feeding \emph{from} a middle odd-odd nucleus to adjacent even-even
ground states. In the figure, as also in Fig.~\ref{fig:GT-transitions} for the
Gamow-Teller transitions, the NME is denoted by $M_{\rm L}$ ($M_{\rm R}$) in case it 
is to the left (right) of the central nucleus.
 
In the early work \cite{Ejiri1968} a systematic schematic analysis of the first-forbidden 
unique $\beta$ decays was performed from the point of view of suppression factors stemming
from the effect of E1 (electric dipole) giant resonance in the final odd-odd nucleus. In
\cite{Towner1971} the suppression mechanism of the
first-forbidden and third-forbidden $\beta$ decays of light nuclei ($A\le 50$) was studied
by using simple shell-model estimates and first-order perturbation theory.
The hindrance was traced to the repulsive $T=1$ (isospin 1) particle-hole force.

In the work \cite{Ejiri2014} 19 first-forbidden unique ground-state-to-ground-state
$\beta$-decay transitions were studied. The interesting transitions are the ones
where both $M_{\rm L}$ and $M_{\rm R}$ NMEs are known experimentally, like in
the case of Fig.~\ref{fig:FF-transitions}. The experimental values of the NMEs
can be deduced by using Eqs.~(\ref{eq:half-life}) and (\ref{eq:kappa}) and by adopting
the free value of the axial-vector coupling strength
% Footnote
\footnote{In \cite{Ejiri2014} the Bohr-Mottelson (BM) formulation \cite{Bohr1969}
of first-forbidden decays is used. The difference between the present and the BM
formulation can be crystallized into the following relations: 
$M(\mathrm{BM})=M_{1\mathrm{u}}/\sqrt{4\pi}$, 
$B(\mathrm{BM})=B_{1\mathrm{u}}/(4\pi g_{\rm A}^2)$,
$f_1(\mathrm{BM})=3f_{1\mathrm{u}}/4$. In addition, since 
$g_{\rm V}(\mathrm{BM})=G_{\rm F}g_{\rm V}$ and $g_{\rm A}(\mathrm{BM})=G_{\rm F}g_{\rm A}$,
one has to make replacements $g_{\rm A}(\mathrm{BM})\to g_{\rm A}$ and
$g_{\rm V}(\mathrm{BM})\to 1$ in order to go from the BM formulation to the
present one.}.
% Footnote
In this case one can use the geometric mean (\ref{eq:G-mean}) of the left and
right NMEs in the analysis, making the analysis more stable.
In \cite{Ejiri2014} a $g_{\rm ph}$- and $g_{\rm pp}$-renormalized Bonn-A G matrix was 
used as the two-nucleon interaction in a pnQRPA framework. The two-quasiparticle and 
pnQRPA NMEs were compared with the ones extracted from the measured comparative
half-lives. Again the relations (\ref{eq:k-bar}) and (\ref{eq:k-bar-g-eff}) can be 
used to obtain the value 
\begin{align} 
\label{eq:g-eff-ave-Ejiri}
	g^{\rm eff}_{\rm A} \approx 0.45\times 1.27 = 0.57 
\end{align}
for the effective axial-vector coupling strength using the pnQRPA wave functions. 
The average of the values of the leading two-quasiparticle NMEs gives in turn
\begin{align} 
\label{eq:g-eff-ave-Ejiri-qp}
	g^{\rm eff}_{\rm A}(\mathrm{2qp}) \approx 0.18\times 1.27
= 0.23 \,,
\end{align}
implying the ratio
\begin{align} 
\label{eq:k-ratio-ff}
k = \frac{\bar{M}_{\rm pnQRPA}}{\bar{M}_{\rm qp}} = 0.4
\end{align}
and thus a drastic nuclear many-body effect when going from the two-quasiparticle 
level of approximation to the pnQRPA level. The 2qp-NME to pnQRPA-NME comparison is the only 
one where a clean separation between the nuclear-medium effects and the nuclear-model 
effects can be achieved, the nuclear-model effect being responsible for the 
(in this case large) shift in the values of the NMEs.

\section{Higher-forbidden unique $\beta$ decays \label{sec:high-forb-u}}

\begin{figure}[htbp!]
\centering
\includegraphics[width=0.6\columnwidth]{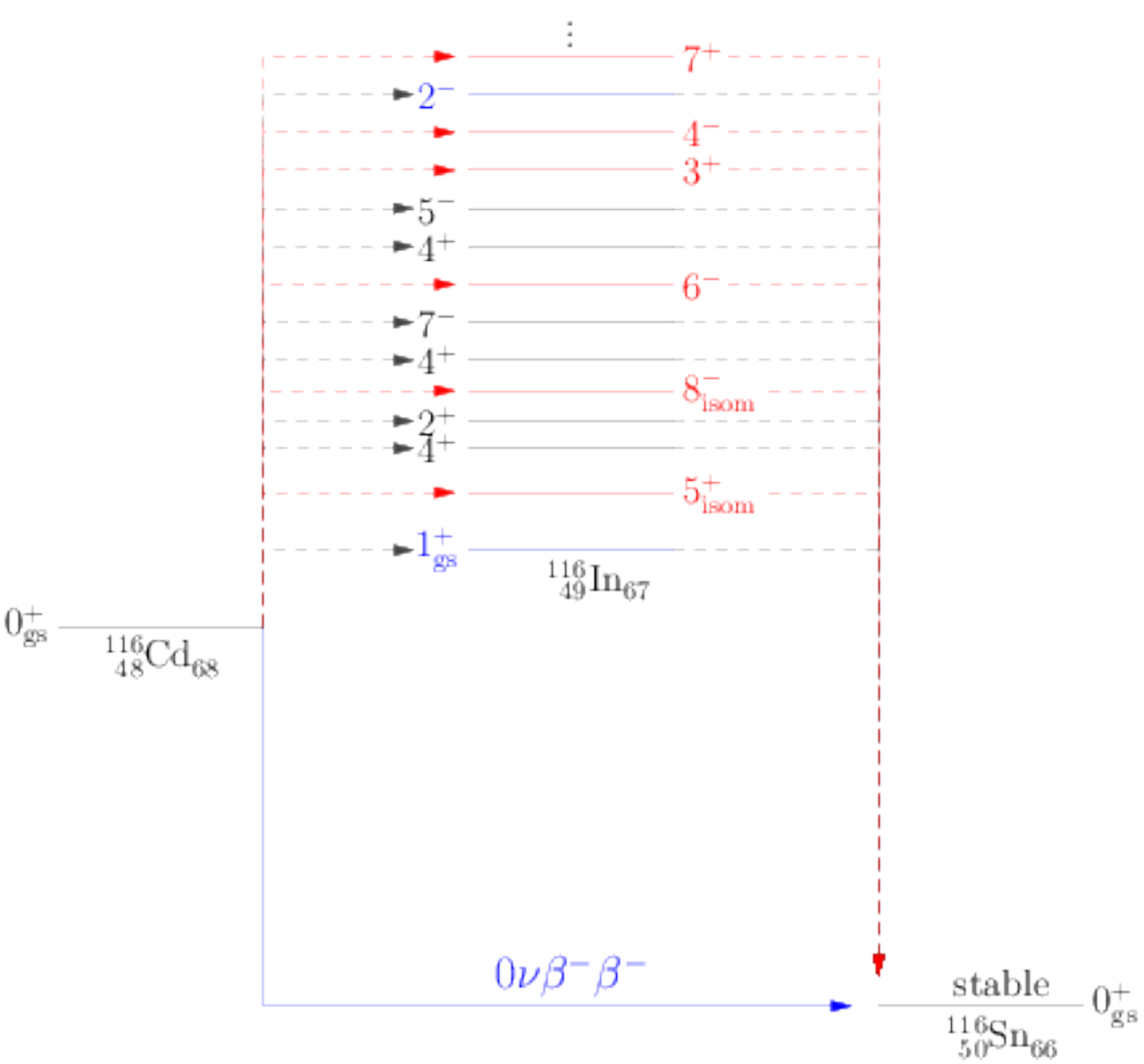}
\caption{The $0\nu\beta\beta$ decay of $^{116}$Cd to $^{116}$Sn via the virtual 
intermediate states in $^{116}$In. The transitions between $^{116}$Cd ($^{116}$Sn)
and $^{116}$In constitute the left-leg (right-leg) transitions.}
\label{fig:0vbb-interm}
\end{figure}

Early studies of the quenching in the second- and third-forbidden unique $\beta$ decays 
were performed in \cite{Towner1971,Warburton1970}. The work of \cite{Towner1971} 
was discussed in Sec.~\ref{subsec:FF-u}. In \cite{Warburton1970} these $\beta$
decays were studied using a simple interacting shell model and 
the unified model (deformed shell model) for six $\beta$ transitions in the 
$A=10,22,26,40$ nuclei. The interest for these studies derived from nuclear-structure 
considerations: how to explain in a nuclear model the hindrance phenomena occurring
in certain measured $\beta$ transitions. Beyond this, the incentive to study the 
Gamow-Teller (Sec.~\ref{sec:GT}), 
first-forbidden unique (Sec.~\ref{subsec:FF-u}), and higher-forbidden unique (this
section) $\beta$ decays stems from their relation to the Gamow-Teller type of 
NME involved in $0\nu\beta\beta$ decays. The $0\nu\beta\beta$ decays proceed via
virtual intermediate states of all multipolarities $J^{\pi}$ due to the multipole
expansion of the Majorana-neutrino propagator (see, e.g., 
\cite{REPORT,Vergados2012,Vergados2016,Kort1,Kort2,Kort3,Suh-Kort,Hyvarinen2015,Hyvarinen2016}). 
Studies of the quenching of
these two-leg (``left-leg'' and ``right-leg'' transitions illustrated in
the schematic Fig.~\ref{fig:0vbb-interm} for the $0\nu\beta\beta$
decay of $^{116}$Cd to $^{116}$Sn via the virtual intermediate states in $^{116}$In) 
virtual transitions is
of paramount importance to, e.g., estimate the sensitivities of the present and
future neutrino experiments to the Majorana-neutrino mass. The possible quenching 
of these intermediate multipole transitions in the GT type of $0\nu\beta\beta$ NME
can be, in a simplistic approach, condensed into an effective axial coupling,
$g^{\rm eff}_{{\rm A},0\nu}$, multiplying the NME:
\begin{equation}
\label{eq:GTGT-0nu}
   M^{(0\nu)}_{\rm GTGT} = (g^{\rm eff}_{{\rm A},0\nu})^2 \sum_{J^{\pi}}
(0^+_f||\mathcal{O}^{(0\nu)}_{\rm GTGT}(J^{\pi})||0^+_i) \,,
\end{equation}
where $\mathcal{O}^{(0\nu)}_{\rm GTGT}$ denotes the transition operator mediating the
$0\nu\beta\beta$ transition through the various multipole states $J^{\pi}$, 
$0^+_i$ denotes the initial ground state, 
and the final ground state is denoted by $0^+_f$ (here, for simplicity, we assume
a ground-state-to-ground-state transition). The effective axial coupling relevant 
for $0\nu\beta\beta$ decay is denoted as $g^{\rm eff}_{{\rm A},0\nu}$ to emphasize that 
its value may deviate from the one determined in single beta and $2\nu\beta\beta$ 
decays. The remarkable feature of Eq.~(\ref{eq:GTGT-0nu}) is that the effective
axial coupling strength is raised to 2$nd$ power making the value of 
$g^{\rm eff}_{{\rm A},0\nu}$ play an extremely important role in determining the 
$0\nu\beta\beta$-decay rate which is (neglecting the smaller double Fermi and 
tensor contributions) proportional to the squared NME and thus to the 4$th$ power 
of the coupling:
\begin{equation}
\label{eq:0vbb}
   0\nu\beta\beta-\mathrm{rate} \sim \left\vert M^{(0\nu)}_{\rm GTGT}\right\vert^2 = 
   g_{{\rm A},0\nu}^4 \left\vert \sum_{J^{\pi}} 
   (0^+_f||\mathcal{O}^{(0\nu)}_{\rm GTGT}(J^{\pi})||0^+_i)\right\vert^2 \,.
\end{equation}

The quenching related to the left-leg and right-leg $\beta$ transitions of 
Fig.~\ref{fig:0vbb-interm} can be studied by using the theoretical machinery of
Sec.~\ref{subsec:F-u-theory}. In \cite{Kostensalo2017a} this machinery was
applied to 148 potentially measurable second-,\- third-, fourth-, fifth-, 
sixth- and seventh-forbidden unique beta transitions. The calculations were done 
using realistic single-particle model spaces and G-matrix-based microscopic two-body 
interactions. The results of \cite{Kostensalo2017a} could shed light on the
magnitudes of the NMEs corresponding to the high-forbidden unique
$0^+\leftrightarrow J^{\pi}=3^+,4^-,5^+,6^-,7^+,8^-$ virtual transitions taking part in
neutrinoless double beta decay, as shown in Fig.~\ref{fig:0vbb-interm}.

\begin{figure}[htb]
\centering
\includegraphics[width=1.0\columnwidth]{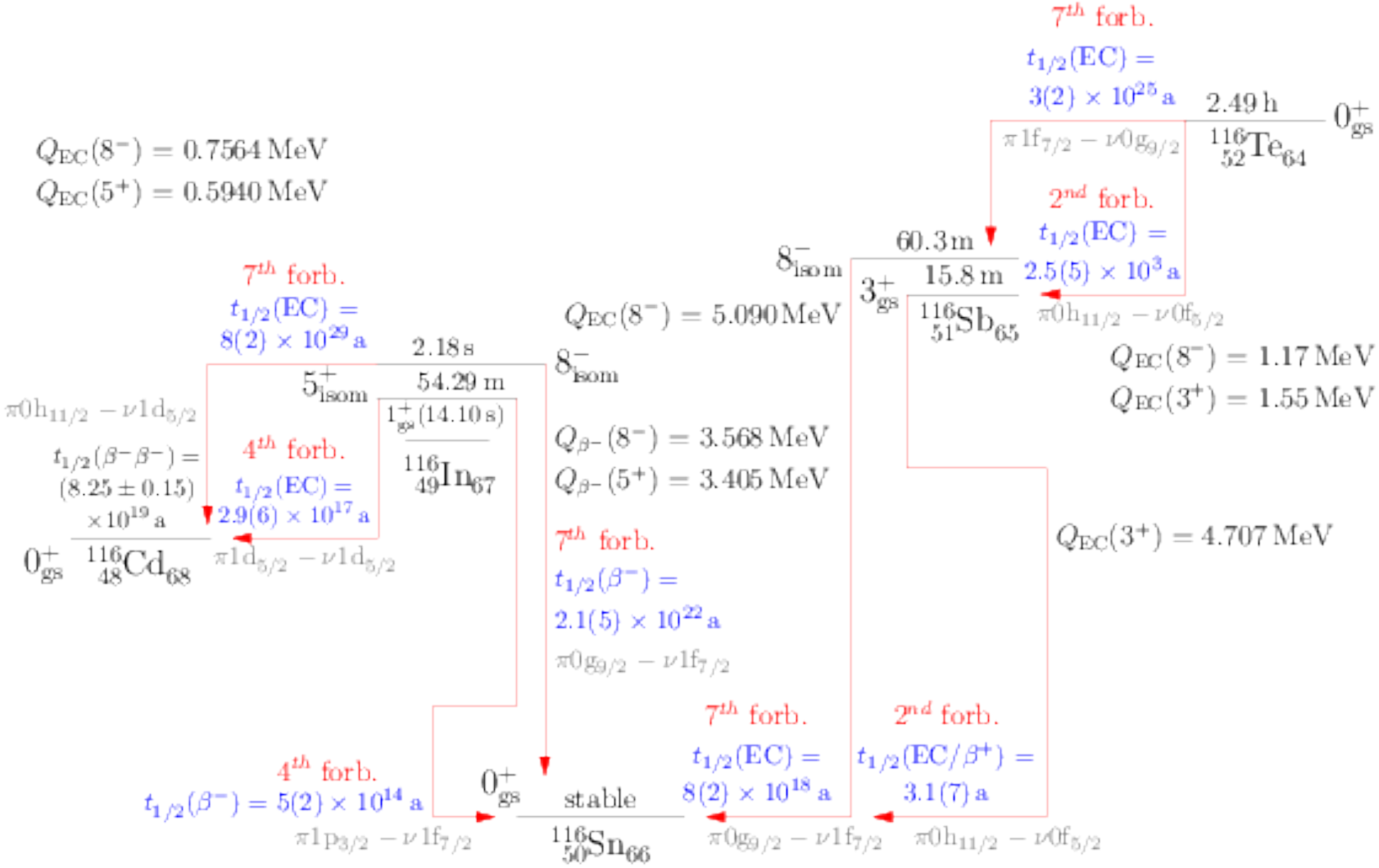}
\caption{Predicted half-lives and their error estimates (in parenthesis) for 
$\beta^-$ and EC (electron-capture) transitions in the isobaric chain $A=116$. 
The spin-parity assignments, decay energies ($Q$ values) and life-times of 
the nuclear ground (gs) and isomeric (isom) states are experimental data and taken 
from \cite{ENSDF}. The $2\nu\beta\beta$ half-life is taken from 
\cite{Barabash2013}. In addition to the half-lives the degree of forbiddenness
and the leading single-particle transition are shown.}
\label{fig:beta-116}
\end{figure}

\begin{figure}[htb]
\centering
\includegraphics[width=1.0\columnwidth]{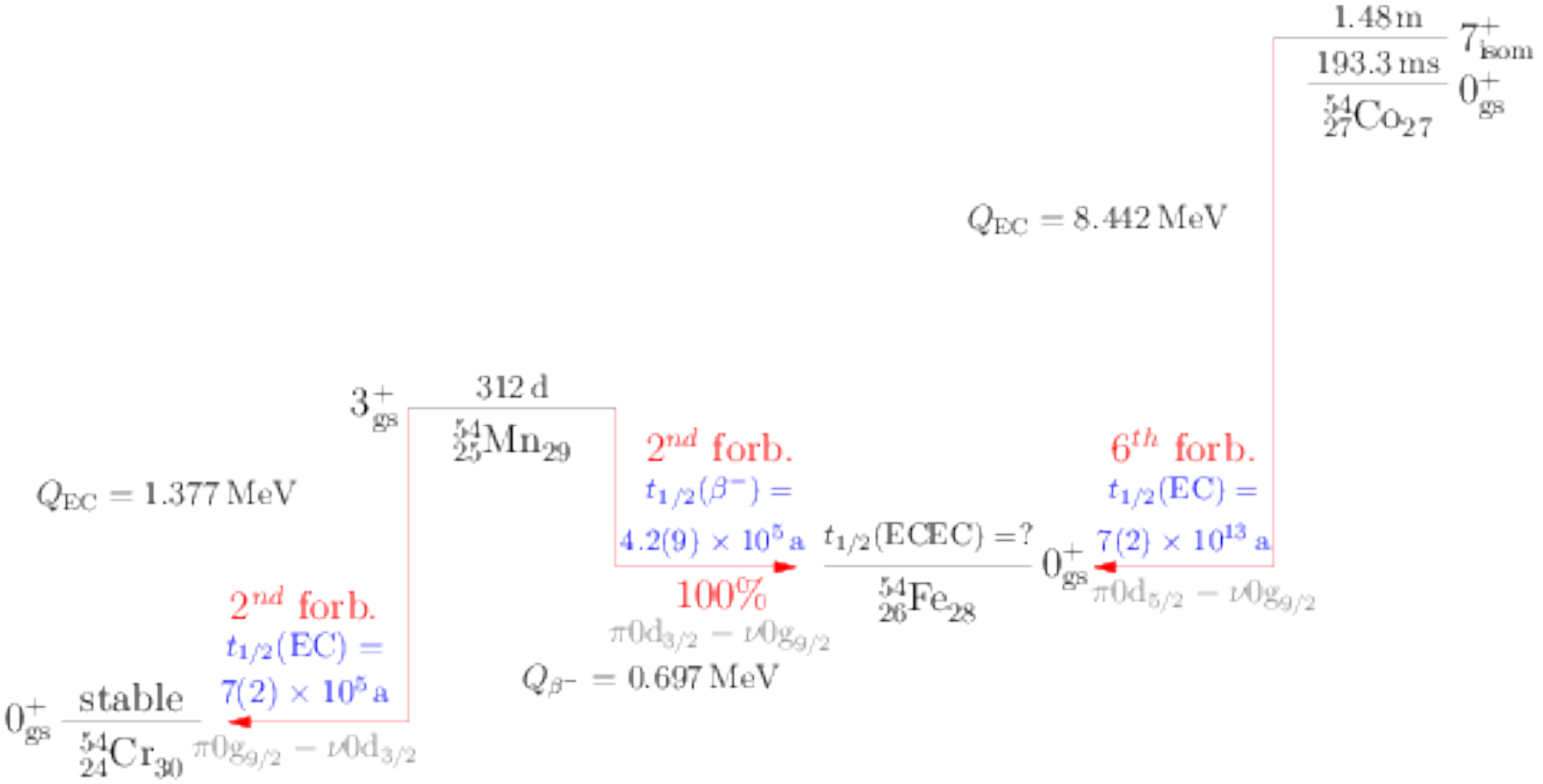}
\caption{The same as Fig.~\ref{fig:beta-116} for the second- and sixth-forbidden
$\beta$ decays in the isobaric chain $A=54$.}
\label{fig:beta-54}
\end{figure}

In \cite{Kostensalo2017a} the ratio $k$, Eq.~(\ref{eq:k-ratio}) below, of
the NMEs, calculated by the pnQRPA, $M_{\rm pnQRPA}$, and a two-quasiparticle model, 
$M_{\rm qp}$, was studied and compared with earlier calculations for the allowed 
Gamow-Teller $1^+$ \cite{Ejiri2015} and first-forbidden spin-dipole (SD) 
$2^-$ \cite{Ejiri2014} transitions. Based on this comparison the \emph{expected}
half-lives of the studied $\beta$-decay transitions were predicted.
An example case of the expected half-lives of second-, fourth-, and 
seventh-forbidden $\beta$ decays is shown in Fig.~\ref{fig:beta-116}. 
The computed NMEs are corrected by the use of the ratio of the geometric means
(\ref{eq:G-mean}) of the experimental and pnQRPA NMEs,
\begin{align} 
\label{eq:k-ratio-exp}
k_{\rm NM} = \frac{\bar{M}_{\rm exp}}{\bar{M}_{\rm pnQRPA}} \,,
\end{align}
extracted from the GT work of \cite{Ejiri2015}, to predict the transition 
half-lives of the figure. In the figure one sees that the expected half-lives
range from 4 years to the astronomical $9\times 10^{29}$ years. It is expected
that the decays to and from isomeric states are not measurable and the
decays between the nuclear ground states are masked by transitions to excited
states with lesser degree of forbiddenness. Only in some cases the high-forbidden
$\beta$ decay exhausts 100\% of the decay rate between two nuclear ground states;
one example being the second-forbidden $\beta^-$ transition
$^{54}\mathrm{Mn}(3^+_{\rm gs})\to\,^{54}\mathrm{Fe}(0^+_{\rm gs})$, with a half-life
$4.2(9)\times 10^5$ years, shown in Fig.~\ref{fig:beta-54}. Even in this case
the measurement will be challenging due to the Gamow-Teller type of
electron-capture feeding of the first excited $2^+$ state of $^{54}$Cr, taking
practically 100\% of the feeding intensity.

The geometric mean of the EC/$\beta^+$ and $\beta^-$ NMEs, defined in 
(\ref{eq:G-mean}), can be generalized to a geometric mean of $n$ NMEs, $M_i$, 
$i=1,2,\ldots n$, of successive $\beta$ transitions with a common mother or 
daughter nucleus: 
\begin{equation}
\label{eq:G-mean-n}
\bar{M} = \left( \prod_{i=1}^n M_i  \right )^{1/n} \,.
\end{equation}
Here the aim, as in the case of (\ref{eq:G-mean}), is to reduce the fluctuations in the
computed NMEs by exploiting the compensating trends of the $\beta^-$ and $\beta^+$/EC 
branches of decay when changing the value of the particle-particle interaction 
parameter $g_{\rm pp}$ of the pnQRPA. One can now define the ratio
\begin{align} 
\label{eq:k-ratio}
k = \frac{\bar{M}_{\rm pnQRPA}}{\bar{M}_{\rm qp}}
\end{align}
of the pnQRPA-calculated mean NME, $\bar{M}_{\rm pnQRPA}$, and the mean two-quasiparticle 
NME, $\bar{M}_{\rm qp}$, computed by using (\ref{eq:G-mean-n}). The ratio $k$ is a measure
of the evolution of the nuclear-model dependent many-body effects on the computed NME.
The ratio (\ref{eq:k-ratio}) is independent of the nuclear-medium effects (the fundamental
quenching of Sec.~\ref{sec:medium-eff}) and gives an idea of how the quenching of
$g_{\rm A}$ depends on the degree of complexity of the adopted nuclear model. 

\begin{figure}[htbp!]
\centering
\includegraphics[width=0.8\columnwidth]{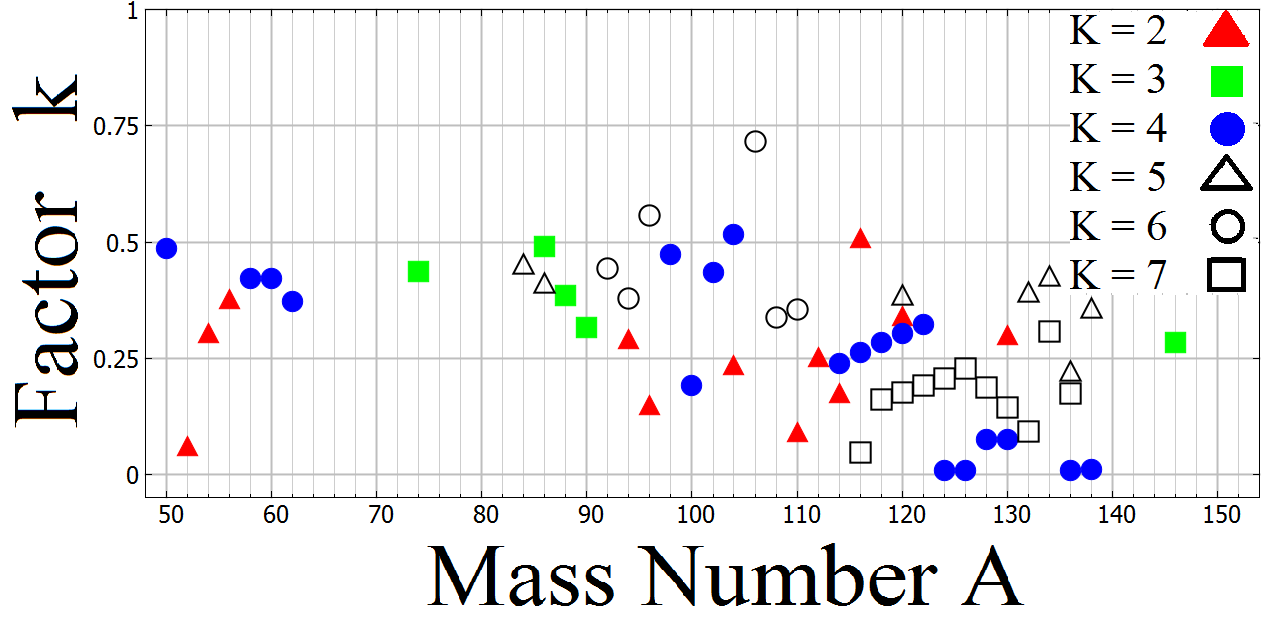}
\caption{Ratio (\ref{eq:k-ratio}) as a function of the mass number $A$ for $\beta$
transitions involving solely non-magic reference nuclei. The degree of forbiddenness
$K$ is indicated by color and shape of the symbol.}
\label{fig:k-nonmagic}
\end{figure}

\begin{figure}[htbp!]
\centering
\includegraphics[width=0.8\columnwidth]{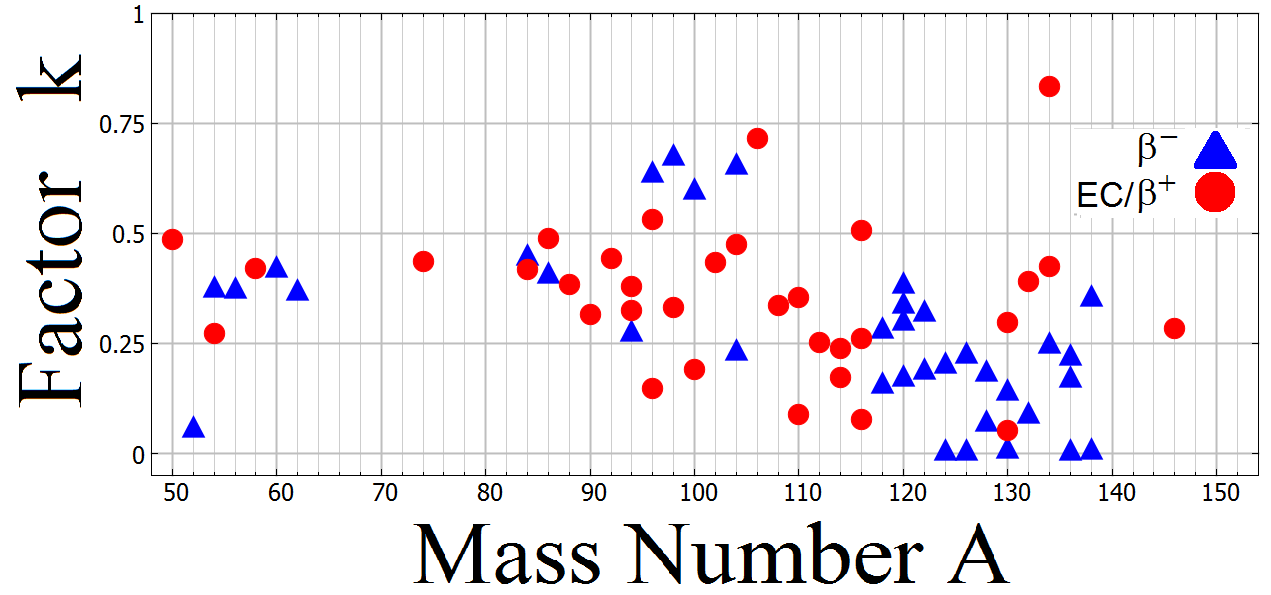}
\caption{Ratio (\ref{eq:k-ratio}) as a function of the mass number $A$ separated to
$\beta^-$ and EC/$\beta^+$ $K$-forbidden transitions of Fig.~\ref{fig:k-nonmagic}.}
\label{fig:k-beta-pm}
\end{figure}

In \cite{Kostensalo2017a} the $\beta$ transitions were divided in two groups: 
GROUP 1 contained only non-magic
even-even reference nuclei (i.e. nuclei where the pnQRPA and the associated BCS
(Bardeen-Cooper-Schrieffer) calculation were performed), whereas GROUP 2 contained
(semi)magic reference nuclei. The transitions in GROUP 2 were left out from the
analysis of the ratio $k$ of (\ref{eq:k-ratio}) since the BCS results tend to 
be unstable at magic shell closures. In Fig.~\ref{fig:k-nonmagic} the ratio $k$ 
is shown for transitions belonging to GROUP 1. The same $k$ distribution is shown
in terms of division to $\beta^-$ and EC/$\beta^+$ decays in Fig.~\ref{fig:k-beta-pm}.

From Fig.~\ref{fig:k-nonmagic} it is visible that the second- and fourth-forbidden 
$\beta$ transitions are distributed to masses below $A=62$ and above masses $A=92$,
whereas the third-forbidden decays occupy the mass range $74\le A\le 90$. The 
sixth-forbidden decays occur within the range $92\le A\le 110$ and the
seventh-forbidden decays occur above $A=116$. The fifth-forbidden decays occur in
a scattered way above $A=84$. From Fig.~\ref{fig:k-beta-pm} one observes that
most of the $\beta^-$ decays are concentrated above mass $A=118$ where also quite
low values of $k$ can be obtained. The EC/$\beta^+$ decays, on the other hand,
are more concentrated in the middle-mass region $82\le A\le 116$.

\begin{figure}[htbp!]
\centering
\includegraphics[width=0.8\columnwidth]{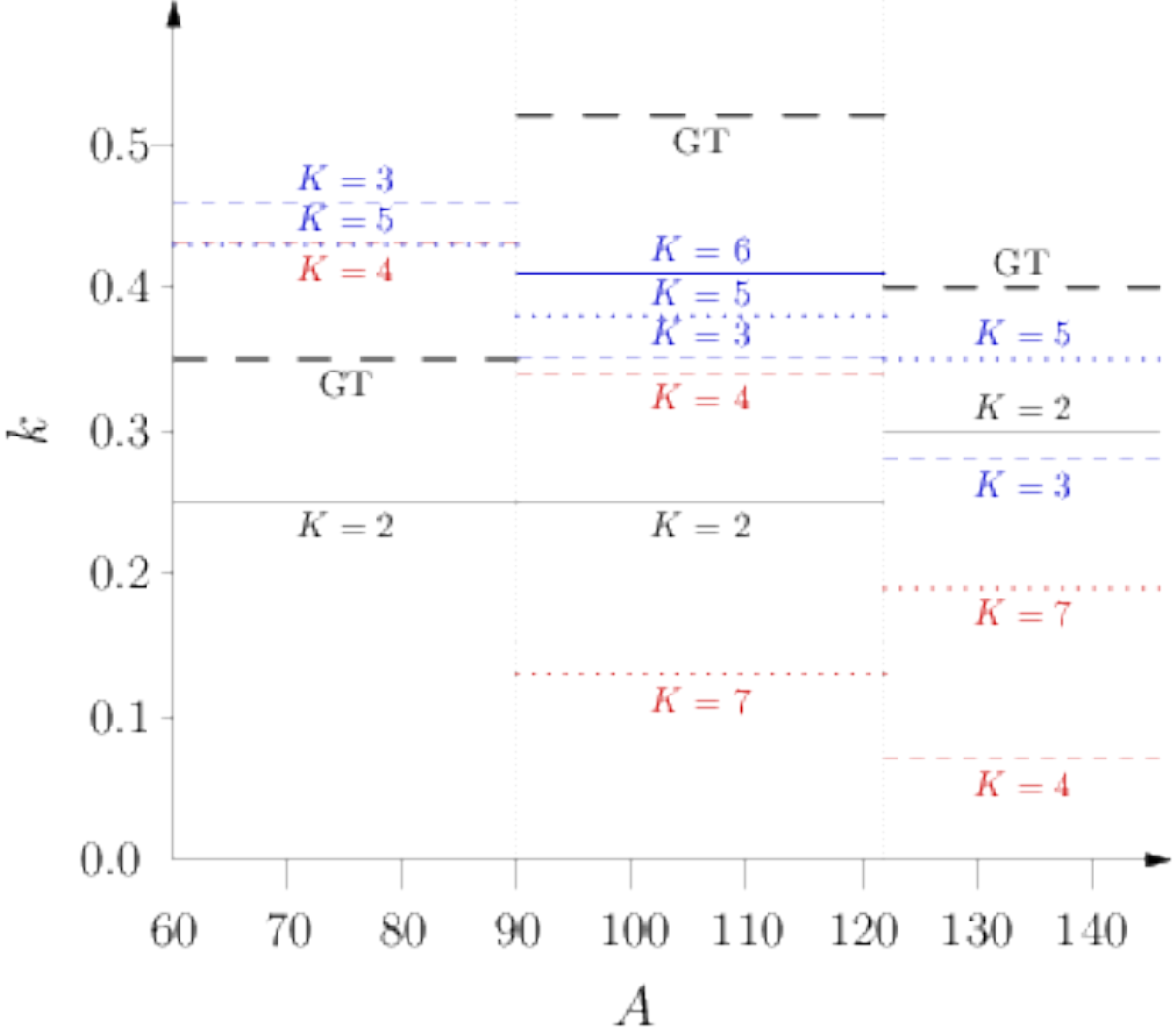}
\caption{Illustration of the values of the ratio $k$ (\ref{eq:k-ratio}), taken
from Table~\ref{tab:k_k}, for the three mass regions $A=50-88$, $A=90-122$, and
$A=122-146$ for different degrees of forbiddenness $K$. GT denotes the ratio $k$ for
Gamow-Teller transitions.}
\label{fig:ratios-k}
\end{figure}

Fig.~\ref{fig:k-nonmagic} suggests that the values of the ratio (\ref{eq:k-ratio}) 
can be classified in terms of three mass regions, namely $A=50-88$ ($k\sim 0.4$),
$A=90-120$ (values of $k$ have a scattered, decreasing trend), and 
$A=122-146$ (a low-$k$ region with $k\sim 0.2$). The ratios $k$ for the three 
mass regions and for various degrees of forbiddenness $K$ are shown in 
Table~\ref{tab:k_k} for $\beta$ transitions belonging to GROUP 1. A comparison is
made to the GT results of \cite{Ejiri2015} and SD results of \cite{Ejiri2014}.
The ratios are also plotted in Fig.~\ref{fig:ratios-k} for illustrative purposes.
In the figure one can see that the trend, in terms of the mass number $A$, is a
bit different for the Gamow-Teller and the forbidden ($K\ge 2$) transitions. For most
of the forbidden transitions, namely $K=3,4,5$, $k$ has a decreasing tendency as 
a function of $A$, in particular for the $k=4$ transitions. For $K=2$ and $K=7$ 
a slightly increasing tendency is observed. It seems, on average, that the quenching
of the forbidden transitions is somewhat stronger than that of Gamow-Teller 
transitions in the mass region $A=90-146$.

The numbers in Table~\ref{tab:k_k} suggest that, in the gross, $k$ is independent of 
the degree of forbiddenness and thus the (low-energy) forbidden unique contributions 
[obeying the simple rule (\ref{eq:fu-rule})] to the
$0\nu\beta\beta$ NME (\ref{eq:0vbb}) should be roughly uniformly quenched. If these
conclusions can be generalized to include also the non-unique $\beta$ transitions,
obeying the rule $(-1)^{\Delta J}\Delta\pi = +1$, one can then speak about
an effective axial coupling, $g^{\rm eff}_{{\rm A},0\nu}$, in front of 
the $0\nu\beta\beta$ NME in (\ref{eq:GTGT-0nu}), at least for low
intermediate excitation energies. The quenching for these low 
intermediate excitation energies could then be deduced from the hatched regions
of Fig.~\ref{fig:gA-ranges1}, implying the effective axial couplings listed in
Table~\ref{tab:g-0vbb} for the three mass regions of interest for 
$0\nu\beta^-\beta^-$-decay calculations in the pnQRPA framework. 
At this point it has to be noted that a ``low'' excitation energy is still 
an undefined notion that has to be investigated in future works.

	\begin{table*}%[H]
\centering
\begin{tabular}{cccccccccc}
\toprule
   $A$ & GT \cite{Ejiri2015} & $K=1$ \cite{Ejiri2014} & 
$K=2$&$K=3$&$K=4$&$K=5$&$K=6$&$K=7$& Avg.\\
  \midrule
	$50-88$ & 0.35 & 0.40 & 0.25 & 0.46 & 0.43 & 0.43 & - & - & 0.39 \\
	$90-122$ & 0.52 & 0.40 & 0.25 & 0.35 & 0.34 & 0.38 & 0.41 & 0.13 & 0.31 \\
	$122-146$ & 0.40 & 0.40 & 0.30 & 0.28 & 0.07 & 0.35 & - & 0.19 & 0.24 \\
	Average & 0.42 & 0.40 & 0.27 & 0.36 & 0.28 & 0.39 & 0.41 & 0.16 & 0.31 \\
\bottomrule
\end{tabular}
  \caption{Ratio (\ref{eq:k-ratio}) for three mass regions and for various
degrees of forbiddenness $K$ for $\beta$ transitions belonging to GROUP 1. The 
results of the Gamow-Teller (GT, see Sec.~\ref{subsec:QRPA}) and first-forbidden 
($K=1$, see Sec.~\ref{subsec:FF-u}) decays are quoted for comparison.} 
\label{tab:k_k}
\end{table*}

\begin{table*}
\centering
\begin{tabular}{cccc}
\toprule
Mass range & $A=76-82$ & $A=100-116$ & $A=122-136$ \\
\midrule
$g^{\rm eff}_{{\rm A},0\nu}$ & $0.7-0.9$ & 0.5 & $0.5-0.7$ \\
\bottomrule
\end{tabular}
   \caption{Values of the low-energy effective axial coupling, $g^{\rm eff}_{{\rm A},0\nu}$ 
of (\ref{eq:GTGT-0nu}), for the three mass regions of interest for 
$0\nu\beta^-\beta^-$-decay calculations in the pnQRPA framework.}
\label{tab:g-0vbb}
\end{table*}

Finally, it should be stressed that the use of the low-energy effective axial 
coupling, $g^{\rm eff}_{{\rm A},0\nu}$, is particular to the pnQRPA many-body framework
and reflects the deficiencies of pnQRPA in calculating the magnitudes of the
NMEs of the allowed and forbidden unique $\beta$-decay transitions. It is not
directly related to the more fundamental quenching of the axial-vector coupling
strength $g_{\rm A}$, related to the meson-exchange currents, delta isobars, two-body
weak currents, etc., discussed in Sec.~\ref{sec:medium-eff}, but it is rather
a nuclear-model effect, discussed in Sec.~\ref{sec:model-eff}.

\section{Quenching of $g_{\rm A}$ in forbidden non-unique $\beta$ decays \label{sec:F-nu}}

The general theory of forbidden beta decays is outlined in \cite{Behrens1982} and
\cite{Schopper1966}. Streamlined version of those is given in \cite{Mustonen2006}.

\subsection{Theoretical considerations \label{sec:F-nu-theory}}

In the forbidden non-unique $\beta$ decay the half-life 
can be given, analogously to (\ref{eq:half-life}), in the form
\begin{equation}\label{eq:half-life2}
t_{1/2} = \kappa/\tilde{C} \,,
\end{equation}
where $\tilde{C}$ is the dimensionless integrated shape function, given by
\begin{equation} \label{eq:shape-f}
\tilde{C} = \int^{w_0}_1 C(w_{e}) p w_{\rm e} (w_0 - w_{e})^2 F_0(Z_f,w_{e}) 
{\rm d}w_{e} \,,
\end{equation}
with the notation explained in Sec.~\ref{subsec:F-u-theory}.
The general form of the shape factor of Eq. \eqref{eq:shape-f} is a sum
\begin{equation} \label{eq:shape-factor}
C(w_e) = \sum_{k_e, k_{\nu}, K} \lambda_{k_e} \left[M_{K}(k_e,k_{\nu})^2 + m_{K}(k_e,k_{\nu})^2 -
\frac{2\gamma_{k_e}}{k_ew_e} M_{K}(k_e,k_{\nu}) m_{K}(k_e,k_{\nu})
\right] \,,
\end{equation}
where the factor $\lambda_{k_e}$ was given in (\ref{eq:lambda}) and $Z_f$ is the
charge number of the final nucleus. The indices $k_e$ and $k_{\nu}$ ($k = 1,2,3... $) 
are related to the 
partial-wave expansion of the electron ($e$) and neutrino ($\nu$) wave functions, 
$K$ is the order of forbiddenness of the transition, and
$\gamma_{k_e} = \sqrt{k_e^2 - (\alpha Z_f)^2}$, $\alpha\approx 1/137$ being the 
fine-structure constant. The nuclear-physics information is hidden in the factors
$M_{K}(k_e,k_{\nu})$ and $m_{K}(k_e,k_{\nu})$, which are complicated combinations of the 
different NMEs and leptonic phase-space factors. For more information on the 
integrated shape function, see \cite{Behrens1982,Mustonen2006}.

The quite complicated shape factor (\ref{eq:shape-factor}) can be simplified in the
so-called $\xi$ approximation when the coulomb energy of the emitted $\beta$ 
particle at the nuclear surface is much larger than the endpoint energy, i.e.
$\xi=\alpha Z_f/2R\gg w_0$, where
$R$ is the nuclear radius. Then the forbidden non-unique transition can be
treated as a unique one of the same $\Delta J$. Applicability of this
approximation has recently been criticized in \cite{Mougeot2015}.

\subsection{First-forbidden non-unique $\beta$ decays \label{subsec:FF-nu}}

For the first-forbidden non-unique $\beta$ decays the shape factor 
(\ref{eq:shape-factor}) has to be supplemented with a $\Delta J = |J_i-J_f|=0$
term $C^{(1)}(w_e)$ \cite{Behrens1982,Schopper1966,Suhonen1993a,Ydrefors2010}, 
where $J_i$ ($J_f$) is the initial-state (final-state) spin of the mother 
(daughter) nucleus. Then the shape factor can be cast in the simple form 
\cite{Behrens1982,Schopper1966,Suzuki2012}
\begin{equation}
\label{eq:C-1st}
C(w_e) = K_0 + K_1w_e + K_{-1}/w_e + K_2w_e^2 \,,
\end{equation}
where the factors $K_n$ contain the NMEs (6 different, altogether) of transition
operators $\mathcal{O}$ of angular-momentum content (rank of a spherical tensor) 
$\mathcal{O}(0^-)$, $\mathcal{O}(1^-)$, and $\mathcal{O}(2^-)$, where the 
parity indicates that the initial and final nuclear
states should have opposite parities according to Table~\ref{tab:forbiddenness}. 
In the leading order these operators contain the pieces \cite{Bohr1969}
\begin{equation}
\label{eq:0-minus-op}
\mathcal{O}(0^-): g_{\rm A}(\gamma^5)\frac{\mbox{\boldmath{$\sigma$}}\cdot 
\mathbf{p}_e}{M_{\rm N}}\ ;\ 
\textrm{i}g_{\rm A}\frac{\alpha Z_f}{2R}(\mbox{\boldmath{$\sigma$}}\cdot \mathbf{r}) \,,
\end{equation}
\begin{equation}
\label{eq:1-minus-op}
\mathcal{O}(1^-): g_{\rm V}\frac{\mathbf{p}_e}{M_{\rm N}} \ ;\ 
g_{\rm A}\frac{\alpha Z_f}{2R}(\mbox{\boldmath{$\sigma$}}\times \mathbf{r}) \ ;\
\textrm{i}g_{\rm V}\frac{\alpha Z_f}{2R}\mathbf{r} \,,
\end{equation}
\begin{equation}
\label{eq:2-minus-op}
\mathcal{O}(2^-): \frac{\textrm{i}}{\sqrt{3}}g_{\rm A}
\left\lbrack\mbox{\boldmath{$\sigma$}}\mathbf{r}\right\rbrack_2\sqrt{\mathbf{p}_e^2 + 
\mathbf{q}_{\nu}^2} \,,
\end{equation}
where $\mathbf{p}_e$ ($\mathbf{q}_{\nu}$) is the electron (neutrino) 
momentum, $\mathbf{r}$ the radial coordinate, and the
square brackets in (\ref{eq:2-minus-op}) denote angular-momentum coupling. The matrix
elements of the operators (\ref{eq:0-minus-op}) and (\ref{eq:1-minus-op})
are suppressed relative to the Gamow-Teller matrix elements by the small momentum 
$\mathbf{p}_e$ of the electron and the large nucleon mass $M_{\rm N}$ or the small value of 
the fine-structure constant $\alpha$. The matrix element
of (\ref{eq:2-minus-op}) is suppressed by the small electron and neutrino momenta.
The axial operator $\mbox{\boldmath{$\sigma$}}\cdot \mathbf{p}_e$ and vector operator
$\mathbf{r}$ trace back to the time component of the axial current $A^{\mu}$ in
(\ref{eq:A-current}) and vector current $V^{\mu}$ in (\ref{eq:V-current}), and the rest of
the operators stem from the space components of $V^{\mu}$ and $A^{\mu}$.
The renormalization of these pieces is discussed next.

The $\xi$ approximation to the first-forbidden non-unique transitions has been 
discussed, e.g., in \cite{Bohr1969,Behrens1982,Schopper1966}.
One of the first analyses of first-forbidden non-unique transitions in this  
approximation was done in \cite{Bohr1969} for nuclei around $^{208}$Pb, 
based on the work of Damgaard and Winther \cite{Damgaard1964}. Assuming certain
dominant single-particle configurations around the double-closed shell
at $A=208$, Bohr and Mottelson obtained two sets of values for the effective vector and 
axial-vector coupling when analyzing the decay rates mediated by
the rank-1 operators $\mathcal{O}(1^-)$ in (\ref{eq:1-minus-op}). 
Combining the two obtained values we obtain
\begin{align} 
\label{eq:g-eff-BM}
	g^{\rm eff}_{\rm A}({\rm sp}) = (0.5-0.6)\times 1.18 = 0.46 - 0.56 \,,
\end{align}
where the symbol sp refers to single-particle estimate for the states involved
in the $\beta$ decays in odd-$A$ nuclei. It is interesting that also an effective
value for the vector coupling was derived:
\begin{align} 
\label{eq:g-eff-BM-V}
	g^{\rm eff}_{\rm V}(\textrm{sp}) = 0.3-0.7 \,.
\end{align}
This deviates quite much from the canonical value $g_{\rm V}=1$ dictated by the CVC
hypothesis \cite{Feynmann1958a}. Hence, strong nuclear-model dependent effects are
recorded in this case. In the case of the axial-vector strength
the numbers of (\ref{eq:g-eff-BM}) can be compared with the ones extracted
from the first-forbidden unique decays in the two-quasiparticle approximation
for odd-odd nuclei. There, in Eq.~(\ref{eq:g-eff-ave-Ejiri-qp}), a value 
$g^{\rm eff}_{\rm A}$(2qp)$\sim 0.2$ was obtained, implying that for the odd-odd
systems the quenching is more drastic than for the odd-mass systems. All in all,
a proper many-body treatment should reduce the quenching markedly, as shown by the factor $k$
in (\ref{eq:k-ratio-ff}), describing the transition from the two-quasiparticle 
approximation to the pnQRPA level in the case of the unique-forbidden $\beta$
transitions.

In \cite{Ejiri1968} a schematic study of the six NMEs corresponding to the operators
(\ref{eq:0-minus-op})--(\ref{eq:2-minus-op}) was performed. The hindrance factors
associated with the NMEs were related to the E1 (electric dipole) giant resonance in a
semi-quantitative way. The nuclear medium effect, in the form of the meson-exchange 
currents, on the $\mbox{\boldmath{$\sigma$}}\cdot \mathbf{p}_e$ part of $\mathcal{O}(0^-)$ in 
(\ref{eq:0-minus-op}) was discussed in \cite{Kubodera1978,Kirchbach1988,Towner1992}. 
This is the well-known (fundamental) enhancement of the $\gamma^5$ NME (axial charge $\rho_5$,
the time component of the axial current,
see Sec.~\ref{sec:medium-eff}), stemming from the renormalization of the
pion-decay constant and the nucleon mass $M_{\rm N}$ in nuclear medium \cite{Kubodera1991}
and exchange of heavy mesons \cite{Kirchbach1992,Towner1992}.
In this review the corresponding coupling strength is coined $g^{\rm eff}_{\rm A}(\gamma^5)$
for short. In \cite{Kirchbach1988} a simple nuclear approach to the meson-exchange
renormalization $g^{\rm eff}_{\rm A}(\gamma^5)=(1+\delta)g_{\rm A}^{\rm free}$ gave the following 
values of $g^{\rm eff}_{\rm A}(\gamma^5)$ (below are given the studied nuclear masses
and the corresponding active single-particle transitions):
\begin{eqnarray} 
\label{eq:g-eff-gamma-Kir}
g^{\rm eff}_{\rm A}(\gamma^5)= 1.90 & A=16 & (1\textrm{s}_{1/2}\to 0\textrm{p}_{1/2}) \nonumber \\ 
g^{\rm eff}_{\rm A}(\gamma^5)= 1.96 & A=18 & (1\textrm{s}_{1/2}\to 0\textrm{p}_{1/2}) \nonumber \\ 
g^{\rm eff}_{\rm A}(\gamma^5)= 1.84 & A=96 & (2\textrm{s}_{1/2}\to 1\textrm{p}_{1/2}) \nonumber \\ 
g^{\rm eff}_{\rm A}(\gamma^5)= 1.78 & A=206 & (2\textrm{p}_{1/2}\to 2\textrm{p}_{1/2})  
\end{eqnarray}
The work of \cite{Kirchbach1988} was extended by \cite{Towner1992} to include 6 nuclear
masses and several single-particle transitions for each mass. The resulting renormalization
by the meson-exchange currents amounted to
\begin{align} 
\label{eq:g-eff-0-minus-Towner}
g^{\rm eff}_{\rm A}(\gamma^5) = 2.0 - 2.3 \quad (A=16-208)
\end{align}
for the masses $A=16-208$. 

The above fundamental renormalization of the axial charge was contrasted with the 
nuclear-model dependent many-body effects by using the framework of the interacting
shell model in several studies in the past. For very low masses, $A=11$ \cite{Millener1982}
and $A=16$ \cite{Warburton1984}, some $40-50$\% enhancement of the axial charge was
obtained leading to $g^{\rm eff}_{\rm A}(\gamma^5) = 1.8 - 1.9$. A further study 
\cite{Warburton1994} of the $A=11-16$ nuclei indicated an enhanced axial charge of
\begin{align} 
\label{eq:g-eff-0-minus-War2}
g^{\rm eff}_{\rm A}(\gamma^5)=2.04\pm 0.04 \,, \quad (A=11-16)
\end{align}
where the uncertainties come solely from the experimental errors, not from the uncertainties
associated with the theoretical analyses. A general study of the
first-forbidden non-unique decays was carried on in \cite{Warburton1988} for
$34\le A\le 44$, and a further comparison \cite{Warburton1991b} with the measured rate of 
the $\beta^-$ decay of $^{50}$K indicated an enhanced value of
\begin{align} 
\label{eq:g-eff-0-minus-War91}
g^{\rm eff}_{\rm A}(\gamma^5) = 1.93\pm 0.09 \,, \quad (A=50)
\end{align}
where the uncertainty is purely experimental.

A thorough shell-model treatment of the mass $A=205-212$ nuclei in the lead region
was carried out in \cite{Warburton1990,Warburton1991a,Warburton1991}. There a 
rather strongly enhanced value of 
\begin{align} 
\label{eq:g-eff-0-minus-War}
g^{\rm eff}_{\rm A}(\gamma^5)=2.55\pm 0.07 \quad (A=205-212)
\end{align}
was obtained for the axial charge. The uncertainty comes from the least-squares fit to 18
measured $\beta$-decay transitions in the indicated mass region.
For the $\mbox{\boldmath{$\sigma$}}\cdot \mathbf{r}$ operator (space component of $A^{\mu}$)
essentially no renormalization (quenching, since space components tend to be 
quenched opposite to the enhancement of the time component, see beginning of 
Sec.~\ref{sec:medium-eff})
was obtained: $g_{\rm A}/g_{\rm A}^{\rm free}(0^-)=0.97\pm 0.06$. The value 
(\ref{eq:g-eff-0-minus-War}) is notably larger than those obtained for the lower masses and
also larger than the lead-region results of Towner (\ref{eq:g-eff-0-minus-Towner}).
However, in \cite{Kubodera1991} the theoretical result
\begin{align} 
\label{eq:g-eff-0-minus-Kub}
g^{\rm eff}_{\rm A}(\gamma^5)=2.5\pm 0.3 \quad (A=205-212)
\end{align}
was obtained by adopting an effective Lagrangian incorporating approximate chiral
and scale invariance of QCD. This seems to confirm the phenomenological result of
\cite{Warburton1991a,Warburton1991}.
For further information see the review \cite{Warburton1994b}.

For the $\mathcal{O}(1^-)$ operator $\mbox{\boldmath{$\sigma$}}\times \mathbf{r}$ 
in (\ref{eq:1-minus-op}) the analyses of
\cite{Warburton1990,Warburton1991} yielded the effective values
\begin{align} 
\label{eq:g-eff-1-minus-War}
g_{\rm A}^{\rm eff}(1^-)\sim 0.6\ ;\quad g_{\rm V}(1^-)\sim 0.6 \quad (\textrm{Warburton})
\end{align}
due to core-polarization effects caused by the limited model space used. In the work
\cite{Rydstrom1990} a shell-model study of the first-forbidden transition 
$^{205}\mathrm{Tl}(1/2^+_{\rm gs})\to\,^{205}\mathrm{Pb}(1/2^-)$ yielded the effective
values
\begin{align} 
\label{eq:g-eff-1-minus-Ryd}
g_{\rm A}^{\rm eff}(1^-)\sim 0.43-0.65\ ;\quad g_{\rm V}(1^-)\sim 0.38-0.85 \,. 
\quad (\textrm{Rydstr\"om}\textit{ et al.})
\end{align}
The shell-model analysis of \cite{Suzuki2012} of the $N=126$ isotones suggests a large 
quenching for $g_{\rm A}^{\rm eff}(1^-)$ but a large quenching of $g_{\rm V}^{\rm eff}(1^-)$ 
is not necessarily needed for most of the studied cases, contrary to 
(\ref{eq:g-eff-1-minus-War}) and (\ref{eq:g-eff-1-minus-Ryd}), in accordance with the CVC
hypothesis \cite{Feynmann1958a}.

In the work \cite{Zhi2013} half-lives of a number of nuclei at the magic neutron numbers
$N=50,82,126$ were analyzed by comparing results of large-scale shell-model 
calculations with experimental data. Both Gamow-Teller and first-forbidden 
$\beta$ decays were included in the analysis. By performing a least-squares fit to
the experimental data the following quenched weak couplings were extracted: For the
enhanced $\gamma^5$ matrix element the value $g^{\rm eff}_{\rm A}(\gamma^5)=1.61$ was obtained
and for the $\mbox{\boldmath{$\sigma$}}\cdot \mathbf{r}$ part the quenching
$g_{\rm A}/g_{\rm A}^{\rm free}(0^-)=0.66$ was obtained. For the $1^-$ part the quenched
values read
\begin{align} 
\label{eq:g-eff-1-minus-Zhi}
g_{\rm A}^{\rm eff}(1^-)\sim 0.48\ ;\quad g_{\rm V}(1^-)\sim 0.65 \,. 
\quad (\mathrm{Zhi}\textit{ et al.})
\end{align}
Interestingly, also for the first-forbidden unique operator $\mathcal{O}(2^-)$ 
of (\ref{eq:2-minus-op}) a quenching
\begin{align} 
\label{eq:g-eff-2-minus-Zhi}
g_{\rm A}^{\rm eff}(2^-)\sim 0.53
\quad (\mathrm{Zhi}\textit{ et al.})
\end{align}
was obtained. This is not far from the result $g_{\rm A}^{\rm eff}\sim 0.57$
[see Eq.~(\ref{eq:g-eff-ave-Ejiri})]
obtained in the analysis of the first-forbidden unique $\beta$ decays in
\cite{Ejiri2014}.

The above considerations for the vector coupling coefficient $g_{\rm V}$ 
are in conflict with the CVC hypothesis \cite{Feynmann1958a} and the findings of 
\cite{Haaranen2016} where the shape of the computed $\beta$-electron spectrum was 
compared with that of the measured one for the fourth-forbidden $\beta^-$ decay 
of $^{113}$Cd. This comparison confirmed an unquenched value $g_{\rm V}=1.0$ for 
the vector coupling coefficient, in accordance with the CVC hypothesis. 
For more discussion of the related method for highly-forbidden $\beta$
decays, see Sec.~\ref{subsec:high-forb-nu}.

\section{Higher-forbidden non-unique $\beta$ decays \label{subsec:high-forb-nu}}

The shape functions of forbidden non-unique beta decays are rather complex 
combinations of different NMEs and phase-space factors. Furthermore, their dependence on 
the weak coupling strengths $g_{\rm V}$ (vector part) and $g_{\rm A}$ (axial-vector part)
is very non-trivial. In fact, the shape factor $C(w_{e})$ (\ref{eq:shape-factor})
can be decomposed into vector, axial-vector and
mixed vector-axial-vector parts in the form \cite{Haaranen2016}
\begin{equation}
C(w_e)= g_{\rm V}^2C_{\rm V}(w_e)+g_{\rm A}^2C_{\rm A}(w_e)+
g_{\rm V}g_{\rm A}C_{\rm VA}(w_e).
\label{eq:decomp}
\end{equation}
Integrating equation (\ref{eq:decomp}) over the electron kinetic energy, we obtain an 
analogous expression for the integrated shape factor (\ref{eq:shape-f})
\begin{equation}
\tilde{C}= g_{\rm V}^2\tilde{C}_{\rm V}+g_{\rm A}^2\tilde{C}_{\rm A}+g_{\rm V}
g_{\rm A}\tilde{C}_{\rm VA},
\label{eq:decomp-tilde}
\end{equation}
where the factors $\tilde{C}_i$ in Eq.~(\ref{eq:decomp-tilde}) are just constants, 
independent of the electron energy.

In \cite{Haaranen2016} it was proposed that the shapes of $\beta$-electron
spectra could be used to determine the values of the weak coupling strengths
by comparing the computed spectrum with the measured one for forbidden
non-unique $\beta$ decays. This method was coined the spectrum-shape method (SSM).
In this study also the next-to-leading-order corrections 
to the $\beta$-decay shape factor were included.
In \cite{Haaranen2016} the $\beta$-electron spectra were studied for the 
4th-forbidden non-unique ground-state-to-ground-state $ \beta^-$ decay branches 
$^{113}\mathrm{Cd}(1/2^+)\to\,^{113}\mathrm{In}(9/2^+)$ and 
$^{115}\mathrm{In}(9/2^+)\to\,^{115}\mathrm{Sn}(1/2^+)$ using the microscopic 
quasiparticle-phonon model (MQPM) \cite{Toivanen1995b,Toivanen1998}
and the interacting shell model. It was verified by both nuclear models
that the $\beta$ spectrum shapes of both transitions are highly sensitive to the 
values of $g_{\rm V}$ and $g_{\rm A}$ and hence comparison of the calculated 
spectrum shape with the measured one opens a way to determine the values of 
these coupling strengths. As a by-product it was found that for all values of
$g_{\rm A}$ the best fits to data were obtained by using the canonical value 
$g_{\rm V} = 1.0$ for the vector coupling strength. This result is in conflict
with those obtained by analyzing first-forbidden non-unique $\beta$ decays in
Sec.~\ref{subsec:FF-nu}, where strongly quenched values of $g_{\rm V}$ were obtained.

\begin{figure}[htbp!]
\centering
\includegraphics[width=0.6\columnwidth]{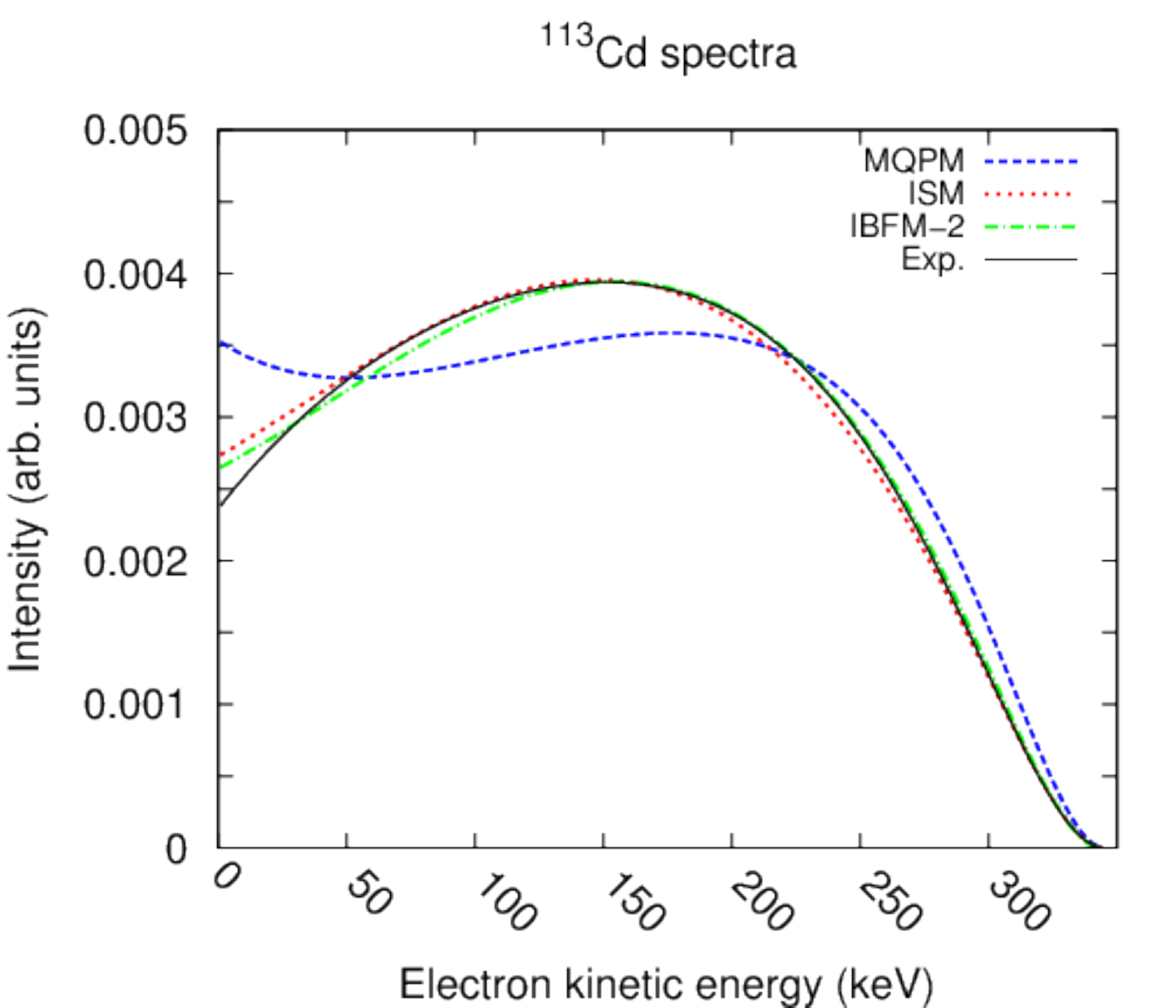}
\caption{Comparison of the computed $\beta$ spectra of $^{113}$Cd with the experiment. 
The next-to-leading-order corrections to the shape factor have been included, and 
only the best matches are shown in the figure. The canonical value 
$g_{\rm V}=1.0$ is used for the vector coupling strength. The areas under the 
curves are normalized to unity.}
\label{fig:113Cd}
\end{figure}

\begin{figure}[htbp!]
\centering
\includegraphics[width=1.0\columnwidth]{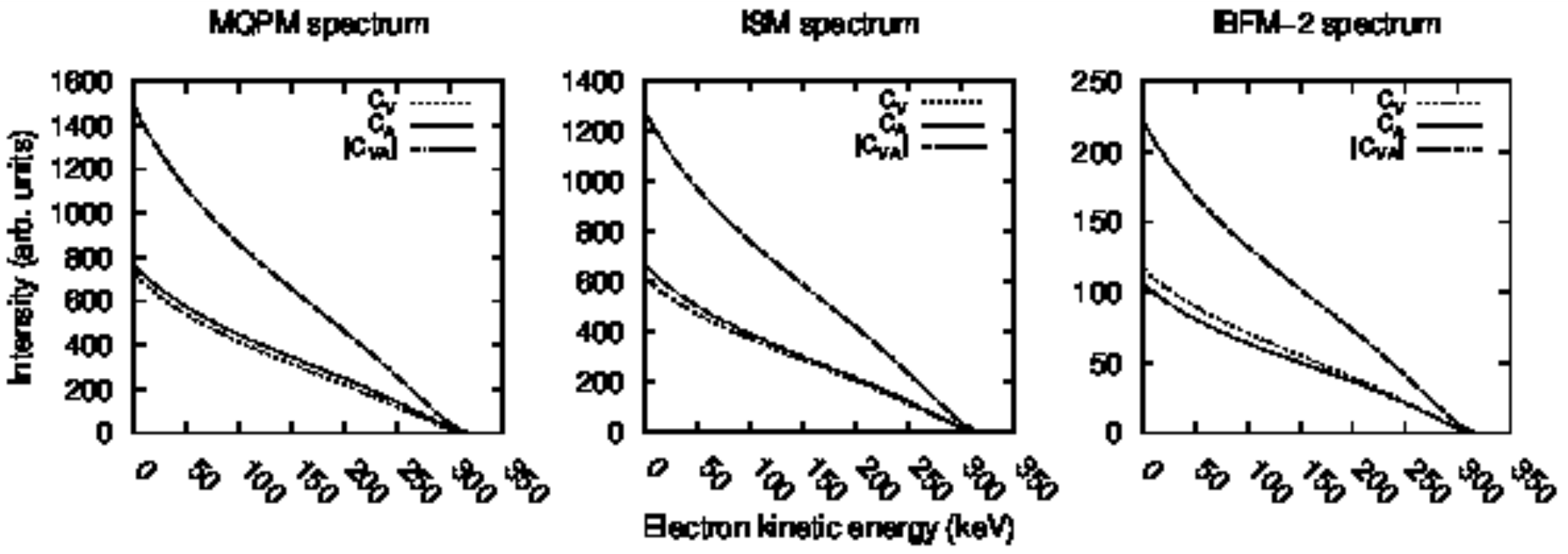}
\caption{Components $C_{\rm V}$, $C_{\rm A}$ and $C_{\rm VA}$ of (\ref{eq:decomp}) for
the electron spectra of the $\beta^-$ decay of $^{113}$Cd as computed
by the three nuclear models discussed in the context of Fig.~\ref{fig:113Cd}. Note
that the contribution of $C_{\rm VA}$ is negative.}
\label{fig:decomp}
\end{figure}

The work of \cite{Haaranen2016} on the $^{113}$Cd and $^{115}$In decays was extended
in \cite{Haaranen2017} to include an analysis made by using a third nuclear model, 
the microscopic interacting boson-fermion model (IBFM-2) \cite{Iachello1991}. At
the same time the next-to-leading-order corrections to the $\beta$-decay
shape factor were explicitly given and their role was thoroughly investigated.
A striking feature of the SSM analysis was that the three models 
yield a consistent result, $g_{\rm A}\approx 0.92$, when the 
SSM is applied to the available experimental $\beta$ spectrum 
\cite{Belli2007} of $^{113}$Cd. The result is illustrated in 
Fig.~\ref{fig:113Cd} where the three curves overlap best at the values
$g_{\rm A}^{\rm eff}=0.92$ (MQPM), $g_{\rm A}^{\rm eff}=0.90$ (ISM), and 
$g_{\rm A}^{\rm eff}=0.93$ (IBM). The agreement of the $\beta$-spectrum shapes computed
in the three different nuclear-theory frameworks  speaks for the robustness of
the spectrum-shape method in determining the effective value of $g_{\rm A}$. 
For completeness, in Fig.~\ref{fig:decomp} are shown the three components (\ref{eq:decomp})
as functions of the electron energy for the three different nuclear models used
to compute the spectrum shapes of $^{113}$Cd in Fig.~\ref{fig:113Cd}. It is seen that
for the whole range of electron energies the two components, $C_{\rm V}(w_e)$ and
$C_{\rm A}(w_e)$ are roughly of the same size whereas the magnitude of the 
component $C_{\rm VA}(w_e)$ is practically the sum of the previous two, but with
opposite sign. Hence, for the whole range of electron energies there is a delicate
balance between the three terms, and their sum is much smaller than the magnitudes
of its constituent components.

The works \cite{Haaranen2016,Haaranen2017} were continued by the work
\cite{Kostensalo2017b} where the evolution of the $\beta$ spectra with changing value
of $g_{\rm A}$ was followed for 26 first-, second-,
third-, fourth- and fifth-forbidden $\beta^-$ decays of odd-$A$ nuclei by 
calculating the associated NMEs by the MQPM. The next-to-leading-order contributions 
were taken into account in the $\beta$-decay shape factor. It was found that
the spectrum shapes of the third- and fourth-forbidden 
non-unique decays depend strongly on the value of $g_{\rm A}$,
whereas the first- and second-forbidden decays were practically 
insensitive to the variations in $g_{\rm A}$. Furthermore, the $g_{\rm A}$-driven evolution 
of the normalized $\beta$ spectra seems to be quite universal, largely
insensitive to small changes of the nuclear mean field and the adopted
residual many-body Hamiltonian. These features were also verified in the follow-up work
\cite{Kostensalo2017c}, where the ISM was used as the nuclear-model framework.
This makes SSM a robust tool for extracting 
information on the effective values of the weak coupling strengths. This
also means that if SSM really is largely nuclear-model independent there is a chance
to access the fundamental renormalization factor $q_{\rm F}$ of Sec.~\ref{sec:medium-eff}
for (highly) forbidden $\beta$ transitions. It is also worth noting that in
the works \cite{Kostensalo2017b,Kostensalo2017c} several new 
experimentally interesting decays for the SSM treatment were discovered. 

\begin{figure}[htbp!]
\centering
\includegraphics[width=0.45\columnwidth]{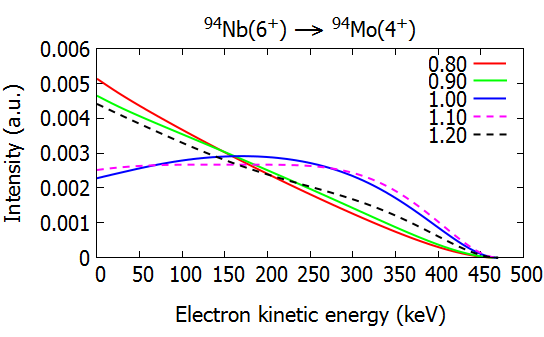}
\includegraphics[width=0.45\columnwidth]{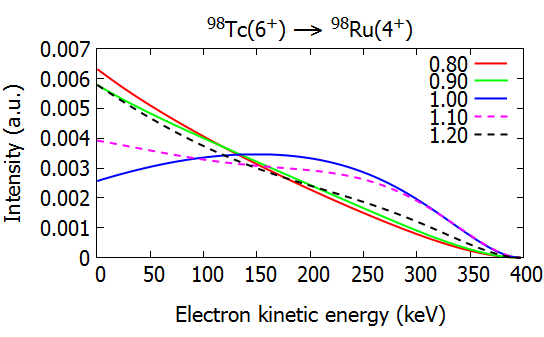}
\caption{Normalized ISM-computed electron spectra for the second-forbidden non-unique 
decays of $^{94}$Nb and $^{98}$Tc. The value $g_{\rm V} = 1.0$ was assumed and the
color coding represents the value of $g_{\rm A}$.}
\label{fig:94Nb-98Tc}
\end{figure}

\begin{figure}[htbp!]
\centering
\includegraphics[width=0.45\columnwidth]{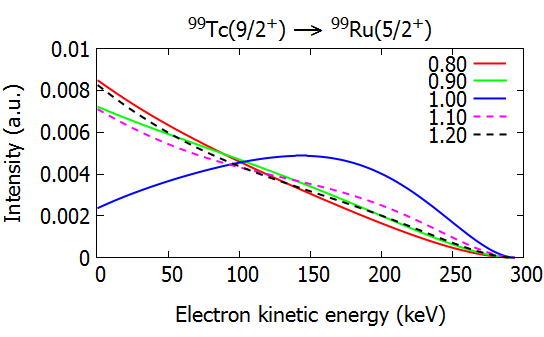}
\includegraphics[width=0.45\columnwidth]{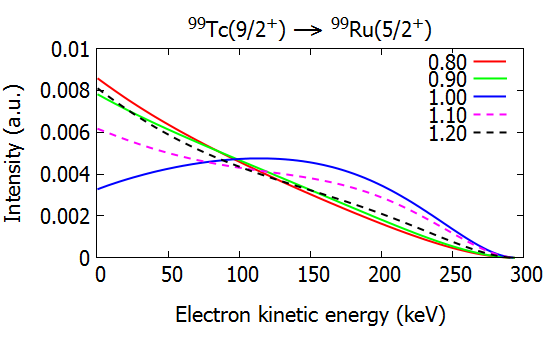}
\caption{Normalized electron spectra for the second-forbidden non-unique 
ground-state-to-ground-state $\beta^-$ decay of
$^{99}$Tc as computed by using the MQPM (left panel) and the ISM (right panel). 
The value $g_{\rm V} = 1.0$ was assumed and the color coding represents the value 
of $g_{\rm A}$.}
\label{fig:99Tc}
\end{figure}

\begin{figure}[htbp!]
\centering
\includegraphics[width=0.6\columnwidth]{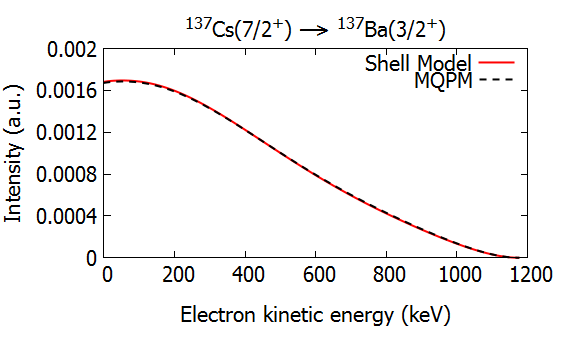}
\caption{Normalized electron spectra for the second-forbidden non-unique 
ground-state-to-ground-state $\beta^-$ decay of
$^{137}$Cs as computed by using the MQPM and the ISM. The value $g_{\rm V} = 1.0$ 
was assumed and the color coding represents the value of $g_{\rm A}$.}
\label{fig:137Cs}
\end{figure}

Results of the investigations of \cite{Kostensalo2017b,Kostensalo2017c} are
summarized in Tables~\ref{tab:C-factors1} and \ref{tab:F-nu-results}, and in
Figs.~\ref{fig:94Nb-98Tc} -- \ref{fig:137Cs}. Figure \ref{fig:94Nb-98Tc}
displays the $\beta$ spectra of the second-forbidden non-unique transitions 
$^{94}\mathrm{Nb}(6+)\to\,^{94}\mathrm{Mo}(4^+)$ (left panel) and
$^{98}\mathrm{Tc}(6^+)\to\,^{98}\mathrm{Ru}(4^+)$ (right panel) calculated by using 
the ISM \cite{Kostensalo2017c}. It is obvious that the shape of 
the spectra depends sensitively on the
value of $g_{\rm A}$ but not as strongly as the transitions associated with
the mother nuclei $^{113}$Cd and $^{115}$In, as shown in the figures of
\cite{Haaranen2016}. It is to be noted that both of the transitions have been
observed experimentally since the branching is 100\%, but the electron spectra 
are not yet available.

In Fig.~\ref{fig:99Tc} a comparison of the MQPM (left panel) and ISM 
(right panel) calculations \cite{Kostensalo2017c} for the $\beta$ spectrum of 
the second-forbidden non-unique decay transition 
$^{99}\mathrm{Tc}(9/2^+)\to\,^{99}\mathrm{Ru}(5/2^+)$ 
is shown. Again there is clear sensitivity to the value of $g_{\rm A}$, at the
level of the $^{94}$Nb and $^{98}$Tc transitions, but the remarkable thing is
that the spectrum shapes computed by the two nuclear models agree almost
perfectly, giving further evidence in favor of the robustness of the SSM.
Again, experimentally, the branching to this decay channel is practically 100\%
so that the $\beta$ spectrum is potentially well measurable.

Finally, In Fig.~\ref{fig:137Cs} the $\beta$ spectrum of the second-forbidden 
non-unique decay-transition $^{137}\mathrm{Cs}(7/2^+)\to\,^{137}\mathrm{Ba}(3/2^+)$ 
is shown. Here the spectrum shape is quite independent of the value of $g_{\rm A}$
and has exactly the same computed shape for the two applied nuclear-model 
frameworks: the MQPM and the ISM \cite{Kostensalo2017c}. The robustness of the
$\beta$-spectrum shape against variations in $g_{\rm A}$ and the calculational 
scheme makes the measurement of this spectrum interesting in terms of testing
the basic framework of high-forbidden non-unique $\beta$ decays. The cause of the
inertia against variations of $g_{\rm A}$ is seen in Table~\ref{tab:C-factors1}
in the decomposition (\ref{eq:decomp-tilde}) of the dimensionless integrated shape 
function $\tilde{C}$ for the decays of both $^{135}$Cs and $^{137}$Cs. It is seen
that for these two decays all the components of $\tilde{C}$ are of the same sign,
thus adding coherently. Hence, changes in the value of $g_{\rm A}$ do not
affect the spectrum shape, contrary to those decays where there is a destructive
interference between the axial-vector and mixed components of (\ref{eq:decomp-tilde}),
like in the cases of Figs.~\ref{fig:113Cd}--\ref{fig:99Tc}, further analyzed in
Table~\ref{tab:C-factors2}.

	\begin{table*}
\centering
\begin{tabular}{lllcccc}
\toprule
   Transition & $K$ & Nucl. model & $\tilde{C}_{\rm V}$ & $\tilde{C}_{\rm A}$ & 
$\tilde{C}_{\rm VA}$ & $\tilde{C}$ \\
\midrule
$^{135}\rm Cs(7/2^+)\rightarrow\,^{135}{\rm Ba}(3/2^+)$ & 2 & MQPM 
& $ 1.133\times 10^{-8}$ & $ 1.656\times 10^{-8}$ & $ 2.737\times 10^{-8}$ 
& $ 5.526\times 10^{-8}$ \\
$^{137}\rm Cs(7/2^+)\rightarrow\,^{137}{\rm Ba}(3/2^+)$ & 2 & MQPM 
& $ 3.217\times 10^{-5}$ & $ 2.654\times 10^{-5}$ & $ 5.822\times 10^{-5}$ 
& $ 1.169\times 10^{-4}$ \\
$^{137}\rm Cs(7/2^+)\rightarrow\,^{137}{\rm Ba}(3/2^+)$ & 2 & ISM 
& $4.211 \times 10^{-6}$ & $2.836 \times 10^{-6}$ & $ 6.879\times 10^{-6}$ 
& $1.392 \times 10^{-5}$ \\
\bottomrule
\end{tabular}
  \caption{Dimensionless integrated shape functions $\tilde{C}$ (\ref{eq:decomp-tilde})
and their vector $\tilde{C}_{\rm V}$, axial-vector $\tilde{C}_{\rm A}$ and mixed 
components $\tilde{C}_{\rm VA}$ for the forbidden non-unique $\beta$ decays of 
$^{135}$Cs and $^{137}$Cs. The forbiddenness $K$ and the nuclear model used to
calculate $\tilde{C}$ is given. For the total 
integrated shape factor $\tilde{C}$ the values of the coupling strengths 
were set to $g_{\rm V}=g_{\rm A}=1.0$.} 
\label{tab:C-factors1}
\end{table*}

Table~\ref{tab:F-nu-results} summarizes the exploratory works of 
\cite{Haaranen2016,Haaranen2017,Kostensalo2017b,Kostensalo2017c} in terms of
listing the studied decay-transition candidates and their potential for
future measurements. Here only the studied non-unique $\beta$-decay transitions
are listed since the unique forbidden transitions are practically 
$g_{\rm A}$-independent even when the next-to-leading-order terms are included
in the $\beta$-decay shape factor \cite{Haaranen2016}. The most favorable cases 
for measurements are the ones that have a \textit{strong} dependence on $g_{\rm A}$
and the branching to the final state of interest is close to 100\%. By these criteria
the best candidates for measurements are the non-unique transitions
$^{94}\mathrm{Nb}(6^+)\to\,^{94}\mathrm{Mo}(4^+)$ (second-forbidden),
$^{98}\mathrm{Tc}(6^+)\to\,^{98}\mathrm{Ru}(4^+)$ (second-forbidden),
$^{99}\mathrm{Tc}(9/2^+)\to\,^{99}\mathrm{Ru}(5/2^+)$ (second-forbidden),
$^{113}\mathrm{Cd}(1/2^+)\to\,^{113}\mathrm{In}(9/2^+)$ (fourth-forbidden), and
$^{115}\mathrm{In}(9/2^+)\to\,^{115}\mathrm{Sn}(1/2^+)$ (fourth-forbidden). Plans
for accurate measurements of (some) of these transitions are on-going in the
DAMA \cite{Tretyak-pc} and COBRA collaborations \cite{Zuber-pc}. It should be noted
that also the transition $^{87}\mathrm{Rb}(3/2^-)\to\,^{87}\mathrm{Sr}(9/2^+)$ could
be of interest for measurements since it has a 100\% branching and the corresponding
$\beta$ spectrum is moderately sensitive to $g_{\rm A}$.

\begin{table}[t!]
\centering
\begin{tabular}{cclccll}
\toprule
Transition & $J_i^{\pi_i}$ (gs) & $J_f^{\pi_f}$ ($n_f$) & Branching & $K$ 
& Sensitivity & Nucl. model \\
\midrule
$^{36}\mathrm{Cl}\to\,^{36}\mathrm{Ar}$ & $2^{+}$ & $0^{+}$ (gs) & \textbf{98\%} & 2 
& None & ISM \\
$^{48}\mathrm{Ca}\to\,^{48}\mathrm{Sc}$ & $0^{+}$ & $4^{+}$ (2) & $\sim$0\% & 4 
& None & ISM \\
$^{48}\mathrm{Ca}\to\,^{48}\mathrm{Sc}$ & $0^{+}$ & $6^{+}$ (gs) & $\sim$0\% & 6 
& None & ISM \\
$^{50}\mathrm{V}\to\,^{50}\mathrm{Cr}$ & $6^{+}$ & $2^{+}$ (1) & $\sim$0\% & 4 
& Weak & ISM \\
$^{60}\mathrm{Fe}\to\,^{60}\mathrm{Co}$ & $0^{+}$ & $2^{+}$ (1) & \textbf{100\%} & 2 
& None & ISM \\
$^{85}\mathrm{Br}\to\,^{85}\mathrm{Kr}$ & $3/2^{-}$ & $9/2^{+}$ (gs) & $\sim$0\% & 3
& Moderate & MQPM \\
$^{87}\mathrm{Rb}\to\,^{87}\mathrm{Sr}$ & $3/2^{-}$ & $9/2^{+}$ (gs) & \textbf{100\%} & 3
& Moderate & MQPM, ISM \\
$^{93}\mathrm{Zr}\to\,^{93}\mathrm{Nb}$ & $5/2^{+}$ & $9/2^{+}$ (gs) & $5\le$\% & 2
& Weak & MQPM \\
$^{94}\mathrm{Nb}\to\,^{94}\mathrm{Mo}$ & $6^{+}$ & $4^{+}$ (2) & \textbf{100\%} & 2
& \textbf{Strong} & NSM \\
$^{96}\mathrm{Zr}\to\,^{96}\mathrm{Nb}$ & $0^{+}$ & $4^{+}$ (2) & $\sim$0\% & 4 
& None & ISM \\
$^{96}\mathrm{Zr}\to\,^{96}\mathrm{Nb}$ & $0^{+}$ & $6^{+}$ (gs) & $\sim$0\% & 6 
& \textbf{Strong} & ISM \\
$^{97}\mathrm{Zr}\to\,^{97}\mathrm{Nb}$ & $1/2^{+}$ & $9/2^{+}$ (gs) & $\sim$0\% & 4
& \textbf{Strong} & MQPM \\
$^{98}\mathrm{Tc}\to\,^{98}\mathrm{Ru}$ & $6^{+}$ & $4^{+}$ (3) & \textbf{100\%} & 2 
& \textbf{Strong} & ISM \\
$^{99}\mathrm{Tc}\to\,^{99}\mathrm{Ru}$ & $9/2^{+}$ & $5/2^{+}$ (gs) & \textbf{100\%} & 2
& \textbf{Strong} & MQPM, ISM \\
$^{101}\mathrm{Mo}\to\,^{101}\mathrm{Tc}$ & $1/2^{+}$ & $9/2^{+}$ (gs) & $\sim$0\% & 4 
& \textbf{Strong} & MQPM \\
$^{113}\mathrm{Cd}\to\,^{113}\mathrm{In}$ & $1/2^{+}$ & $9/2^{+}$ (gs) & \textbf{100\%} & 4  
& \textbf{Strong} & MQPM, ISM, IBFM-2 \\
$^{115}\mathrm{Cd}\to\,^{115}\mathrm{In}$ & $1/2^{+}$ & $9/2^{+}$ (gs) & $\sim$0\% & 4  
& \textbf{Strong} & MQPM \\
$^{115}\mathrm{In}\to\,^{115}\mathrm{Sn}$ & $9/2^{+}$ & $1/2^{+}$ (gs) & \textbf{100\%} & 4  
& \textbf{Strong} & MQPM, ISM, IBFM-2 \\
$^{117}\mathrm{Cd}\to\,^{117}\mathrm{In}$ & $1/2^{+}$ & $9/2^{+}$ (gs) & $\sim$0\% & 4  
& \textbf{Strong} & MQPM \\
$^{119}\mathrm{In}\to\,^{119}\mathrm{Sn}$ & $9/2^{+}$ & $1/2^{+}$ (gs) & $\sim$0\% & 4  
& \textbf{Strong} & MQPM \\
$^{123}\mathrm{Sn}\to\,^{123}\mathrm{Sb}$ & $11/2^{-}$ & $1/2^{+}$ (4) & $\sim$0\% & 5 
& Weak & MQPM \\
$^{126}\mathrm{Sn}\to\,^{126}\mathrm{Sb}$ & $0^{+}$ & $2^{+}$ (5) & \textbf{100\%} & 2 
& None & ISM \\
$^{135}\mathrm{Cs}\to\,^{135}\mathrm{Ba}$ & $7/2^{+}$ & $3/2^{+}$ (gs) & \textbf{100\%} & 2 
& None & MQPM \\
$^{137}\mathrm{Cs}\to\,^{137}\mathrm{Ba}$ & $7/2^{+}$ & $3/2^{+}$ (gs) & 5.4\% & 2 
& None & MQPM, ISM \\
$^{125}\mathrm{Sb}\to\,^{125}\mathrm{Te}$ & $7/2^{+}$ & $9/2^{-}$ (3) & 7.2\% & 1 
& None & MQPM \\
$^{141}\mathrm{Ce}\to\,^{141}\mathrm{Pr}$ & $7/2^{-}$ & $5/2^{+}$ (gs) & 31\% & 1 
& Weak & MQPM \\
$^{159}\mathrm{Gd}\to\,^{159}\mathrm{Tb}$ & $3/2^{-}$ & $5/2^{+}$ (1) & 26\% & 1
& None & MQPM \\
$^{161}\mathrm{Tb}\to\,^{161}\mathrm{Dy}$ & $3/2^{+}$ & $5/2^{-}$ (1) & $\sim$0\% & 1  
& None & MQPM \\
$^{169}\mathrm{Er}\to\,^{169}\mathrm{Tm}$ & $1/2^{-}$ & $3/2^{+}$ (1) & 45\% & 1 
& None & MQPM \\
\bottomrule
\end{tabular}
\caption{List of the studied high-forbidden \emph{non-unique} $\beta^-$-decay transitions
and their sensitivity to the value of $g_{\rm A}$. Here $J_i$ ($J_f$) is the
angular momentum of the initial (final) state, $\pi_i$ ($\pi_f$) the parity of the
initial (final) state, and $K$ the degree of forbiddenness. The initial state is
always the ground state (gs, column 2) of the mother nucleus and the final state 
is either the ground state (gs) or the $n_f:th$, $n_f=1,2,3,4,5$, excited state (column 3)
of the daughter nucleus. Column 4 gives the branching to this particular decay 
channel [with boldface if (almost) 100\%],
column 5 indicates the sensitivity to the value of $g_{\rm A}$ (with boldface if strong), 
and the last column lists the nuclear models which have been used 
(thus far) to compute the $\beta$-spectrum shape.}
\label{tab:F-nu-results} 
\end{table}

In Table~\ref{tab:C-factors2} the dimensionless integrated shape functions $\tilde{C}$ 
(\ref{eq:decomp-tilde}) have been decomposed into their
vector $\tilde{C}_{\rm V}$, axial-vector $\tilde{C}_{\rm A}$ and mixed vector-axial-vector
components $\tilde{C}_{\rm VA}$ for the experimentally most promising forbidden
non-unique $\beta$ decays of forbiddenness $K$ of Table~\ref{tab:F-nu-results}.
In the table also the nuclear model used to calculate $\tilde{C}$ is given. A 
characteristic of the numbers of Table~\ref{tab:C-factors2} is that the magnitudes
of the vector, axial-vector, and mixed components are of the same order of magnitude, 
and the vector and axial-vector components have the same sign whereas the
mixed component has the opposite sign. This makes the three components largely
cancel each other and the resulting magnitude of the total dimensionless integrated shape 
function is always a couple of orders of magnitude smaller than its components. Thus
the integrated shape function becomes extremely sensitive to the value of $g_{\rm A}$,
as seen in Fig.~\ref{fig:94Nb-98Tc} for the decays of $^{94}$Nb and $^{98}$Tc, and in
Fig.~\ref{fig:99Tc} for the decay of $^{99}$Tc. 

	\begin{table*}
\centering
\begin{tabular}{lllcccc}
\toprule
   Transition & $K$ & Nucl. model & $\tilde{C}_{\rm V}$ & $\tilde{C}_{\rm A}$ & 
$\tilde{C}_{\rm VA}$ & $\tilde{C}$ \\
\midrule
$^{94}\rm Nb(6^+)\rightarrow\,^{94}{\rm Mo}(4^+)$ & 2 & ISM
& $1.598 \times 10^{-8}$ & $1.469 \times 10^{-8}$ & $ -3.058\times 10^{-8}$ 
& $1.03 \times 10^{-10}$ \\
$^{98}\rm Tc(6^+)\rightarrow\,^{98}{\rm Ru}(4^+)$ & 2 & ISM
& $ 2.723\times 10^{-8}$ & $ 2.544\times 10^{-8}$ & $ -5.254\times 10^{-8}$ 
& $1.21 \times 10^{-10}$ \\
$^{99}\rm Tc(9/2^+)\rightarrow\,^{99}{\rm Ru}(5/2^+)$ & 2 & ISM
& $2.240 \times 10^{-9}$ & $2.130 \times 10^{-9}$ & $ -4.361\times 10^{-9}$ 
& $8.78 \times 10^{-12}$ \\
$^{113}\rm Cd(1/2^+)\rightarrow\,^{113}{\rm In}(9/2^+)$ & 4 & MQPM
& $ 1.925\times 10^{-19}$ & $ 2.094\times 10^{-19}$ & $ -4.002\times 10^{-19}$ 
& $ 1.38\times 10^{-21}$ \\
$^{113}\rm Cd(1/2^+)\rightarrow\,^{113}{\rm In}(9/2^+)$ & 4 & ISM
& $ 1.678\times 10^{-19}$ & $ 1.825\times 10^{-19}$ & $ -3.494\times 10^{-19}$ 
& $ 9.90\times 10^{-22}$ \\
$^{113}\rm Cd(1/2^+)\rightarrow\,^{113}{\rm In}(9/2^+)$ & 4 & IBM-2
& $ 3.228\times 10^{-20}$ & $ 3.007\times 10^{-20}$ & $ -6.106\times 10^{-20}$ 
& $ 1.28\times 10^{-21}$ \\
$^{115}\rm In(9/2^+)\rightarrow\,^{115}{\rm Sn}(1/2^+)$ & 4 & MQPM
& $ 6.503\times 10^{-18}$ & $ 6.126\times 10^{-18}$ & $ -1.256\times 10^{-17}$ 
& $ 6.49\times 10^{-20}$ \\
$^{115}\rm In(9/2^+)\rightarrow\,^{115}{\rm Sn}(1/2^+)$ & 4 & ISM
& $ 3.146\times 10^{-18}$ & $ 3.851\times 10^{-18}$ & $ -6.939\times 10^{-18}$ 
& $ 5.74\times 10^{-20}$ \\
$^{115}\rm In(9/2^+)\rightarrow\,^{115}{\rm Sn}(1/2^+)$ & 4 & IBM-2
& $ 5.531\times 10^{-19}$ & $ 5.444\times 10^{-19}$ & $ -1.065\times 10^{-18}$ 
& $ 3.25\times 10^{-20}$ \\
\bottomrule
\end{tabular}
  \caption{Dimensionless integrated shape functions $\tilde{C}$ (\ref{eq:decomp-tilde})
and their vector $\tilde{C}_{\rm V}$, axial-vector $\tilde{C}_{\rm A}$ and mixed 
components $\tilde{C}_{\rm VA}$ for the experimentally most promising forbidden
non-unique $\beta$ decays of forbiddenness $K$. Also the nuclear model used to
calculate $\tilde{C}$ is given. For the total 
integrated shape factor $\tilde{C}$ the values of the coupling strengths 
were set to $g_{\rm V}=g_{\rm A}=1.0$.} 
\label{tab:C-factors2}
\end{table*}

For the beta spectrum of the decays of
$^{113}$Cd and $^{115}$In there are calculations available
in three different nuclear-theory frameworks as shown in Fig.~\ref{fig:113Cd} and
Tables~\ref{tab:F-nu-results} and \ref{tab:C-factors2}. 
As visible in Table~\ref{tab:C-factors2}, an interesting 
feature of the components of the integrated shape functions $\tilde{C}$ is that the 
MQPM and ISM results are close to each other whereas the numbers produced by IBM-2 
are clearly smaller. Surprisingly enough, the total value of $\tilde{C}$ is roughly
the same in all three theory frameworks. This is another indication of the 
robustness of the SSM.

There are indirect ways to access the quenching of high-forbidden $\beta$-decay
transitions. One of them is to study electromagnetic decays of analogous structure.
In \cite{Jokiniemi2016} magnetic hexadecapole (M4) $\gamma$ transitions in odd-$A$
medium-heavy nuclei were studied by comparing the single-quasiparticle NMEs against
the MQPM-computed NMEs to learn about the quenching in the analogous third-forbidden
unique $\beta$ decays (parity change with angular-momentum content 4). The MQPM
calculations suggest a strong quenching $g_{\rm A}\sim 0.33g^{\rm free}_{\rm A}\sim 0.4$
for these transitions. This strong quenching could be an artifact of the MQPM
framework since there the excitations of an odd-$A$ nucleus are formed by coupling
BCS quasiparticles to excitations of the neighboring even-even reference nucleus.
Thus the predicted M4 giant resonance in the odd-$A$ nucleus might not be strong 
enough to draw low-lying M4 strength to higher excitation energies, around 
the giant-resonance region.

\section{Quenching of $g_{\rm A}$ in $2\nu\beta\beta$ decays \label{sec:2n-beta-beta}}

\begin{figure}[htbp!]
\centering
\includegraphics[width=0.8\columnwidth]{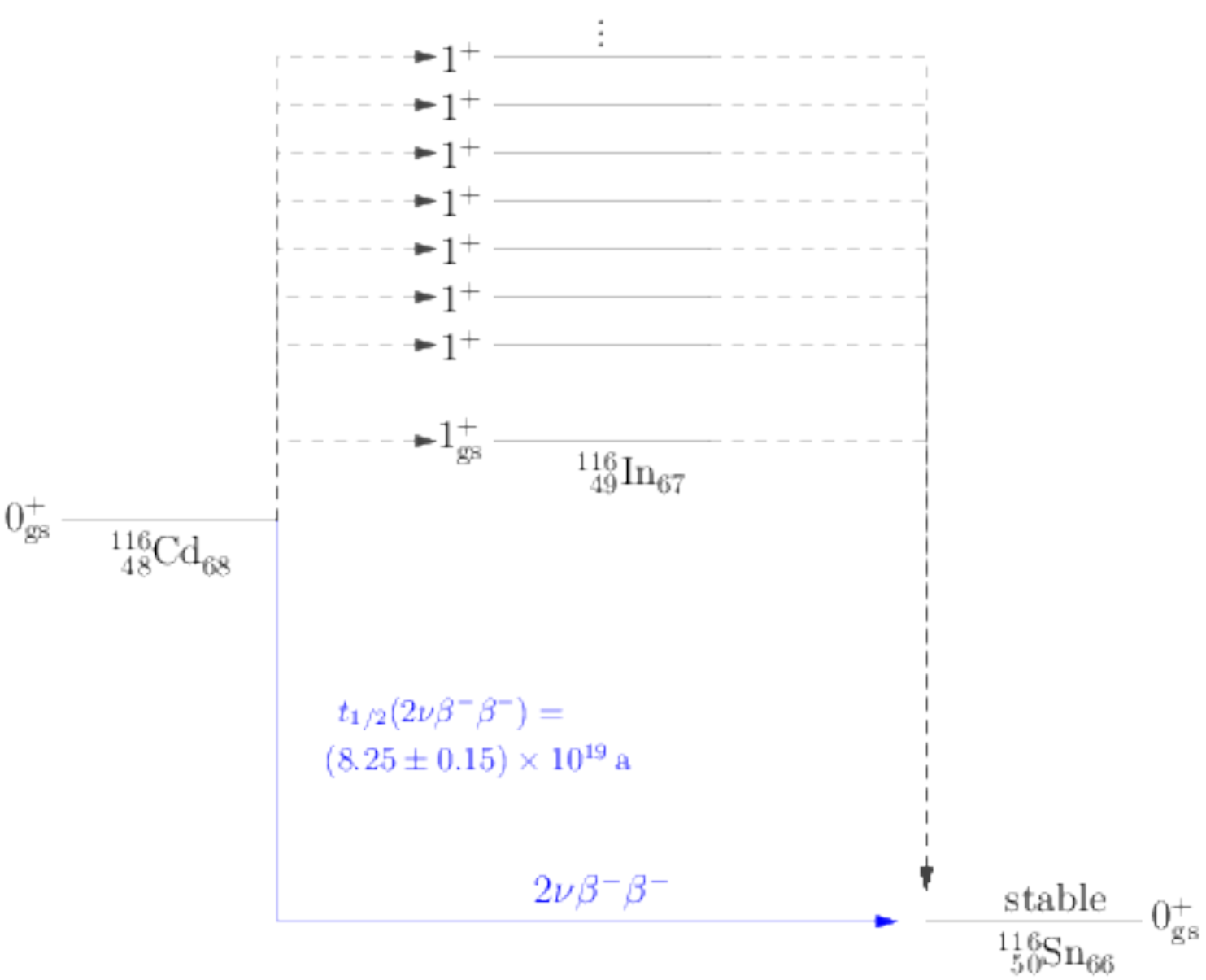}
\caption{The $2\nu\beta\beta$ decay of $^{116}$Cd to $^{116}$Sn via the $1^+$ virtual 
intermediate states in $^{116}$In. The transitions between $^{116}$Cd ($^{116}$Sn)
and $^{116}$In constitute the left-leg (right-leg) transitions.}
\label{fig:2vbb-interm}
\end{figure}

The $2\nu\beta\beta$ decay rate can be compactly written as
\begin{equation}
\label{eq:2vbb}
	\left[ t_{1/2}^{(2\nu)}(0_i^+ \to 0_f^+) \right]^{-1} = 
	g_{\rm A}^4 G_{2\nu} \left\vert M^{(2\nu)}\right\vert^2 \,,
\end{equation}
where $G_{2\nu}$ represents the leptonic phase-space factor (without including 
$g_{\rm A}$) as defined in \cite{Kotila2012}. The initial ground state is 
denoted by $0^+_i$ and the final ground state by $0^+_f$. The $2\nu\beta\beta$ 
NME $M^{(2\nu)}$ can be written as
\begin{align}
\label{eq:NMEb}
	M^{(2\nu )} = \sum_{m,n} 
	\frac{M_{\rm L}(1^+_m) M_{\rm R}(1^+_n)}{D_m} \,,
\end{align}
where the quantity $D_m$ is the energy denominator and the NMEs $M_{\rm L}$ 
and $M_{\rm R}$ correspond to left-leg and right-leg virtual Gamow-Teller 
transitions depicted in Fig.~\ref{fig:2vbb-interm}.
The summation is in general over all intermediate $1^+$ states, not just the first
one as implied by the very schematic Fig.~\ref{fig:0vbb-interm}. On the other
hand, the summation in (\ref{eq:NMEb}) can be dominated by one transition,
usually through the lowest $1^+$ state if it happens to be the ground state of
the intermediate nucleus. In this case one speaks about 
\emph{single-state dominance}. This dominance has been addressed in several
works, e.g. in \cite{Civitarese1998,Bhattacharya1998,Civitarese1999}.

The $2\nu\beta\beta$ decay rate (\ref{eq:2vbb}) and $0\nu\beta\beta$ decay rate 
(\ref{eq:0vbb}) share the same strong dependence on $g_{\rm A}$.
It is thus essential to study the renormalization of $g_{\rm A}$ in
beta and $2\nu\beta\beta$ decays before entering studies of the $0\nu\beta\beta$ decay. 
These studies touch only the $1^+$ contribution to the $0\nu\beta\beta$ decay. 
However, it is known that contributions from higher multipoles are also very
important for the $0\nu\beta\beta$ decay (see Sec.~\ref{sec:high-forb-u}).  
It is challenging to relate the results emanating from the $\beta$ and 
$2\nu\beta\beta$ decay studies to the value of the $0\nu\beta\beta$ NME:
the former two involve momentum transfers of a few MeV whereas the latter involves 
momentum exchanges of the order of 100 MeV through the virtual Majorana neutrino.
The high exchanged momenta in the $0\nu\beta\beta$ decay allow for the possibility 
that the effective value of $g_{\rm A}$ acquires momentum dependence, 
as discussed in Sec.~\ref{sec:medium-eff}. In addition, the high exchanged momenta
induce substantial contributions from the higher $J^{\pi}$ states to the $0\nu\beta\beta$ 
decay rate \cite{Hyvarinen2016}. The renormalization of $g_{\text{A}}$ for these 
higher-lying states could be different from the renormalization for the low-lying states, 
the subject matter of this review.

After this preamble we now proceed to discuss the possible renormalization of the
axial-vector coupling strength [at zero-momentum limit $q\to 0$ in 
(\ref{eq:dipole})] as obtained from the combined $\beta$-decay and 
$2\nu\beta\beta$-decay analyses performed in different theoretical approaches. 
It is important to
be aware that in all the studies of the present section it is impossible to
disentangle between the fundamental, nuclear-matter affected, and the 
many-body, nuclear-model affected, contributions to the renormalization of $g_{\rm A}$.

\subsection{Quasiparticle random-phase approximation \label{subsec:pnQRPA-bb}}

\begin{figure}[htbp!]
\centering
\includegraphics[width=0.60\columnwidth]{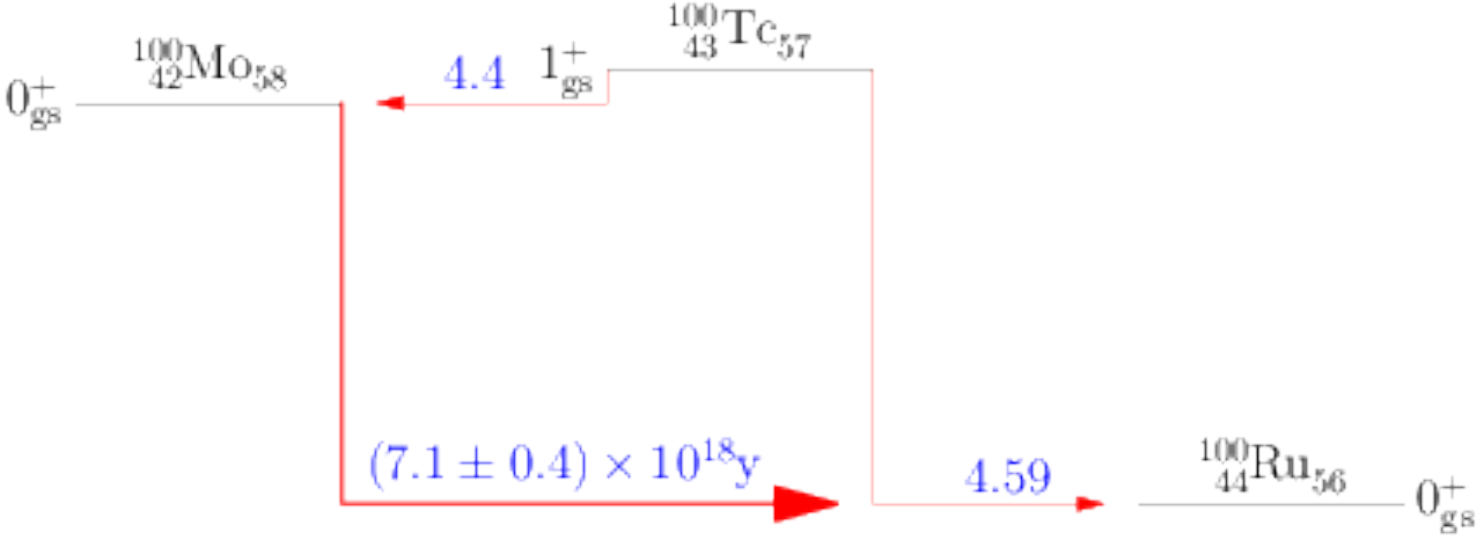}
\includegraphics[width=0.60\columnwidth]{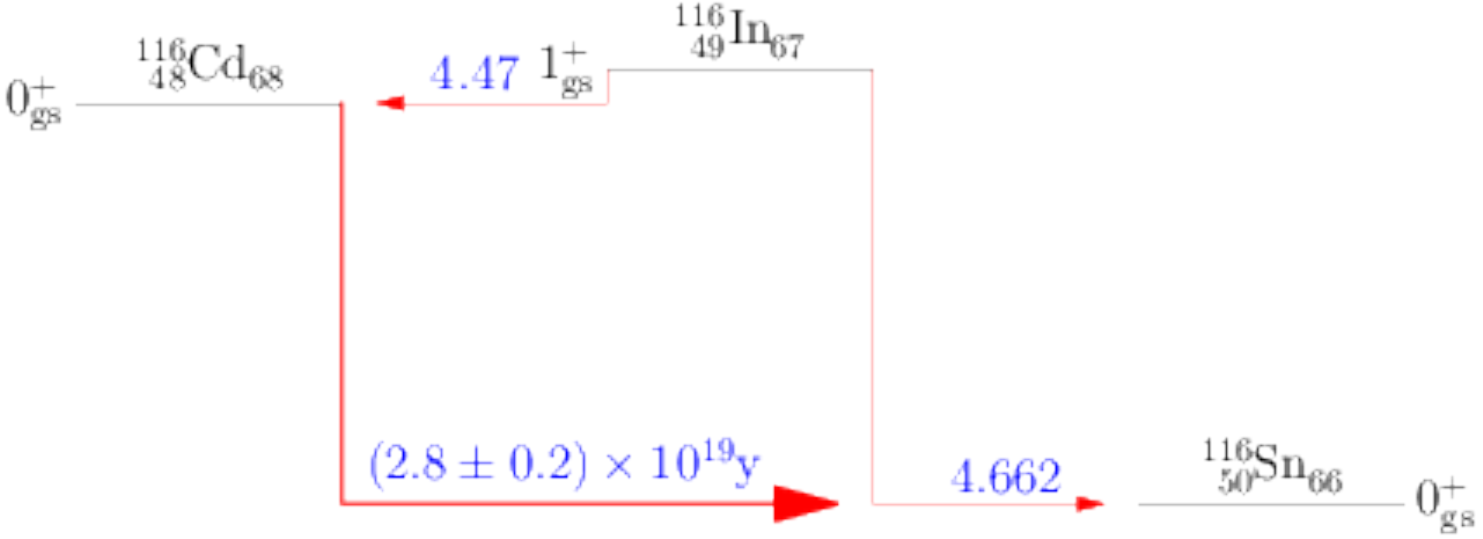}
\includegraphics[width=0.60\columnwidth]{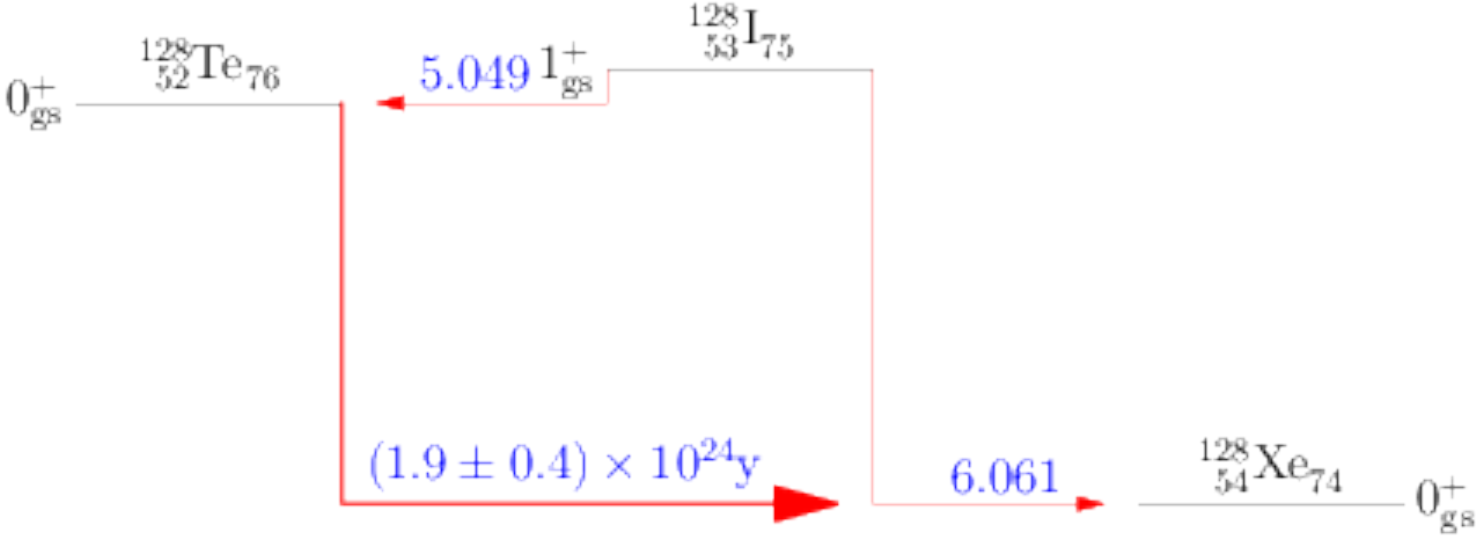}
\caption{Schematic plot of $\beta$ and $\beta\beta$ decays in the three isobaric 
chains ($A=100,116,128$) with data on both decay modes. Given are the 
$2\nu\beta\beta$ half-life and $\log ft$ values of the left- and right-branch
$\beta$ decays.}
\label{fig:bb-2vbb}
\end{figure}

The simultaneous analysis of both $\beta$ and $2\nu\beta\beta$ decays opens up
new vistas in attempts to pin down the effective value of the weak axial-vector
coupling strength. Indeed, analysis of these two decay modes is possible for
few nuclear systems where both the $\beta$-decay data \cite{ENSDF} and 
$2\nu\beta\beta$-decay data \cite{Barabash2013,Barabash2010} are available.
The involved transitions, with the available data, are depicted schematically 
in Fig.~\ref{fig:bb-2vbb}. The aim in using the three pieces of data available
for the three isobaric systems is to gain information on the effective value of
$g_{\rm A}$ and the value of the particle-particle interaction parameter $g_{\rm pp}$
of pnQRPA in the mass regions $A=100,116,128$.

The first work to address the quenching in both $\beta$ and $2\nu\beta\beta$
decays was \cite{Faessler2008} where both the beta-decay and 
$2\nu\beta\beta$ decay data were analyzed for the $A=100,116$ systems
in the framework of the pnQRPA using the method of least squares to fit the pair 
($g_{\rm pp}$,$g_{\rm A}$) to the available three pieces of data, namely the
$\log ft$ values of the left- and right-branch $\beta$ decays, and the 
$2\nu\beta\beta$ half-life (see Fig.~\ref{fig:bb-2vbb}).
Realistic model spaces (large and small basis) and a phenomenologically renormalized 
microscopic G-matrix-based Hamiltonian was used in the investigations. 
In \cite{Faessler2008} the best fit values $g_{\rm A}^{\rm eff}=0.74$ ($A=100$) and 
$g_{\rm A}^{\rm eff}=0.84$ ($A=116$) were obtained in the large single-particle 
model space. Furthermore, it is interesting to note that in the first 
version \cite{Faessler-draft} of the paper \cite{Faessler2008} also
results for the $A=128$ system were included. There the result 
$g_{\rm A}^{\rm eff}=0.39$ ($A=128$) was quoted. These values of $g_{\rm A}^{\rm eff}$
have been quoted in Table~\ref{tab:ga-ranges} and plotted in
Fig.~\ref{fig:gA-plus-curves} in Sec.~\ref{subsec:IBM-bb}.

In \cite{Suhonen2013c} and \cite{Suhonen2014} realistic single-particle bases and
a G-matrix-based microscopic interaction was used to analyze the $A=100,116,128$ 
systems of $\beta$ and $\beta\beta$ decays. A slightly different approach to the
one of \cite{Faessler2008,Faessler-draft} was adopted:
by taking the left and right branches of $\beta$-decay data of Fig.~\ref{fig:bb-2vbb}
one can fix the pair ($g_{\rm pp}$,$g_{\rm A}(\beta)$) by reproducing the available
$\log ft$ values. By using the just determined value of $g_{\rm pp}$ one can 
compute the $2\nu\beta\beta$ NME and half-life and compare with the experimental
half-life. This comparison produces a new value of $g_{\rm A}^{\rm eff}$, which can
be denoted as $g_{\rm A}(\beta\beta)$. In an ideal case the two effective values
of $g_{\rm A}$, namely $g_{\rm A}(\beta)$ and $g_{\rm A}(\beta\beta)$, are the same but the
over-constrained nature of the problem tends to yield different values to these
parameters. The thus obtained values of $g_{\rm A}(\beta)$ and $g_{\rm A}(\beta\beta)$
are quoted in Table~\ref{tab:ga-ranges} and plotted in
Fig.~\ref{fig:gA-plus-curves} in Sec.~\ref{subsec:IBM-bb}.

\subsection{Interacting shell model and interacting boson model \label{subsec:IBM-bb}}

\begin{figure}[htbp!]
\centering
\includegraphics[width=0.6\columnwidth]{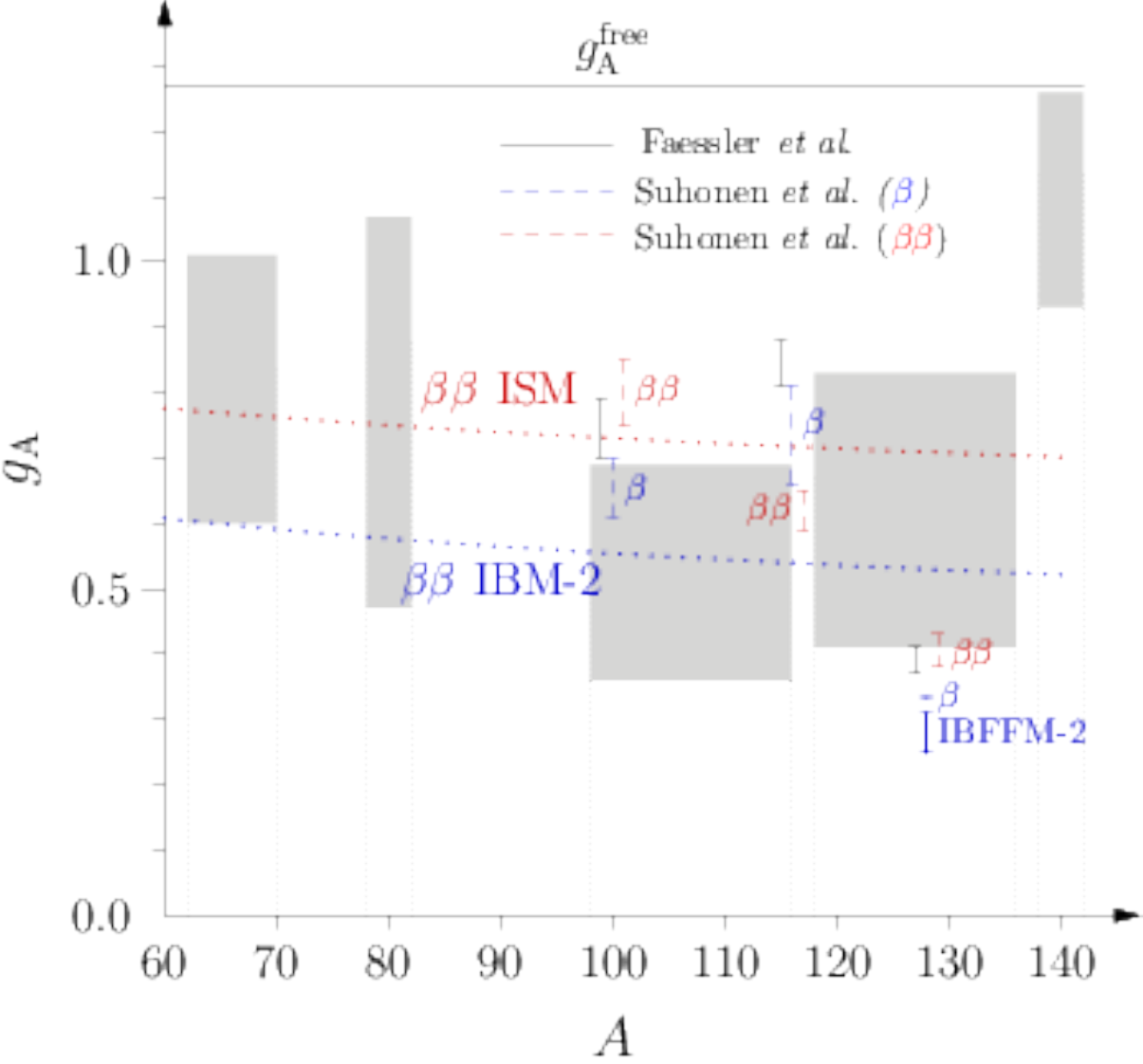}
\caption{Sketch of the effective values of $g_{\rm A}$ taken from 
Table~\ref{tab:ga-ranges} against the light-hatched ranges of $g^{\rm eff}_{\rm A}$ 
in the 5 mass regions of Fig.~\ref{fig:gA-ranges1} (the combined pnQRPA results). 
The curves are from the $2\nu\beta\beta$ analysis
of \cite{Barea2013} [curves (\ref{eq:curves})] and the vertical segments display
the results of the combined $\beta$ and $2\nu\beta\beta$ analyses of
\cite{Faessler2008,Faessler-draft} (solid line) and \cite{Suhonen2014}
(dashed lines). Also the (combined) IBFFM-2 result of \cite{Yoshida2013} is depicted.}
\label{fig:gA-plus-curves}
\end{figure}

A monotonic behavior of $g_{\rm A}(\beta\beta)$ was parametrized in 
\cite{Barea2013} by analyzing the magnitudes of $2\nu\beta\beta$ NMEs 
produced by the microscopic interacting boson model (IBM-2) \cite{Iachello1987} 
and the interacting shell model (ISM). 
In this study the obtained $g_{\rm A}$-versus-$A$ slopes were very flat, having
the analytic expressions
\begin{equation}
\label{eq:curves}
 g_{\rm A}^{\rm eff}(\textrm{IBM-2}) = 1.269A^{-0.18} \ ; \quad 
g_{\rm A}^{\rm eff}(\textrm{ISM}) = 1.269A^{-0.12} \,.
\end{equation}
These curves have been plotted in Fig.~\ref{fig:gA-plus-curves} together
with the results obtained in the pnQRPA analyses of $\beta\beta$ decays in
Sec.~\ref{subsec:pnQRPA-bb}.
The results of these analyses, together with the the original
numbers for $g_{\rm A}(\beta\beta)$ produced in the IBM-2 calculations of
\cite{Barea2013} are quoted in Table~\ref{tab:ga-ranges}. The IBM-2 numbers are
given in the last column of the table and the
first two lines refer to the use of the single-state
dominance (SSD) hypothesis in the IBM-2 calculations. Based on the
analysis in \cite{Suhonen2014} this assumption is approximately valid since 
the magnitudes of the first $1^+$ contribution and
the final $2\nu\beta\beta$ NME are practically the same 
for the decays of $^{100}$Mo and $^{116}$Cd.
The last number of the IBM-2 column in Table~\ref{tab:ga-ranges} refers to
the assumption of closure approximation (CA) in the IBM-2 calculation. It is
well established \cite{REPORT,Tomoda1991}
that such an approximation does not work for the
$2\nu\beta\beta$ decays and thus this number could be dubious. Indeed,
in a later publication \cite{Yoshida2013} a more consistent theoretical framework was 
used (the interacting boson-fermion-fermion model, IBFFM-2 \cite{Brant1988})
and in the case of $A=128$ values of $g_{\rm A}$ were obtained that differ
notably from the ones obtained in \cite{Barea2013}. The IBFFM-2 numbers,
based on analysis of both the $\beta$ and $2\nu\beta\beta$ decay, are 
presented in columns 5 and 6 of Table~\ref{tab:ga-ranges}.
One can see that the IBFFM-2 values of $g_{\rm A}(\beta)$ and 
$g_{\rm A}(\beta\beta)$ are quite close to those of the pnQRPA-based calculations.
The combined $\beta$ and $\beta\beta$ results of \cite{Yoshida2013} have also been
depicted in Fig.~\ref{fig:gA-plus-curves}.

Recent ISM calculations \cite{Neacsu2015,Horoi2016} for the $2\nu\beta\beta$ 
NMEs of $^{130}$Te and $^{136}$Xe, and a subsequent comparison with the experimental
NMEs (updated comparison performed in \cite{Horoi2016}) suggest a mild 
quenching and a rather large value of for the effective coupling strength:
\begin{align}
\label{eq:Horoi}
g^{\rm eff}_{\rm A}(A=130-136)=0.94 \,.
\end{align}
This result was already discussed in Sec.~\ref{subsec:ISM} and it was included in
Table~\ref{tab:ISM} of that section. The result (\ref{eq:Horoi}) was also
illustrated in Fig.~\ref{fig:gA-ranges2} of Sec.~\ref{subsec:QRPA}.

\begin{table}
\centering
  \begin{tabular}{ccccccc}
\toprule
 $A$ & \multicolumn{3}{c}{pnQRPA} & 
\multicolumn{2}{c}{IBFFM-2 \cite{Yoshida2013}} & IBM-2 \cite{Barea2013} \\
   \cmidrule(lr{0.75em}){2-4}\cmidrule(lr{0.75em}){5-6}
    & $g_{\rm A}(\beta + \beta\beta)$ \cite{Faessler2008,Faessler-draft} & $g_{\rm A}(\beta)$ 
\cite{Suhonen2014} & $g_{\rm A}(\beta\beta)$ \cite{Suhonen2014} & 
$g_{\rm A}(\beta)$ & $g_{\rm A}(\beta\beta)$ & $g_{\rm A}(\beta\beta)$ \\ 
\midrule
  $100$ & $0.70-0.79$ & $0.61-0.70$ & $0.75-0.85$ & - & - & 0.46(1) [SSD] \\
  $116$ & $0.81-0.88$ & $0.66-0.81$ & $0.59-0.65$ & - & - & 0.41(1) [SSD] \\
  $128$ & $0.37-0.41$ & $0.330-0.335$ & $0.38-0.43$ & $0.25-0.31$ & 
0.293 & 0.55(3) [CA] \\
\bottomrule
  \end{tabular}
  \caption{Extracted values of $g_{\rm A}$ for three isobaric chains hosting a
$2\nu\beta\beta$ transition. The values are obtained in the pnQRPA, in the 
IBFFM-2, and in the IBM-2 theory frameworks. In the last column 
SSD denotes single-state dominance, CA denotes closure approximation, and the
errors in parentheses stem from the error limits of the adopted data. The
intervals in column 2 correspond to the $1\sigma$ errors quoted in
\cite{Faessler2008,Faessler-draft} and the ranges in the third and fourth columns
stem from the experimental errors of the adopted data. The range in the fifth
column stems from the different obtained values for the $\beta^-$ and
$\beta^+$/EC branches, respectively. \label{tab:ga-ranges} } 
\end{table}

From Fig.~\ref{fig:gA-plus-curves} one sees that the results of the pnQRPA analyses of
Faessler \textit{et al.} \cite{Faessler2008,Faessler-draft} and Suhonen
\textit{et al.} \cite{Suhonen2014} are consistent with each other, and are in
agreement with the $2\nu\beta\beta$ results of the ISM (the upper dotted curve in 
Fig.~\ref{fig:gA-plus-curves}) for the masses $A=100,116$. 
For the mass $A=128$ both the pnQRPA and IBFFM-2 results deviate strongly from 
the ISM result, coming closer to the IBM-2 results. 
Both the ISM and IBM-2 curves follow, in average, the trend of the pnQRPA results of the
$\beta$-decay analyses of Sec.~\ref{subsec:QRPA} (the light-hatched regions of 
Fig.~\ref{fig:gA-plus-curves}), except for the very heavy masses, $A\ge 138$.
The differences in the results of the $\beta$-decay and $2\nu\beta\beta$-decay analyses 
are not drastic but they still exist. The differences may stem from the fact that
not only one $1^+$ state takes part in most of the $2\nu\beta\beta$ decays. 
Contributions from the $1^+$ states above the lowest $1^+$ state, sometimes the ground 
state, interfere with each other and the contribution coming from the lowest one. These
interferences have been discussed, e.g., in \cite{Civitarese1998,Civitarese1999}.

\section{Spin-multipole strength functions, giant resonances and the renormalization of 
$g_{\rm A}$ \label{sec:spin-multipole}}

\begin{figure}[htbp!]
\centering
\includegraphics[width=0.8\columnwidth]{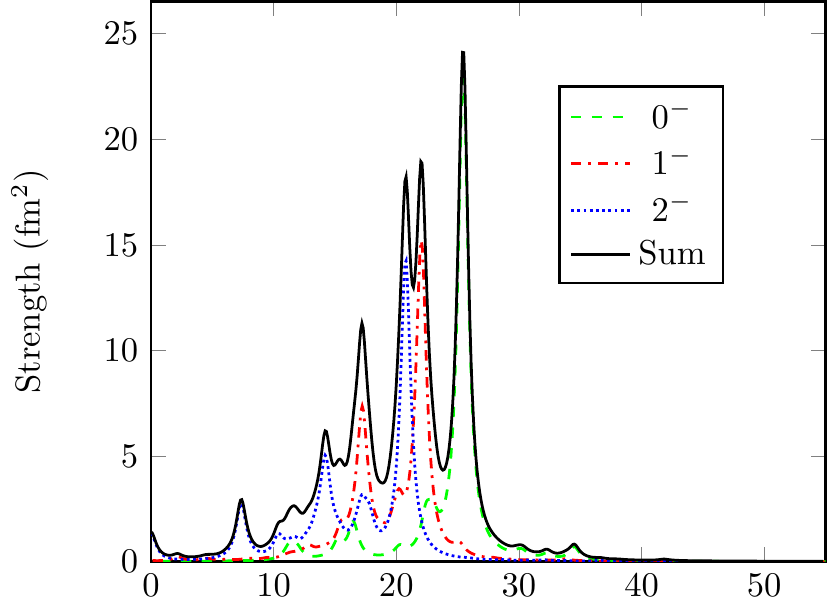}
\caption{Isovector $\beta^-$ spin-dipole strengths for $0^-$, $1^-$, and $2^-$ states
in $^{76}$As as excited from the $0^+$ ground state of $^{76}$Ge. The solid envelope
curve represents the sum of the three multipole distributions. The excitation energy
is relative to the $2^-$ ground state of $^{76}$As.}
\label{fig:spin-dipole-minus}
\end{figure}

As discussed in Secs.~\ref{sec:GT} and \ref{sec:F-u}
the low-lying Gamow-Teller and higher isovector spin-multipole strengths, 
in particular the $2^-$ strength, are quenched against nuclear-model calculations. 
The low-lying spin-multipole strength represents the low-energy tail of the
corresponding spin-multipole giant resonance (SMGR). Usually only the low-energy
part, with excitation energies $E\le 5\,\mathrm{MeV}$, of the spin-multipole 
\emph{strength function} is experimentally known, and only for low multipoles,
like for Gamow-Teller strength \cite{Frekers2013} or spin-dipole $2^-$ strength
\cite{Frekers2017}. These strength functions have been measured using charge-exchange
reactions at low momentum transfers, like the (p,n), ($^{3}$He,$t$), (n,p), and
($d$,$^{2}$He) reactions \cite{Goodman1980,Taddeucci1987,Akimune1997}. As an
example, in Fig.~\ref{fig:spin-dipole-minus} is shown the strength for 
isovector spin-dipole excitations from the $0^+$ ground state of $^{76}$Ge to
the $0^-$, $1^-$, and $2^-$ states in $^{76}$As. The centroid energies of the 
corresponding giant resonances are roughly 24 MeV ($0^-$), 20 MeV ($1^-$), 
and 18 MeV ($2^-$) \cite{Jokiniemi2017}.

The measured strength functions for Gamow-Teller transitions
can be pestered by the isovector spin monopole
(IVSM) contributions at high energies \cite{Bes2012,Civitarese2014}. The location
(\ref{eq:GTGR}) of the Gamow-Teller giant resonance, GTGR, dictates partly the amount of
strength remaining at low energies [at zero-momentum limit $q\to 0$ in 
(\ref{eq:dipole})], and thus the quenching of the axial-vector 
coupling strength $g_{\rm A}$ in model calculations \cite{Delion2017}. 
These calculations have mostly
been performed in the framework of the pnQRPA which represents well the centroids
of the strong Gamow-Teller peaks, and extensions of the pnQRPA to two- plus
four-quasiparticle models, like the proton-neutron microscopic anharmonic vibrator 
approach (pnMAVA) \cite{Kotila2009,Kotila2010}, does not alter the picture very much.

Like in the case of the Gamow-Teller strength, also the location of the SMGRs 
affect the low-lying strength of, e.g. isovector spin-dipole ($J^{\pi}=0^-,1^-,2^-$,
see Fig.~\ref{fig:spin-dipole-minus}) and spin-quadrupole
($J^{\pi}=1^+,2^+,3^+$) excitations \cite{Jokiniemi2017,Auerbach1984}. This is why
measurements of such giant resonances could help in solving the quenching problems
associated to $g_{\rm A}$ at low energies.

\section{Effective $g_{\rm A}$ from nuclear muon capture \label{sec:mucapture}}

\begin{figure}[htbp!]
\centering
\includegraphics[width=0.6\columnwidth]{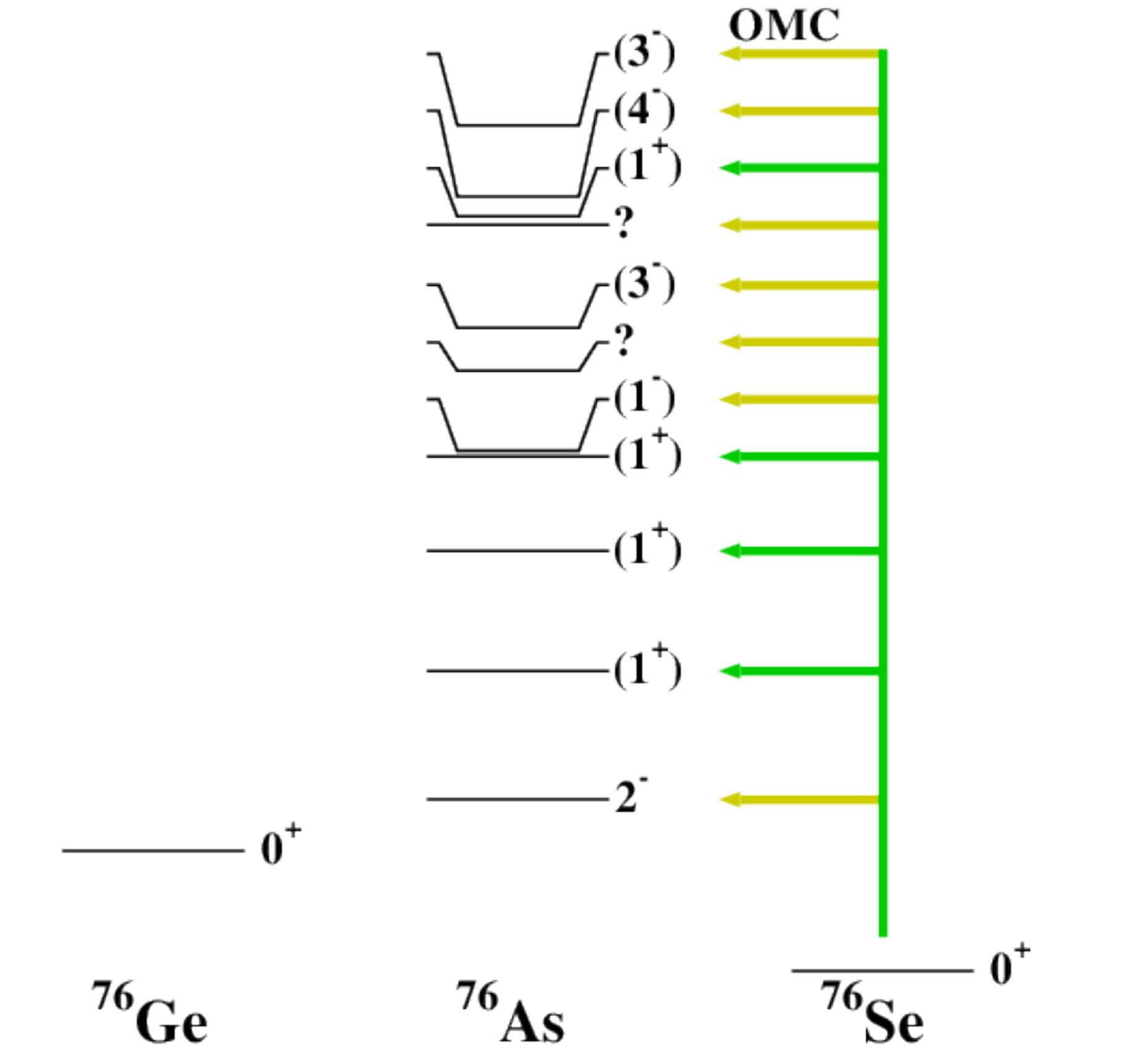}
\caption{Double $\beta$ triplet $^{76}$Ge, $^{76}$As and $^{76}$Se with the nuclear
states shown in the intermediate nucleus $^{76}$As. These states are populated by 
the ordinary muon capture (OMC) transitions from $^{76}$Se.}
\label{fig:mucapt}
\end{figure}

The (ordinary, non-radiative) nuclear muon capture is a transition between nuclear isobars such
that
\begin{align}
\label{eq:mucapt}
\mu^- + (A,Z,N) \to \nu_{\mu} + (A,Z-1,N+1) \,,
\end{align}
where a negative muon is captured from an atomic $s$ orbital and as a result
the nuclear charge decreases by one unit and a muon neutrino is emitted. The process is 
schematically depicted in Fig.~\ref{fig:mucapt} for the capture on $^{76}$Se, with the
final states in $^{76}$As. Here also the nucleus $^{76}$Ge is depicted since it 
$\beta\beta$ decays to $^{76}$Se. Properties of the $\mu$-mesonic atoms have been treated
theoretically in \cite{Ford1962} and experimentally in e.g. 
\cite{Miller1972,Suzuki1987,Gorringe1994,Gorringe1999}. 
Due to the heavy mass of the muon ($m_{\mu}=105$ MeV) the
process has a momentum exchange of the order of $q\sim 100$ MeV and is thus similar
to the neutrinoless $\beta\beta$ decay where a Majorana neutrino of a similar momentum is
exchanged. This means that contrary to $\beta$ decays all the terms of the hadronic 
current (\ref{eq:h-current-eff}) are activated and that the contributions from the 
forbidden transitions $J>1$ are not suppressed relative to the allowed ones, just
like in the case of $0\nu\beta\beta$ decays. Since the induced
currents in (\ref{eq:h-current-eff}) are activated the theoretical expressions for the
individual capture transitions are rather complex 
\cite{Primakoff1959,Morita1960,Gillet1965,Partha1978,Partha1981} whereas the total capture
rates are much easier to calculate \cite{Luyten1963,Luyten1965}. 

Most of the theoretical attempts to describe the muon capture to individual nuclear 
states have concentrated on very light nuclei, $A\le 20$ 
\cite{Miller1972,Gillet1965,Duck1962,Gmitro1990,Jonkmans1996,Siiskonen1999a,Govaerts2000} or to the mass region $A=23-40$
\cite{Gorringe1994,Gorringe1999,Partha1978,Partha1981,Kolbe1994,Brudanin1995,Johnson1996,Siiskonen1998,Siiskonen1999b,Siiskonen1999c,Kort2000}.
Also studies in the $1s-0d$ and $1p-0f$ shells have been performed \cite{Eramzhyan1998,Kort2004}.
Heavier nuclei, involved in $\beta\beta$-decays, have been treated in \cite{Kort2002,Kort2003}.
Interestingly enough, the muon-capture transitions can be used to probe the right-leg
virtual transitions of $0\nu\beta\beta$ decays \cite{Kort2004,Kort2002,Kort2003}, but they
can also give information on the in-medium renormalization of the axial current 
(\ref{eq:A-current}) in the form
of an effective $g_{\rm A}$ \cite{Gorringe1999,Kolbe1994,Johnson1996} and an effective induced 
pseudoscalar coupling $g_{\rm P}$ (in fact the ratio $g_{\rm P}/g_{\rm A}$)
\cite{Gorringe1994,Gillet1965,Partha1978,Partha1981,Gmitro1990,Jonkmans1996,Kolbe1994,Brudanin1995,Johnson1996,Siiskonen1999b,Siiskonen1999c}
at high (100 MeV) momentum transfers, relevant for studies of the virtual transitions
of the $0\nu\beta\beta$ decays. A recent review 
on the renormalization of $g_{\rm P}$ is given in \cite{Gorringe2004}.

More experimental data on partial muon-capture rates to nuclear states are needed for 
heavier nuclei in order to access the renormalization of $g_{\rm A}$ and $g_{\rm P}$ for 
momentum transfers of interest for the $0\nu\beta\beta$ decay. The present (see. e.g.
\cite{Cook2016}) and future experimental muon-beam installations should help 
solve this problem.

\section{Conclusions \label{sec:conclusions}}

The quenching of the weak axial-vector coupling strength, $g_{\rm A}$, is an important
issue considering its impact on the detectability of the neutrinoless double beta
decay. The quenching appeared in old shell-model calculations as a way to reconcile
the measured and calculated $\beta$-decay rates and strength functions. Later
such quenching was studied in other nuclear-model frameworks, like the quasiparticle
random-phase approximation and the interacting boson model. The quenching of
$g_{\rm A}$ can be observed in allowed Gamow-Teller decays as also in forbidden $\beta$
decays. The origins of the quenching seem to be both the nuclear-medium effects and
deficiencies in the nuclear many-body approaches, but a clean separation of these
two aspects is formidably difficult. Different quenchings have been obtained in 
different calculations, based on different many-body frameworks. There is not yet
a coherent approach to the quenching problem and many different separate studies have been
performed. However, when analyzed closer, the obtained quenching of $g_{\rm A}$ is 
surprisingly similar in  different many-body schemes for different physical processes
(e.g. for Gamow-Teller $\beta$ transitions, for electron spectra of forbidden nonunique 
$\beta$ decays) in the mass range from light to medium-heavy nuclei. 

Different ways to access the quenching have been proposed, like comparisons with
Gamow-Teller $\beta$-decay and two-neutrino double-$\beta$-decay data. In a promising
new method, the spectrum-shape method, the comparison of the computed and
measured electron spectra of high-forbidden non-unique $\beta$ decays is proposed. The
robustness of the method is based on the observations that the
computed spectra seem to be relatively insensitive to the adopted mean-field and 
nuclear models. Measurements of such electron spectra for certain key transitions
are encouraged. Also the relation of the quenching problem to the low-lying
strength for Gamow-Teller and higher isovector spin-multiple excitations is worth
stressing, as also the relation to the corresponding giant resonances, accessible in
present and future charge-exchange-reaction experiments. The development of high-intensity
muon beams makes measurements of nuclear muon-capture rates easier and enables access to
the renormalization of the axial current at momentum exchanges relevant for the
neutrinoless $\beta\beta$ decay.

\section*{Acknowledgements}
This work was supported by the Academy of Finland under the Finnish 
Center of Excellence Program 2012-2017 (Nuclear and Accelerator Based 
Program at JYFL)

%\section*{References}

\end{document}